\documentclass[useAMS,usenatbib]{mn2e}

\usepackage{graphicx}
\usepackage{txfonts}
\usepackage{natbib}
\usepackage{longtable,lscape}
\usepackage{color}

\newcommand\pasa{PASA}
\newcommand\aj{AJ}
\newcommand\apj{ApJ}
\newcommand\apjs{ApJS}

\newcommand\aap{A\&A}
\newcommand\mnras{MNRAS}
\newcommand\apjl{ApJ}
\newcommand\pasp{PASP}
\newcommand\nat{Nature}
\newcommand\aaps{A\&AS}
\newcommand\araa{ARA\&A}

\def\teff{\mbox{T$_{\rm eff}$}}
\def\logg{\mbox{log~{\it g}}}
\def\vmicro{\mbox{$\xi_{\rm t}$}}
\def\kmsec{\mbox{km~s$^{\rm -1}$}}

\def\deg{\mbox{$\rm {^{\circ}}$}}

\title[]{The halo+cluster system of the Galactic globular cluster NGC\,1851.\thanks{Based on data collected at the European Southern Observatory with the FLAMES/GIRAFFE spectrograph, under the programs 088.A-9012 and 084.D-0470. Based also on observations made with MPG 2.2m telescope at the La Silla Paranal Observatory under the programs 085.A9028(A) and 088.A9012(A).}
}

\author[A.\, F.\, Marino et al.]
{A.\, F.\, Marino$^{1}$\thanks{E-mail:amarino@mso.anu.edu.au},
A.\,P.\, Milone$^{1}$,
D. Yong$^{1}$,
A. Dotter$^{1}$,
G. Da Costa$^{1}$,
M. Asplund$^{1}$,
\newauthor
H. Jerjen$^{1}$,
D. Mackey$^{1}$,
J. Norris$^{1}$,
S. Cassisi$^{2}$,
L. Sbordone$^{3}$,
P.B. Stetson$^{4}$,
A. Weiss$^{5}$,
\newauthor
A. Aparicio$^{6,7}$,
L.\,R.\, Bedin$^{8}$,
K. Lind$^{9}$,
M. Monelli$^{6,7}$,
G. Piotto$^{8,10}$,
R. Angeloni$^{11,12}$,
\newauthor
R. Buonanno$^{2}$
\\
$^{1}$Research School of Astronomy \& Astrophysics, Australian National University, Mt Stromlo Observatory, via Cotter Rd, Weston, ACT 2611, Australia \\
$^{2}$INAF-Osservatorio Astronomico di Teramo, via M.\, Maggini, 64100 Teramo, Italy \\
$^{3}$Zentrum f\"ur Astronomie der Universit\"at Heidelberg, Landessternwarte, K\"onigstuhl 12, 69117 Heidelberg, Germany\\
$^{4}$Herzberg Institute of Astrophysics, National Research Council Canada, 5071 West Saanich Road, Victoria, BC V9E 2E7\\
$^{5}$Max-Planck-Institut f\"ur Astrophysik Karl-Schwarzschild-Str.\, 1 85741 Garching bei M\"unchen Germany \\              
$^{6}$Instituto de Astrof\'isica de Canarias, La Laguna, Tenerife, Spain\\
$^{7}$Departamento de Astrof\'isica, Universidad de La Laguna, Tenerife, Spain\\
$^{8}$INAF-Osservatorio Astronomico di Padova, Vicolo dell'Osservatorio 5, Padova I-35122, Italy\\
$^{9}$Institute of Astronomy, University of Cambridge, Madingley Road, Cambridge, CB3 0HA, UK \\
$^{10}$Dipartimento di Fisica e Astronomia `Galileo Galilei', Universit\`a di Padova, Vicolo dell'Osservatorio 3, Padova, I-35122, Padova, Italy.\\
$^{11}$Department of Electrical Engineering, Center for Astro-Engineering, Pontificia Universidad Cat\'{o}lica de Chile, Av. Vicu\~{n}a Mackenna 4860, 782-0436 Macul, Santiago, Chile\\
$^{12}$The Milky Way Millennium Nucleus, Av. Vicu\~{n}a Mackenna 4860, 782-0436 Macul, Santiago, Chile\\
}

\begin{document}

\date{Draft Version Maj, 2013}

\pagerange{\pageref{firstpage}--\pageref{lastpage}} \pubyear{2013}

\maketitle

\label{firstpage}

\begin{abstract}  
NGC\,1851 is surrounded by a stellar 
component that extends more than ten times
beyond the tidal radius.
Although the nature of this stellar structure is not known, it has been suggested 
to be
a sparse halo of stars or
associated with a stellar stream.
We analyse the nature of this intriguing stellar component
surrounding NGC\,1851 by investigating its
radial velocities and
chemical composition,
in particular in comparison
with those of the central cluster analysed in a homogeneous manner.
In total we observed 23 stars in the halo with radial velocities consistent with NGC\,1851, and for 15 of them we infer [Fe/H] abundances. 
Our results show that: 
{\it (i)} stars dynamically linked to NGC\,1851 are present at least up to
$\sim$2.5 tidal radii, supporting the presence of a halo of stars surrounding the cluster;
{\it (ii)} apart from the NGC\,1851 radial velocity-like stars, our observed velocity distribution agrees with that expected from Galactic models, suggesting that no other sub-structure (such as a stream) at different radial velocities is present in our field;
{\it (iii)} the chemical abundances for the
$s$-process elements Sr and Ba are consistent with the $s$-normal stars observed in NGC\,1851;
{\it (iv)} all halo stars have metallicities, and abundances for the other studied elements Ca, Mg and Cr, consistent with those exhibited by the cluster.
The complexity of the whole NGC\,1851 cluster+halo system may agree
with the scenario of a
tidally-disrupted dwarf galaxy in which NGC\,1851 was originally embedded. 
\end{abstract}


\section{Introduction}\label{sec:intro}

Recent discoveries on multiple stellar populations in globular
  clusters (GCs) have revealed that some of these old stellar systems
  show chemical inhomogeneities, not just in the light elements involved in the
  hot H-burning, but also in heavier elements and in
  the overall metallicity. 
To explain the large metallicity dispersion in $\omega$~Centauri it has
been suggested that this GC is the remnant of a dwarf galaxy disrupted
through tidal interactions with the Milky Way, rather than a true GC
(e.g. Norris et al.\,1996; Bekki \& Freeman 2003; Bekki \& Norris
2006). 
In this scenario, $\omega$~Centauri would be the dense nucleus of a
dwarf galaxy, which was cannibalised by the Milky Way. 
The recent discoveries of more $\omega$~Centauri-like GCs, with
internal metallicity variations (e.g. Marino et al.\,2009; Da Costa et
al.\,2009; 2014; Yong et al.\,2014), support the
hypothesis that at least these ``anomalous'' GCs were as massive as
small galaxies able of retaining fast supernovae ejecta. 
The relatively high fraction of metal-richer stars in these GCs also supports the idea that they were more massive. 
The possible GC-dwarf galaxy connection may have  
consequences for near-field cosmology and the hierarchical assembly of
our Galaxy. At least the GCs with internal variations in metallicity
may contribute to the inventory of original satellites, along with
existing dwarf spheroidals and ultra-faints, to alleviate 
the 'missing satellites' problem of the $\Lambda$-Cold Dark Matter
scenario (e.g., Kauffmann et al. 1993, Klypin et
al. 1999, Moore et al. 1999).

The hypothesis that at least some GCs may constitute the surviving nuclei of tidally disrupted dwarf galaxies implies that the Milky Way has stripped the less bound external stars from these systems during successive passages through the Galactic potential, leaving only their compact nuclei.
A snapshot of this phenomenon may be M\,54, as it shows an intrinsic Fe dispersion (Bellazzini et al. 2008, Carretta et al. 2010) and lies at the centre
of the Sagittarius dwarf galaxy that is being tidally disrupted by the Milky Way (Ibata et al.\,1994).

While nearly all the Galactic GCs have chemical variations in the
light elements involved in the hot H-burning (such as C, N, O, Na, see e.g., Kraft 1994; Gratton et al. 2004),
only a few of them are known to possess spreads in Fe and
$s$-element abundances ($\omega$~Cen, M22, NGC~1851, M2, NGC\,5824, NGC\,3201, 
e.g., Smith et al. 2000; Marino et al. 2009, 2011a; Yong \& Grundahl 2008; 
Lardo et al.\ 2013; Yong et al.\ in prep.; Da Costa et al.\ 2009, 2013; Simmerer et al.\ 2013). 
It is intriguing that the GCs with metallicity and $s$-element variations are generally the more massive ones. 
The complexity of the multiple stellar populations in these
  objects is puzzling and we do not have yet a coherent picture to
  explain the formation of their different generations of stars.

In this context, NGC~1851 is one of the most intriguing targets. 
Much effort has been dedicated
to this GC after the discovery of a prominent bimodal sub-giant branch (SGB, Milone et al.\ 2008). 
The formation scenario for these two SGB components is still under debate. Observational constraints for the sequence of events that led to the formation of these stellar groups can be inferred from their chemical compositions and their radial distributions in the cluster. The radial profile of the two SGB components has been found to not change significantly within 8\arcmin\ from the cluster center (Milone et al.\ 2009). 
In contrast, 
Zoccali et al.\ (2009) did not observe the faint SGB out to $\sim$2.4\arcmin\ in the southwest quadrant. 

By analysing high-resolution UVES spectra for NGC~1851 red giants (RGB), Yong
\& Grundahl (2008) discovered that this cluster hosts two groups of
stars with different content of $s$-process elements 
(see also Villanova et al. 2010; Carretta et al. 2010). These two
stellar groups have been found to define two RGB sequences 
following on from the two different SGBs 
(Han et al. 2009; Lee et al. 2009; Lardo et al. 2012).
The photometric split on the SGB observed by Milone et al.\ (2008) has been theoretically interpreted as due to either a difference in age of $\sim$1~Gyr or  to a possible dichotomy in the C+N+O (Cassisi et al.\ 2008; Ventura et al.\ 2009; Sbordone et al.\ 2011).  The latter scenario has been supported by spectroscopic studies showing
that $s$-enriched stars are also enhanced in their overall C+N+O
content 
(Yong et al.\ 2009; and in prep.).
We note, however, that Villanova et al.\ (2010) did not find an abundance spread
for C+N+O.
To date, variations in the overall C+N+O have been found also in other
two ``anomalous'' GCs, 
M\,22 (Marino et al.\ 2011b, 2012a, Alves Brito
et al.\ 2012) and $\omega$~Centauri (Marino et al.\ 2012b).

Interestingly, NGC\,1851 is
surrounded by a diffuse stellar halo with a radius of more than 250 pc 
(67\arcmin\ from the cluster center) and
a mass of about 0.1\% of the dynamical mass of NGC~1851 (Olszewski
et al. 2009).
The extension of this stellar structure is far beyond the tidal radius 
predicted by 
the King model (King 1962), that is the distance from the cluster center 
where cluster stars are expected to drastically disappear 
due to tidal interactions. 

The origin of this halo remains unknown, although various hypotheses exist.
It could be the consequence of
isolated cluster evaporation through tidal or disk shocking that may have originated a stellar tail. Such
processes are believed responsible for the streams observed in several GCs, such as
Pal\,5 (Odenkirchen et al. 2001; Koch et al. 2004).
Observations of NGC\,1851 are contradictory: 
while photometrically there is no evidence for tidal streams (Olszewski et al.\ 2009),
the presence of a possible tail of stars with radial velocity around $\sim$150~\kmsec\
has been reported by Sollima et al.\ (2012).

Alternatively, the huge halo of NGC\,1851 could have formed from the destruction of a dwarf galaxy in
which the cluster may have once been embedded.
Bekki \& Yong (2012) outlined a possible self consistent and dynamically
plausible scenario for the formation of NGC\,1851's multiple populations
and its stellar halo.
In their scenario, two GCs in a dwarf galaxy merge (owing to the low
velocity dispersion of the host dwarf) and form a new nuclear star cluster
surrounded by field stars  of the host dwarf.
The host dwarf galaxy is stripped through tidal interaction with the
Milky Way leaving the stellar nucleus which is observed as NGC\,1851.
Thus, the two stellar populations in NGC~1851 originate in the two GCs
that merged to form the nucleus.
Bekki \& Yong (2012) predict that NGC~1851's stellar halo contains three
stellar populations: two from the original GCs that merged to form the
nucleus and the remaining population is from field stars surrounding
stellar nuclei.

In the present study we investigate the nature of this intriguing stellar system, the GC NGC\,1851 plus its halo, by deriving radial velocities and, for the first time, chemical abundances for the halo stars. 
The chemical and dynamical properties of the halo stars will be compared 
with the ones observed within the tidal radius of the cluster.

    \begin{centering}
    \begin{figure*}
     \includegraphics[width=16cm]{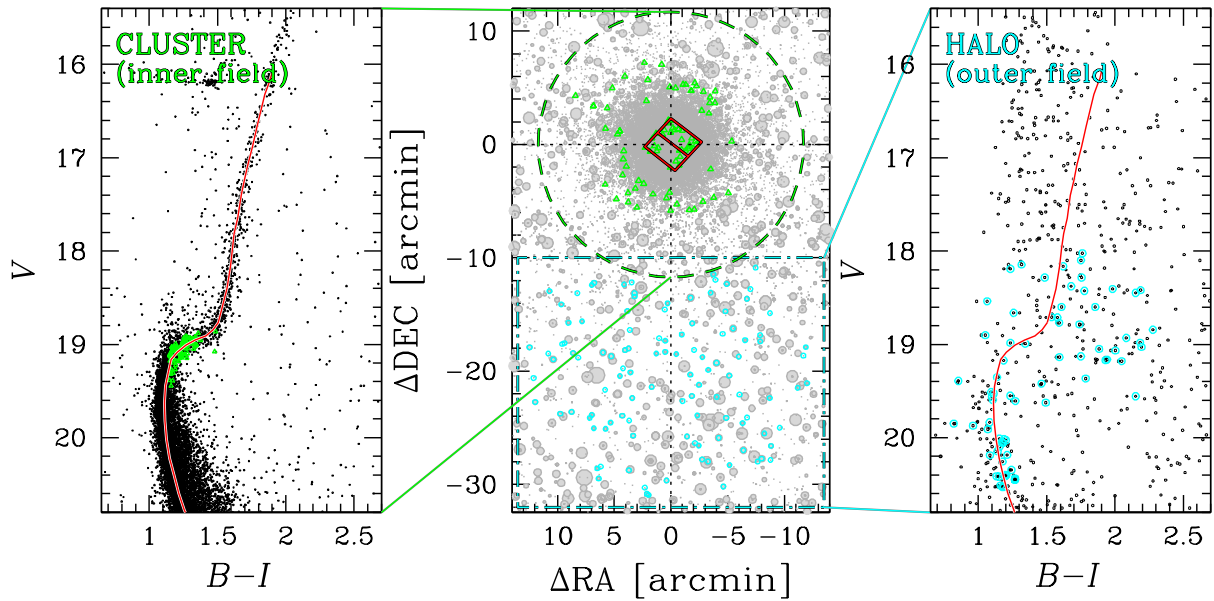}
     \caption{Location of the spectroscopic targets in RA and DEC (central panel). The NGC\,1851 stars in the cluster (inner field) have been plotted in green, and are within the tidal radius (green dashed circle). Stars in the halo have been plotted in cyan and are contained in the outer WFI field delimited by the cyan-dashed line. 
The footprint of the ACS/WFC field has been delimited in red. 
In the two sided panels we represent the $V$ versus $(B-I)$ CMDs for stars in the inner and outer fields, with the corresponding position of our spectroscopic targets. Red lines are the fiducials of the cluster CMD.}
     \label{fig:CMDall}
    \end{figure*}
    \end{centering}

\section{Data}\label{sec:data}

Basic information for NGC\,1851 can be found in Harris (1996, 2010
edition, and references therein).
Its distance from the Sun is $\sim$12.1~kpc.
The King tidal radius of NGC\,1851 was estimated to be 
6.7\arcmin\ by McLaughlin \& van der Marel (2005),
which is lower than the value of 11.7\arcmin\ in the previous
  Harris compilation derived by Trager et al.\,(1993). 
Given the large uncertainties related to the determination of the 
tidal radius, we assume in this work the conservative value of 11.7\arcmin.
Assuming the $\rm {M_{V}}$ from Harris (2010 edition) and a typical $\rm {M/L_{V}}=1.6$,
we get a mass of $M \approx 10^{5.45}~{M_{\odot}}$, which places NGC\,1851 among the most massive GCs.
We assumed for the center of the cluster the coordinates (RA; DEC)$_{\rm J2000}$=(05:14:06.76; $-$40:02:47.6) 
from Goldsbury et al.\,(2011).

\subsection{The photometric dataset\label{sec:phot_data}}

In this paper we used four distinct photometric data sets.
First, we used Stetson (2000) ground-based $B$, $V$, $R$ and $I$ photometry.
This photometric catalog has been 
established from about 550 images taken at different telescopes,
  i.e. the Max Planck 2.2m, the CTIO 4m, 1.5m, and 0.9m telescopes,
and the Dutch 0.9m telescope in La Silla.
These data have been reduced by using procedures for the photometric and astrometric data reduction described by Stetson\,(2005), and
have already been used in Milone et al.\,(2009). 
We refer the reader to the Sect.~2.3 of Milone et al.\,(2009) for further information on this dataset.
In the present work we have complemented the Stetson catalog with images collected with the Wide Field Imager (WFI) of the Max Planck 2.2m telescope at La Silla (WFI@2.2m) through the $U$ filter under the SUMO campaign. Details on the WFI data in the $U$ band, analysed here for the first time, are provided in Tab.~\ref{tab:photometry}. In summary $UBVRI$ photometry has been used for stars in a region between $\sim$10\arcmin\ to the south and 13\arcmin\ to the north, and between $\sim$15\arcmin\ to the west and 15\arcmin\ to the east, relatively to the center of NGC\,1851.

Secondly, to study stars in the halo of NGC\,1851, we collected $BVI$ images with WFI@2.2m of a field between $\sim$10\arcmin\ and 35\arcmin\ to the south of the cluster center. 
Photometry and astrometry for this dataset have been obtained by using the
program img2xym\_WFI and the procedure described by Anderson et
al.\,(2006).
Details of the WFI@2.2m dataset which has been analysed for the first time in this paper, are listed in Tab.~\ref{tab:photometry}.

Third and finally, to investigate the most crowded central regions we use {\it Hubble Space Telescope} ($HST$) 
F606W and F814W photometry obtained with the Wide Field Channel of
the Advanced Camera for Survey (WFC/ACS) and F275W photometry collected with the
Ultraviolet and Visual Channel of the Wide Field Camera 3 (UVIS/WFC3).
The WFC/ACS photometry comes from GO-10775 (PI.\,A.\,Sarajedini, see
Sarajedini et al.\,2007 and Anderson et al.\,2008) and is presented in Milone
et al.\,(2008). The UVIS/WFC3 photometry comes from GO\,12311 (PI.\,G.\,Piotto)
and is presented in Piotto et al.\,(2012, see also Bellini et al.\,2010 for
details on the data reduction). 

We adopt the following terminology for the various fields for which different photometries are available: (1) `central field' for the field of 3\arcmin$\times$3\arcmin\ covered by the $HST$ photometry; (2) `inner field' is all the field  inside the tidal radius, e.g.,  $HST$ photometry where available, Stetson+SUMO photometry otherwise; (3) `outer field' is the field outside the tidal radius.

Left and right panels of Fig.~\ref{fig:CMDall} show the inner and outer field $V$ versus $(B-I)$ CMDs. 
On each CMD we superimpose the fiducial line of the cluster. 
The inner and outer field CMDs have not been de-reddened. However, the reddening across the NGC\,1851 field of view is very low, $E(B-V)=0.02$, and the differential reddening should not be significant as it is much lower than internal uncertainties in the ground-based photometry. The reddening variations across a field of 2\deg$\times$2\deg\ around the cluster center predicted from the Schlegel et al.\ (1998) maps is also low with minimum and maximum values for $E(B-V)$ 0.0287 and 0.0351, respectively. A maximum reddening variation of 0.0064 mag does not produce significant color changes to the inner and outer field CMDs.
The location on the sky of the ground-based photometry fields, as well as the central $HST$ field, relative to the cluster center, 
are represented in the middle panel: the central $HST$ field is delimited in red; the region that we define 'inner field' is located within the tidal radius, represented as a green circle; while the region defined as 'outer field' is comprised within the cyan square.

Ground-based photometric data have been used to estimate atmospheric parameters, as described in Sect.~\ref{sec:atm}. 
Hence, it is important to have an estimate of the internal photometric uncertainties. 
According to the photometric catalogs from Stetson\,(2000), the average $\sigma$(mag) for a star with $V \sim$19.5 in the $B$, $V$, $R$, $I$ bands is $\sim$0.003, $\sim$0.002, $\sim$0.012, $\sim$0.003 mag, respectively. Since each star has been typically observed in tens of images, the formal error should be significantly smaller. In the external field we obtained $B$, $V$, $I$ formal errors of $\sim$0.005, $\sim$0.004, $\sim$0.006 mag, respectively. Such small values are lower limits of the true photometric errors as demonstrated by the fact that the color spread of all the sequences of the CMDs is significantly larger than few milli-magnitudes. Photometry is indeed affected by a number of additional uncertainties. Spatial variations of the photometric zero point along the field of view, introduced by small inaccuracy in the PSF model, by the sky or bias determination, or by small reddening variations, are a very common property of any photometric catalog (see Anderson et al.\,2008 for a discussion on this issue). To estimate the error of target stars we started by measuring the color spread of MS stars as described in Milone et al.\,(2009, see their Sect.~6). Briefly, we have verticalized the MS of NGC\,1851 by subtracting to the color of each star the color of the MS at the same $V$ magnitude, then we have determined the histogram distribution of the color difference ($\Delta$color), and, finally, we have fitted the histogram with a Gaussian. We have repeated this procedure for $(B-I)$, $(V-I)$, $(R-I)$, and $(V-I)$ colours. We assumed the $\sigma$ of the least-squares best-fitting Gaussian as our estimate of the color error. Specifically, to estimate the errors associated to SGB stars in the central field, we used the color spread of MS stars with $19.2<V<19.4$. We obtained $\sigma_{B-I}=0.030$, $\sigma_{B-R}=0.031$, $\sigma_{B-V}=0.024$,  $\sigma_{V-I}=0.024$. We thus assumed 0.02 as typical magnitude uncertainty for stars in the internal field. In the case of stars in the external field, due to the small number of MS stars, we used a larger magnitude interval ($19.4<V<20.4$) and accounted for field-star contamination by subtracting from the observed histograms of $\Delta$color distribution, the corresponding histogram distributions for field stars. The later has been determined by using the Besan\c con Galactic model (Robin et al.\,2003) for stars within the same area as the external field. We found $\sigma_{B-I}=0.035$, $\sigma_{B-V}=0.031$,  $\sigma_{V-I}=0.032$, thus assumed a typical magnitude uncertainty of 0.02 mag in each filter. These uncertainties will be considered when discussing the impact of photometric errors on the atmospheric parameters (see Sect.~\ref{sec:atm}).
\begin{table*}
\caption{Description of the photometric images used for the first time in this work. \label{tab:photometry}}
\begin{tabular}{lccccccc}
\hline\hline
Telescope & Camera &  Filter & Exposure Time & Field & Date & Program & PI \\
\hline
Max Planck 2.2m, La Silla & WFI & $U$  & 8$\times \sim$800s & inner &Feb., 22-26, 2012 & 088.A9012(A) (SUMO) & A.\,F.\,Marino \\
Max Planck 2.2m, La Silla & WFI & $B$  & 2$\times \sim$300s$+$6$\times \sim$150s & outer & Feb., 26-28, 2012 & 088.A9012(A) (SUMO)& A.\,F.\,Marino \\
Max Planck 2.2m, La Silla & WFI & $V$  & 8$\times \sim$300s & outer & Nov., 13, 2013 & 085.A9028(A) & R.\,Gredel \\
Max Planck 2.2m, La Silla & WFI & $I$  & 3$\times \sim$120s$+$4$\times \sim$250s & outer & Feb., 26-28, 2012 & 088.A9012(A) (SUMO)& A.\,F.\,Marino \\
\hline
\end{tabular}
\end{table*}

\subsection{The spectroscopic dataset}\label{sec:spectradata}

Our spectroscopic data consist of FLAMES/GIRAFFE 
spectra (Pasquini et al.\ 2002) 
observed under the program 090.D-0687A (PI: A.\ P.\ Milone) 
taken with no  simultaneous calibration lamp.
The low resolution LR02 GIRAFFE setup was employed, which covers 
a spectral range of $\sim$600~\AA\ 
from 3964~\AA\ to 4567~\AA, and provides a resolving power 
$R \equiv \lambda/\Delta\lambda \sim$6,400. 
All our target halo stars, for which we aim to obtain kinematical and chemical information, were observed in the same FLAMES plate in 25 different exposures of 46 minutes,
for a total integration time of $\sim$19 hours. 

The large amount of observing time, the multi-object capability 
of FLAMES and the low resolution were crucial 
to observe mostly very faint stars. 
In fact, we wanted to find the largest possible number of stars 
associated with NGC\,1851 in a field of 30\arcmin$\times$30\arcmin\ outside
the tidal radius of the cluster, mostly populated by field stars.
Hence, our observations concentrated on the fainter but more densely 
populated regions of the CMD where the resolution of the low-resolution 
GIRAFFE settings ($\sim$6000) is the limit to get decent signal in a 
reasonable amount of observing time.

The signal-to-noise ratio (S/N) of the fully reduced combined spectra 
varies from star to star. 
It not only depends on the luminosity of the targets but also on the 
efficiency of the fibers.
Among the stars for which we inferred chemical abundances, the maximum S/N is $\sim50$ per pixel
for the more luminous stars that are starting to ascend the RGB; the S/N decreases for MS stars. 
We impose a limit of S/N $\sim15$ on the spectra of the outer field from which we infer
chemical abundances. Of course, the number of analysed spectral lines, and hence elements, 
increases with the S/N.

To supplement our halo star sample, we analysed data from the archive
for the internal field of NGC\,1851.
This sample for the internal field 
has already been analysed by
Gratton et al.\ (2012, hereafter G12). We decided to re-analyse these data to ensure an optimal comparison
sample as it consists of SGB stars for NGC\,1851 observed with
the same FLAMES/GIRAFFE setup (LR02) as our NGC\,1851-halo stars.
A homogenous comparison of the chemical contents of the halo stars with those
obtained for the internal field of NGC\,1851 (within the
  tidal radius) is crucial to understand if
the halo of NGC\,1851 shares similar abundances with the cluster.
The fully reduced spectra for the stars in the internal field have S/N of around 50.

The position on the sky and on the $V$-$(B-I)$ CMD of our spectroscopic targets is shown in Fig.~\ref{fig:CMDall}.
In the following we will refer to the stars outside the tidal radius of
NGC\,1851 as NGC\,1851-halo or external field stars; 
the internal field stars (within the tidal radius) will be
called NGC\,1851-cluster stars; 
while all the other stars in the external field that do not share the radial velocity (RV) of NGC\,1851 will be simply considered field stars (see Sect.~\ref{sec:rv}).
A list of all the analysed stars (cluster$+$halo) and their basic photometric data is provided in Tab.~\ref{tab:phot_data}. 

All the data were reduced in the same manner. 
The reduction, involving bias-subtraction, flat-field correction, 
wavelength-calibration, and sky-subtraction,
was done with the dedicated pipeline BLDRS v0.5.3\footnote{See {\sf http://girbld-rs.sourceforge.net}}. 
Radial velocities for both external and internal field stars were derived
using the IRAF@FXCOR task, which cross-correlates 
the object spectrum with a template. 
For the template we used a synthetic spectrum obtained through the 
spectral synthesis code SPECTRUM 
(Gray \& Corbally 1994)\footnote{
See {\sf http://www.phys.appstate.edu/spectrum/spectrum.html}
for more details.}.
This spectrum was computed with a model stellar atmosphere interpolated 
from the Castelli \& Kurucz (2004) grid,
adopting parameters 
(effective temperature, surface gravity, microturbolence, [Fe/H]) = (6000~K, 3.5, 1.0~\kmsec, $-$1.20).
The cross-correlation function was fitted with a Gaussian
  profile. We used the entire observed wavelength range to
  cross-correlate the spectra with the template, that includes
  hydrogen lines.
  The choice of including hydrogen lines, that can increase internal uncertainties, was due to the low-S/N of some
  single-exposure spectra for the halo stars. For the stars with
  higher S/N we have weaker lines available for the cross-correlation,
  but to ensure homogeneity in the RV determination, we analyse all
  the spectra at the same manner and use the entire available spectral range. 
 In any case, the dispersion of the RV measurements from different
 exposures for each star is indicative of our internal error.
(see Sect.~\ref{sec:rv}).

Observed radial velocities were finally corrected to the heliocentric system. 

In total we gathered spectra for 110 candidate NGC\,1851-halo stars,
spanning a
wide range in both magnitude and color. 
In the selection of the targets we maximize the number of
stars close to the NGC\,1851 photometric sequences in order to increase the
chance to observe NGC\,1851 halo stars. We have also observed
stars at larger distances from the NGC\,1851 sequence to explore the
possibility of having stars dynamically linked to the cluster, but
with possible different chemistry.

   \begin{figure}
    \includegraphics[width=8.6cm]{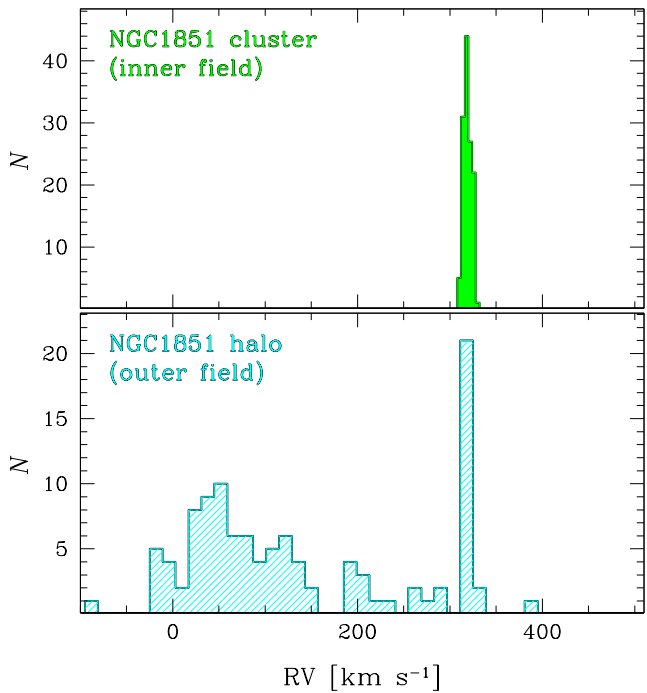}
    \caption{\textit{Lower panel:} histogram distribution of radial velocities
      for the 110 stars in the outer field. 
      \textit{Upper panel:} histogram distribution of radial velocities 
      for inner-field stars.}
    \label{fig:rv}
   \end{figure}

    \begin{centering}
    \begin{figure*}
     \includegraphics[width=12cm]{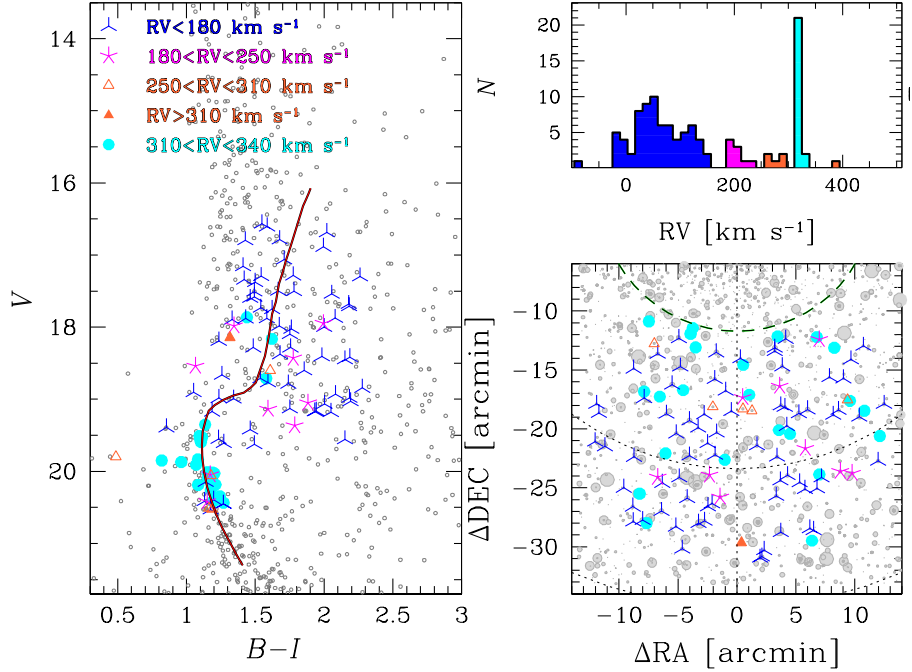}
     \caption{\textit{Left panel:} $V$ versus $(B-I)$ CMD of stars in the outer
       field.      Spectroscopic targets are represented with coloured
     symbols, according to their radial velocity as indicated on the upper
     left and in the RV histogram distribution on the upper right panel. 
     \textit{Right panel:} spatial distribution of target stars in the halo.
     We mark with dashed green circle the tidal radius of
     NGC\,1851. The dotted circles correspond to two and three
     times the tidal radius. }
     \label{fig:CMD}
    \end{figure*}
    \end{centering}

\begin{table*}
\caption{Coordinates, basic photometric data and radial velocities for the NGC\,1851-halo and cluster (inner field) stars. The full versions will be available online as supplementary material. \label{tab:phot_data}}
\begin{tabular}{lccccccccccrr}
\hline\hline
ID           & RA & DEC & $B$    & $V$ & $R$ & $I$ &     & $F606W$ & $F814W$         &  & RV [\kmsec]  & SGB\\ \hline
   \multicolumn{13}{c}{HALO}  \\
   T119        &  05:14:08.280	   &  $-$40:16:48.000        &  19.909    & 18.843 &  --  & 17.631 &    & --   &  --  &  &     98.924  & field \\       
   T198        &  05:13:47.160	   &  $-$40:14:14.200        &  19.399    & 18.711 &  --  & 17.820 &    & --   &  --  &  &    319.500  & uncertain \\      
   T105        &  05:13:56.870	   &  $-$40:18:34.600        &  18.935    & 17.872 &  --  & 16.855 &    & --   &  --  &  &     50.489  & field \\       
   \multicolumn{13}{c}{CLUSTER}  \\
  SGB-a.10       &  05:14:14.390  & $-$40:02:56.101   &  19.543  &  19.032  &  18.648  &   18.322  &   &  18.805  &  18.273  & &    319.579     &    bSGB	          \\
  SGB-a.12       &  05:14:13.250  & $-$40:03:22.201   &  19.496  &  18.957  &  18.604  &   18.234  &   &  18.803  &  18.283  & &    325.908     &    bSGB	          \\
  SGB-a.13       &  05:14:12.630  & $-$40:03:05.501   &  19.483  &  18.946  &  18.334  &   18.220  &   &  18.798  &  18.271  & &    326.645     &    bSGB	          \\
\hline
\end{tabular}
\end{table*}

\section{Radial velocities}\label{sec:rv}

Radial velocities have been obtained as explained in Sect.~\ref{sec:spectradata} 
from the individual exposures, and then simply averaged to get the final values
for each star listed in Table~\ref{tab:phot_data}.
The internal errors associated with our mean RV values depend on the luminosity of the stars.
They are higher for dwarfs with lower S/N. 

In the inner field the average error\footnote{The error is assumed to
  be equal to the rms of the RVs obtained from each exposure divided
  by the square root of the number of exposures minus one.} from the
distribution of the RV values obtained from different exposures is
0.49$\pm$0.02~\kmsec\ (rms=0.20~\kmsec). Only two stars (SGB-b-St.24
and SGB-b-St.4), both in the fSGB, have errors that are more than a
3~$\sigma$ level larger than the mean value. However, as these two
stars are among the fainter stars in our inner field sample, their
larger errors are likely due to the lower S/N of their spectra, rather
than to binarity. 
The lack of outliers with large RV rms values
  for the stars analysed in the inner field, suggests that there is
  no evidence for binaries in this sample.  
We note that we do not expect a large fraction of binaries in our
inner field sample. Indeed, we note that the fraction
of MS-MS binaries measured in the ACS field outside the
half-mass radius is 1.6$\pm$0.6\% (Milone et al.\ 2012). 

In the outer field (total sample of 110 stars), 
the rms values in the RV distributions span a wide
range. In some cases, the larger values are due to the low S/N of the
low-luminosity stars spectra; for some cases with good quality
spectra the large rms may reflect binarity. However, we prefer to not
enter into much details regarding this issue because 
we are interested exclusively in the subsample of outer field stars with RV
compatible with NGC\,1851.
For this particular subsample of stars the
errors in RVs range from 0.47 to 7.8~\kmsec, decreasing with the S/N. 
In most cases the error is as large as $\gtrsim$2~\kmsec so we 
cannot draw conclusions on the possible presence of binaries in our halo sample, especially for the fainter stars.
Concluding, the internal error in RV for the outer field stars is
  typically high, preventing a secure assessing of the presence of
  binaries in this sample. We only can note that, if these halo
  stars belong to NGC\,1851, due to the equipartition of energy, the
  binary fraction should be lower than in the central field.

In Fig.~\ref{fig:rv} we summarise our results for RVs obtained for stars
in the inner (upper panel) and the outer field (lower panel). 
Inner-field stars are clustered around an average RV value of
$+$319.5$\pm$0.5~\kmsec\ (rms=4.4~\kmsec), which is similar to
previous estimate for NGC\,1851 SGB stars of 318.2$\pm$0.5~\kmsec\
  (rms=4.3~\kmsec) and 320.0$\pm$0.4~\kmsec\ (rms=4.9~\kmsec), from
  Gratton et al.\,(2012) and Scarpa et al.\,(2011), respectively. 
The RVs of the internal field stars suggest that 
all the analysed targets are likely
cluster members as also supported by their position on the CMD (Fig.~\ref{fig:CMDall}).

Due to the high contamination from the field,
in contrast with what is observed in the inner region, the RV histogram of stars
in the external field is complex. 
Most of the stars have RV$\lesssim 200$~\kmsec\ and define a broad distribution peaked at  RV$\approx 50$~\kmsec; then we observe 
a narrow peak around the same RV of NGC\,1851, 14 stars at intermediate values, and one 
star with a very high radial velocity of RV$=$381.3~\kmsec. 
The peak around the mean RV of NGC\,1851 comprises
23 stars, and their average RV is $+$318.4$\pm$1.0~\kmsec\ (rms=4.9~\kmsec).

Since NGC\,1851 has a distinct radial velocity in this Galactic sight-line, members of its
stellar halo should share its unambiguous kinematic signature. 
Hence, the presence of stars with RVs compatible with the cluster strongly supports the
presence of a halo extending beyond the tidal radius of NGC\,1851.
To further support the existence of this halo,
in Fig.~\ref{fig:CMD} we analyse the spatial distribution and the position in
the CMD of stars with different radial velocities. 
We have defined
four groups of stars corresponding to different RV intervals. 
The RV histograms for the four
groups of stars are plotted in the upper-right panel of Fig.~\ref{fig:CMD}.
The location of these stars in our analysed field is represented in the lower-right panel of Fig.~\ref{fig:CMD}. 

We note that the majority  of stars with cluster-like RVs are
distributed along the fiducial line of NGC\,1851 (red line in the left-panel), 
in contrast with
 most of the stars with RV$<250$~\kmsec, which span a broad interval of
 color. This finding further supports the possibility that the group of cyan
 stars belongs to a halo surrounding NGC\,1851. The color and magnitude of
 four out of five stars with $250<$RV$<310$~\kmsec are consistent with those of
 NGC\,1851. 
However, we anticipate that a similar number of stars with these RVs and in this region 
of the CMD is expected from Galactic models (see the following section for details).

\subsection{Comparison with a Galactic model}\label{subs:rvS10}
To investigate whether the peak at RV$\approx$320~\kmsec\ observed in Fig.~\ref{fig:rv} is associated with the halo of NGC\,1851 or not, we compare our observations with the Galactic model by Robin et al.\,(2003). 

\begin{centering}
    \begin{figure*}
     \includegraphics[width=12cm]{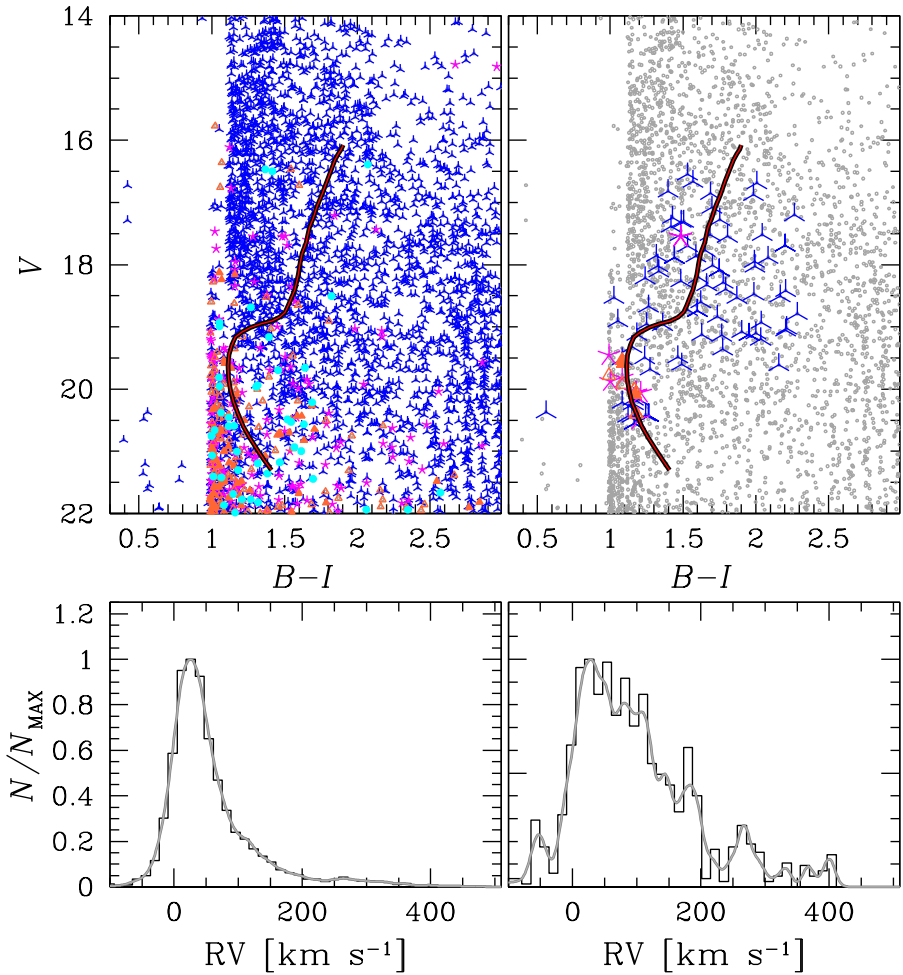}
     \caption{\textit{Left panels:} $V$ versus $(B-I)$ CMD predicted
       by the Galactic model by Robin et al.\ (2003) for all the stars
       in a 60\arcmin$\times$60\arcmin\ field of view centered at the
       same coordinates as the field studied in this paper (top). We
       used the same symbols introduced in Fig.~\ref{fig:CMD} to
       represent stars in different radial-velocity intervals. The
       histogram and the kernel-density distribution of radial
       velocities predicted by the Galactic model for the same stars
       are shown in the lower-left panel. \textit{Right panels:} 
         In gray we represent the same CMD of the upper-left panel;
         stars that have $V$ and $(B-I)$ similar to the observed
         stars, belonging to the ``equivalent sample'' (see Sect.~\ref{subs:rvS10}
         for details) have been represented according with symbols and
         colours corresponding to their radial velocity (see Fig.~\ref{fig:CMD}) . }
     \label{fig:model}
    \end{figure*}
    \end{centering}

The upper-left panel of Fig.~\ref{fig:model} shows the synthetic $V$ versus $(B-I)$ CMD for the $\sim$6,100 stars that, according to the Besan\c con model (Robin et al.\,2003), are located in a  60\arcmin$\times$60\arcmin\ region with the same coordinates as the field analysed here.
The histogram and the kernel-density distributions of radial velocities for the stars are shown in the lower-left panel. 
The model generates population realisations based on probability as a function of the specific position in the parameter space. 

 A visual inspection of the CMD in the upper-left panel reveals that, as expected, stars with different velocities populate different regions of the CMD. Low-velocity stars (RV$<$180~\kmsec), mainly dwarfs, define a broad sequence, mainly populated by disk stars, which is characterised by a large color spread ($1 \lesssim$($B-I$)$\lesssim 4$) and extends up to bright luminosities ($V<14$). Most of the stars with high velocities (RV$>$180~\kmsec) populate a narrower sequence  ($1 \lesssim$($B-I$)$\lesssim 2$) and have, on average, lower luminosities.

However, we note that the sample of stars analysed here is located in a limited region of the CMD. In fact, to properly compare radial velocities of the observed stars and those from the Galactic model we need to select a sample of stars in the synthetic CMD with almost the same color and magnitude as the observed stars.

To this aim, we associated to each observed star, the star in the Galactic-model CMD at the smallest `distance' as suggested by Gallart et al.\,(2003, see their Sect.~4) and Moni Bidin et al.\,(2011). 

This distance is assumed as $d=\sqrt{( k \times (B-I))^{2} + V^{2}}$ where $k$ is a factor enhancing the difference in color with respect to the magnitude difference. Our procedure to estimate the factor $k$ appropriate for our dataset comprises different steps: 
\begin{itemize}
\item{We have defined in the $V$ versus $(B-I)$ CMD a grid of points spaced in color and  in magnitude by $\Delta (B-I)$=0.025 mag and $\Delta V$=0.05 mag, respectively. }
\item{ We started from the analysis of the observed CMD. For each grid-point (i,j) and each observed star (l), we have estimated the probability ($P^{\rm l}_{\rm OBS, i,j}$) of having that star in a box within $\Delta (B-I)$/2 and $\Delta V$ from the point i,j. This probability has been estimated by assuming for each star the errors in color and magnitude estimated for real stars. For each grid point we have determined $P_{\rm OBS, i,j}=\sum_{l=1}^{N} P^{\rm l}_{\rm i,j}$. } 
\item {Then, we have generated the equivalent sample by assuming a given value of $k$, and determined for these stars $P_{\rm k, i,j}$ by following the same procedure described above for observed stars. The assumed values of $k$ range from $k$=0 to $k$=10 in steps of 0.1.}
\item {For each value of $k$, we calculated $\chi(k)=\sum(P_{\rm k, i,j}-P_{\rm OBS, i,j})$. The $k$ value that, for our dataset, returns the minimum $\chi$ is $k$=2.2.}
\end{itemize}
For the determination of the equivalent sample we then assume $k$=2.2, however we have verified that the conclusions of our paper are identical for any  1$<k<$10.

In the upper-right CMD of Fig.~\ref{fig:model} we highlight
only the selected sample of stars  while the rest of the stars is represented with small grey dots. 
To avoid that our conclusions are affected by low numbers, we have increased the number of selected stars by a factor of 100. To do this we have generated other 99 Galactic models for stars from Robin et al.\,(2003) in a 60\arcmin$\times$60\arcmin\ region with the same coordinates as the external field of NGC\,1851. For each of them we have extracted a sample of stars as described above. In the following we use the whole collection of the 100 samples of stars (hereafter ``equivalent sample'').

The corresponding RV distribution for these stars in the equivalent sample is shown in the lower-right panel.
We conclude that when the velocities of all the stars from the Galactic model are analysed, the distribution has a single peak at RV=25~\kmsec\ and about 82\% of stars have RV$<$100~\kmsec\ (lower-left panel); 
in the case of the equivalent sample (lower-right panel) the main peak is shifted to higher velocities, at RV$\sim$50~\kmsec, and the fraction of stars with RV$<100$~\kmsec\ decreases to $\sim$65\%. There is some hint of less prominent peaks at RV$\approx 100$~\kmsec, RV$\approx 180$~\kmsec, and RV$\approx 270$~\kmsec.

A comparison between the RV
distribution determined in this paper and that expected from the Galactic model is provided in the upper panel of Fig.~\ref{fig:confrontoII}. 
To properly compare the two distributions, we have normalised each histogram to the total number of stars with RV$<$225~\kmsec, which is the value marked by the vertical dotted line of Fig.~\ref{fig:confrontoII}.
By considering all our analysed stars,
the probability that the observed RVs and those from the model come from the same parent distribution is almost null ($P<$10$^{-5}$), not depending on the adopted normalisation.
 The most striking difference between the two distributions is the lack of stars with $310<$RV$<340$~\kmsec in the model. 
Indeed while 23 stars have been observed in this interval of RV, 
the Besan\c con model predicts that only $\sim$1.5 stars have such kinematics.
This provides further support of an halo of stars, with the same kinematic as NGC\,1851 but located between 1 and $\sim$2.5 tidal radii from the cluster center.
If we neglect these stars and consider only stars with RV$<$300\kmsec, the observed and expected RV distributions are very similar as confirmed by the Kolmogorov-Smirnov (KS) test which provides a probability $P$=0.85 to come from the same parent distribution.
Regarding the small contamination from field stars that can occur at the RV of NGC\,1851 ($<$2 stars, as suggested from the models), we note that two out of the 23 stars (T066 and T073) in our sample of NGC\,1851 RV-like halo stars that do not lie on the fiducial sequence of the cluster.
The chemical abundance analysis for these two stars has been inferred in a similar way as the other halo stars. However, keeping in mind that we cannot exclude that these objects are field stars, when necessary, we will highlight the presence of these two stars along the paper.

\subsection{Comparison with Sollima et al.\ (2012)}\label{subs:rvS10}
The first radial velocity study of stars in the halo of NGC\,1851 was conducted by Sollima et al.\ (2012, hereafter S12) based on VIMOS/VLT spectra.
The 107 stars analysed by S12 are between 12\arcmin\ and 33\arcmin\ north-east from the center of NGC\,1851; so, even if at a similar distance from the center of the cluster, the field analysed by S12 observed a different quadrant, and none of their targets is in common with our sample. 

\begin{centering}
    \begin{figure}
     \includegraphics[width=8cm]{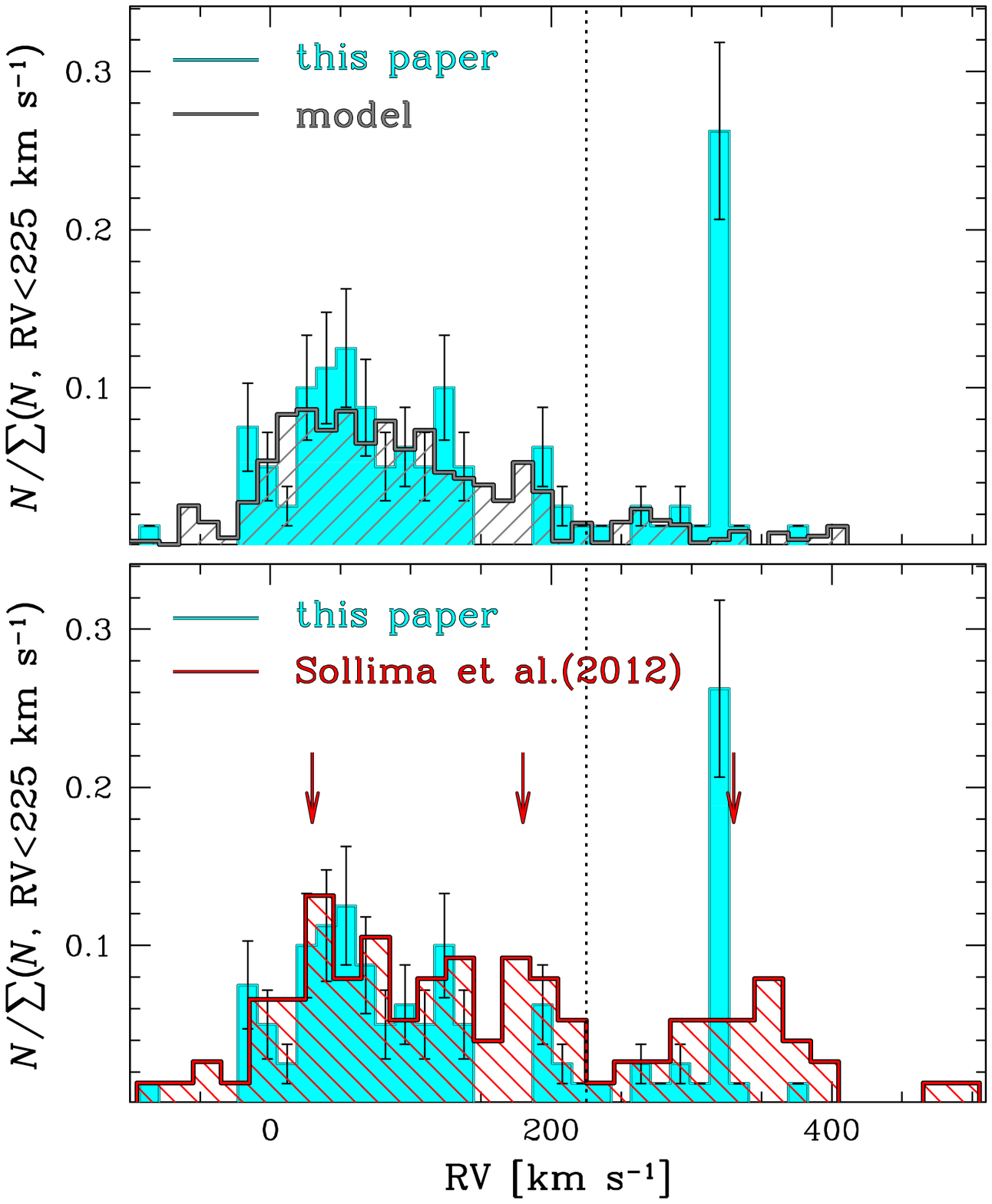}
     \caption{ \textit{Upper panel}: Comparison of the radial-velocity distribution observed in this paper for the halo of NGC\,1851 (cyan histogram) and the distribution predicted by the Galactic model by Robin et al.\ (2003, grey-dashed histograms). 
    \textit{Lower panel}: Comparison between the radial-velocity distribution derived for the halo in this paper (cyan) and by S12 (red-dashed). The red arrows indicate the three peaks found in S12, attributed to the field, a possible stream, and the halo associated with NGC\,1851.
     }
     \label{fig:confrontoII}
    \end{figure}
    \end{centering}

In the lower panel of Fig.~\ref{fig:confrontoII} we compare the RV histogram distributions from this paper (cyan) and from S12 (red), obtained by binning in intervals of 20~\kmsec. Both histograms are normalised to the total number of stars with RV$<$225~\kmsec\ (vertical dotted line of Fig.~\ref{fig:confrontoII}).
 We emphasise that caution must be used when comparing the two RV distributions as the stars studied by S12 and  those analysed in this paper have not been selected homogeneously.

S12 
identified three main peaks in their RV distribution. They associated most of their stars to foreground disc stars peaked around RV$\sim$30~\kmsec, then they found a peak at RV$\sim$330~\kmsec\ corresponding to the bulk motion of NGC\,1851, and an overdensity of stars at RV$\sim$180~\kmsec. The position of the three peaks inferred by S12 is highlighted in the lower panel of Fig.~\ref{fig:confrontoII} by red arrows.
The comparison of the two observed distributions shows that the uncertainties associated with the RV measurements are much higher in S12, due to the lower resolution data they used.
Their estimated internal uncertainties, of $\sim$15~\kmsec\ (see S12
for details) reflects in a much broader distribution of stars about the peak at the RV of NGC\,1851.
On the basis of their RV distribution, S12 suggested that, apart the bulk of stars at the same mean RV of the cluster, the
overdensity at $\sim$180~\kmsec\ could be associated with a cold stream having $\sigma_{\rm v}<$20~\kmsec. However, they pointed out that additional studies are needed to confirm this possibility. 

Our RV distribution
exhibits a sharp peak at the same motion as NGC\,1851, and a broad peak around RV$\approx$50~\kmsec\ in close analogy to that observed by S12. Our RV distribution does not show any peak of stars at RV$\approx$180~\kmsec, but we observed two small groups of stars at the similar velocities of RV$\approx$110 and $\approx$200~\kmsec. 
We recall here that the RV distribution observed from the dataset
analysed in this paper for stars with RV$<$300~\kmsec\ is fully
consistent with that predicted by the Galactic model by Robin et al.\
(2003), that also predicts a minor peak at RV$\sim$180~\kmsec\, thus excluding any evidence of a stellar stream associated to NGC\,1851 in our field of view.

\section{Chemical abundance analysis}\label{sec:abundances}

Our sample is almost entirely composed of low luminosity dwarfs and
sub-giants.
In fact, during the target selection we wanted to maximize the number
of possible NGC\,1851 halo stars, with stars as close as possible
to the cluster sequences along the CMD. 
As the RGB is relatively poorly populated,  we selected mostly MS stars.
In the end our sample was successful in the identification of 23 stars
belonging to the NGC\,1851 halo. 
Among the 23 stars with NGC\,1851-like RVs, we were able
to estimate metallicities for 15 objects, including 7 stars with both Sr and Ba estimates, 
thus providing the first
elemental abundances for the halo surrounding NGC\,1851;
the other stars had insufficient S/N.

The chemical abundances that we were able to estimate for the halo stars were then
compared with those obtained for the central field of NGC\,1851. To
have all the measurements in the same absolute scale, 
we have analysed 
both the external and the internal field 
in an uniform manner (e.g., same code and linelist).

\subsection{Model atmospheres}\label{sec:atm}

As the moderate resolution and wavelength range of our spectra do not
allow us to determine atmospheric parameters from Fe lines, we used
our photometry to estimate effective temperatures (\teff) and surface
gravity (\logg) of our stars.
Photometry includes $BVRI$ for the center field and $BVI$ for the
halo field. 
The photometry is adjusted by assuming a reddening
value, in this case $E(B-V)$=0.02, and then computing an extinction
for each bandpass using the Cardelli et al. (1989) extinction curve
and adopting $R_{\rm v}$=3.1. The reddening is assumed constant across
the field, which should be reasonable for such a low-reddening system, 
and the results are only weakly dependent on variations in $E(B-V)$ of 0.01. 

Effective temperatures and their uncertainties are estimated via
Markov chain Monte Carlo (MCMC) simulations using the parallel
\texttt{emcee} package (Foreman-Mackey et al. 2010). The
\texttt{emcee} code takes as input model parameters and a probability
function that compares models to individual data points. The
log-probability function is the sum of the squares of the differences
between observed and modeled magnitudes, divided by the error, in each
filter.

The modeled magnitudes are obtained by interpolating within grids of bolometric 
corrections for all photometric bands ($BVRI$ for the inner field, $BVI$ 
for the outer field) and assuming that all cluster stars are equidistant. 
Essentially we construct an H-R diagram from the input CMD, and 
map a given star to the point on the H-R diagram
that best matches the observed photometry.
The bolometric corrections are derived from PHOENIX
model atmospheres (Hauschildt et al.\ 1999a,b) using the $BVRI$ bandpasses
defined by Bessell \& Murphy (2012). 
These calculations assume [Fe/H]=$-1.3$ 
and [$\alpha$/Fe]=+0.2 broadly consistent with the finding reported in this 
paper. However, it is important to note that the effective temperatures 
derived in this way have only weak sensitivity to [Fe/H] because of the 
broadband, optical nature of the photometry.

The MCMC simulations provide estimates of effective temperature, 
bolometric luminosity, and surface gravity.  The surface gravity is 
very roughly determined
because
of its weak dependence on broadband photometry.
The resulting estimate of \teff\ for a given star is obtained from the Markov chain, 
which samples the posterior probability density function, by computing the mean and 
standard deviation. A star with a large standard deviation also tends to have a mean 
that differs considerably from the median, suggesting that the distribution is 
asymmetric. However, such stars are rare in the present data set ($<10$\% in both fields).

Surface gravities are obtained from the apparent $V$ magnitudes,
corrected for differential reddening, the \teff,\ apparent bolometric
luminosities obtained from the MCMC simulations above and 
an apparent distance modulus of $(m-M)_{\rm V}$~=15.47 (Harris 1996, 2010 edition). 
We assume that all stars lie at essentially the same distance, and masses 
resulting from one single best-fitting isochrone\footnote{We note that the masses on the two SGB populations of NGC\,1851 may be different, but, at a given magnitude, this difference is of the order of $\sim$0.01~$M_{\odot}$. A possible systematic error in the adopted masses of that order has negligible effects ($<$0.01~dex) on the \logg\ values and on the derived abundance.} 
For microturbulent velocities (\vmicro) 
 we adopted the latest version of the appropriate relation used in the Gaia-ESO survey
 (Gilmore et al. 2012; Bergemann et al.\ in prep.), assuming a
   metallicity of [Fe/H]=$-$1.18~dex for NGC\,1851.
The dispersion of the recommended \vmicro\ values for the GES UVES spectra around the adopted relation is about 0.20~\kmsec, which is a reasonable internal uncertainty to be associated with our adopted values. 

We note that our technique to derive effective temperatures
is independent of the projection of stars on any isochrone, 
as there is no isochrone used in the MCMC process.
The model-dependency of our technique is related to the different set of colours used (e.g., Castelli \& Kurucz or MARCS instead of PHOENIX), and to the fact that we used a fixed metallicity and distance.
On the other hand, our \teff\ values do not depend on projecting on
isochrone projection.

If projecting on theoretical sequences has the advantage to minimize
the impact of photometric errors on the derived atmospheric
parameters, on the other hand in GCs like NGC\,1851 with multiple SGB
and RGB sequences, one single isochrone is not able to represent all
the observed populations.
We prefer to use the actual photometric data for each star, instead of
using multiple isochrones, to not force each star to a given
isochrone; this allows us to avoid errors due to mismatch between the
observed stars and the two SGB, that may occur due to photometric
errors, and in regions of the CMD (like the upper MS) 
where it is not possible to assign the observed stars to different
populations just on photometric information.
Furthermore, we do not want to impose a {\it two discrete SGBs}
scenario, as we cannot exclude that the real situation may be more complex.

Table~\ref{tab:abundances} lists the adopted stellar parameters for the NGC\,1851-like
RV stars in the halo and the central field stars.
The uncertainties on the \teff\ values given by the MCMC simulations
are listed in the third column of Tab.~\ref{tab:abundances}.
The median of these errors is 94$\pm$2~K (rms=17~K), and we adopt this value as
an estimate of the internal error associated with our \teff.

\begin{figure}
    \includegraphics[width=7.0cm]{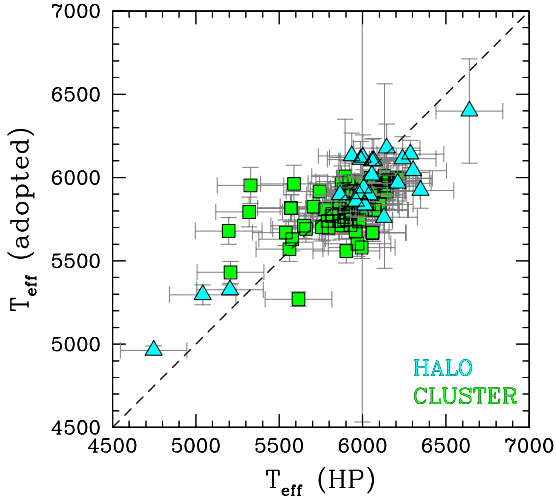}
    \caption{Adopted \teff\ values for the
      NGC\,1851-halo (cyan triangles) and central field (green squares) targets as a
      function of the values obtained from the HP index. }
    \label{fig:HP}
   \end{figure}

An independent temperature estimate was obtained
from the H$\delta$ line index
(HP2), calibrated as a function of \teff\  (Ryan et
al. 1999), and the quadratic relationship provided in Norris et al.\
(2013) obtained for metal-poor dwarfs, subgiants and giants. 
The adopted \teff\ as a function of the values derived from the HP
index has been shown in Fig.~\ref{fig:HP} for the NGC\,1851-halo and
NGC\,1851 central field stars. 
The halo and cluster stars adopted \teff\ values compare similarly to those
obtained from the HP index, with mean differences
$\Delta$\teff(HP$-$adopted)=55$\pm$39~K (rms=184~K) for the NGC\,1851
halo, and  $\Delta$\teff(HP$-$adopted)=73$\pm$21~K (rms=191~K) for the
central field stars. 
This comparison confirms that the temperature scale adopted for the
NGC\,1851 halo stars agrees with the one adopted for the NGC\,1851
cluster stars. 

Errors in \teff\ and mass of $\pm$94~K and $\pm$0.05~$\rm {M_{\odot}}$, affect
the \logg\ values by $\pm$0.03 and $\pm$0.02 dex, respectively. 
Internal uncertainties in the bolometric luminosity of $\pm$0.01 have
small effects on \logg\ values: $\mp$0.01~dex.
All these effects, added in quadrature, contribute to a very small
internal error in \logg\ of $\approx$0.04~dex.
We emphasise that this is just the formal internal error in \logg\, while real uncertainties in this parameter may be much larger.
Internal uncertainties of $\pm$94~K and $\pm$0.04~dex in \teff\ and
\logg\ affect \vmicro\ by only $\pm$0.03 \kmsec.
Although the internal error in \vmicro\ due to \teff\ and \logg\ is formally
small, we assume for this parameter a more realistic internal error
that is the rms of the UVES GES data around the used relation,
i.e., we used an error $\sim$0.20~\kmsec.
We will take into account these uncertainties in the atmospheric
parameters for the estimation of the errors associated with  the chemical
abundances.

\begin{figure*}
    \includegraphics[width=5.86cm]{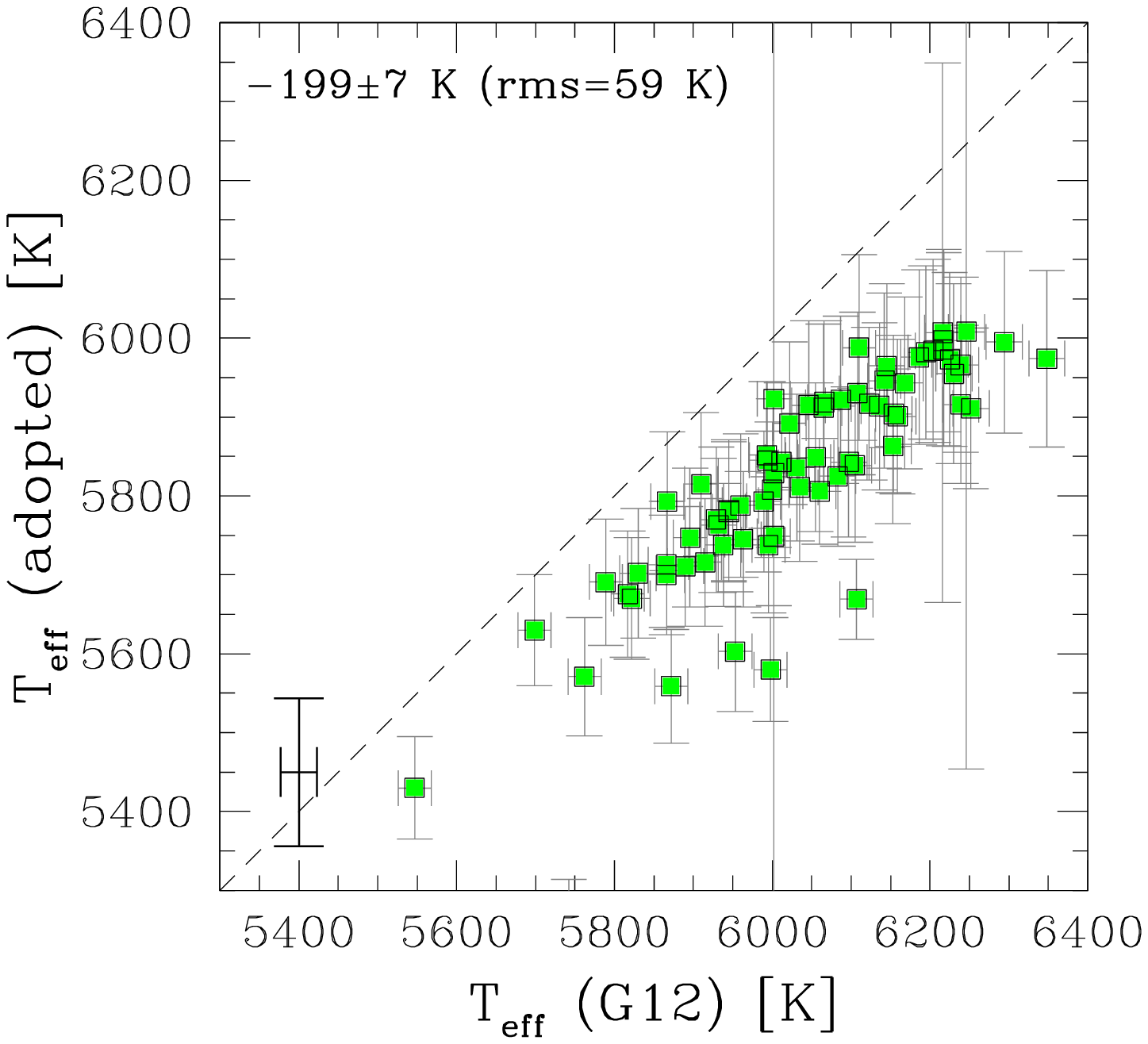}
    \includegraphics[width=5.86cm]{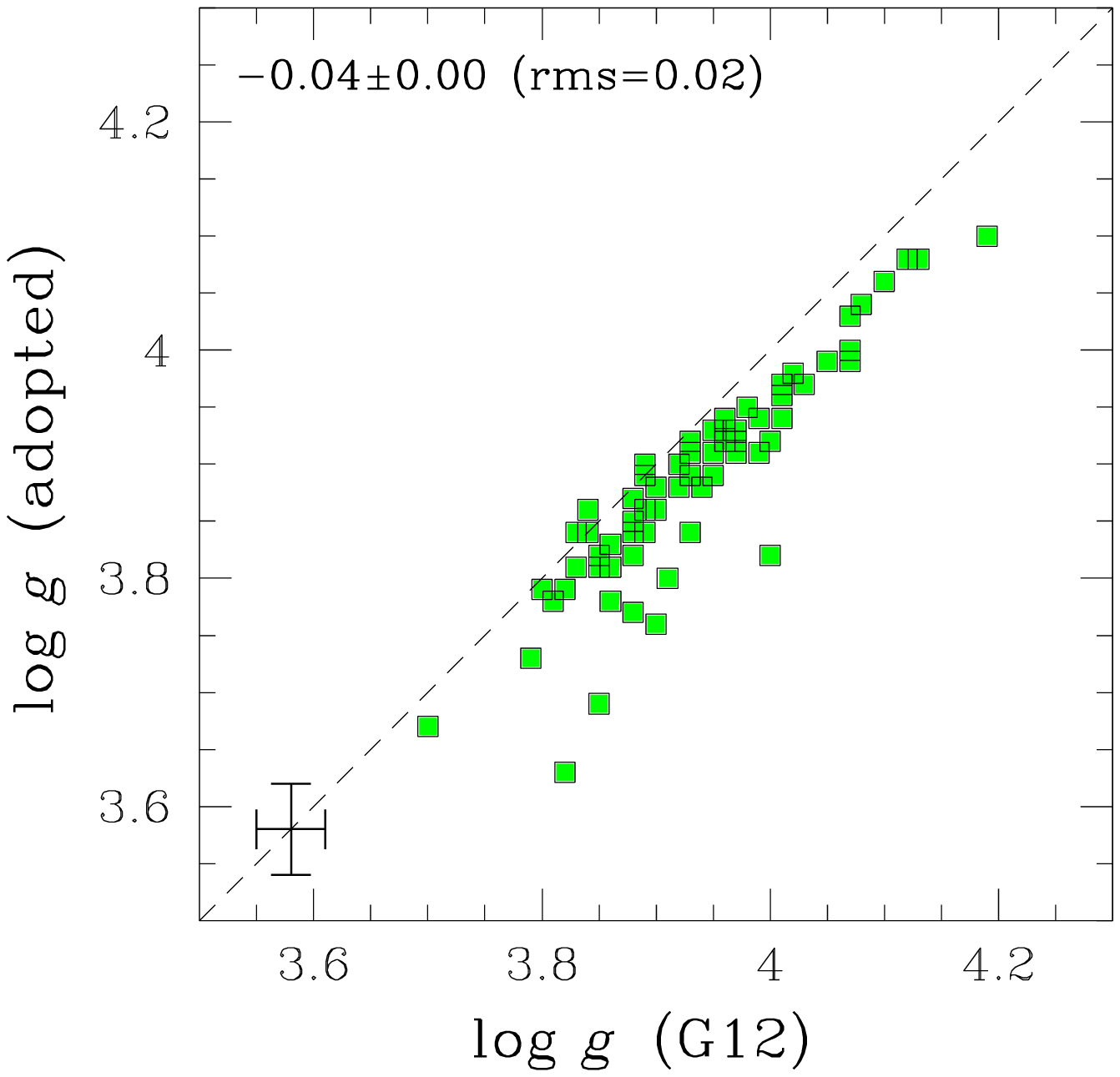}
    \includegraphics[width=5.86cm]{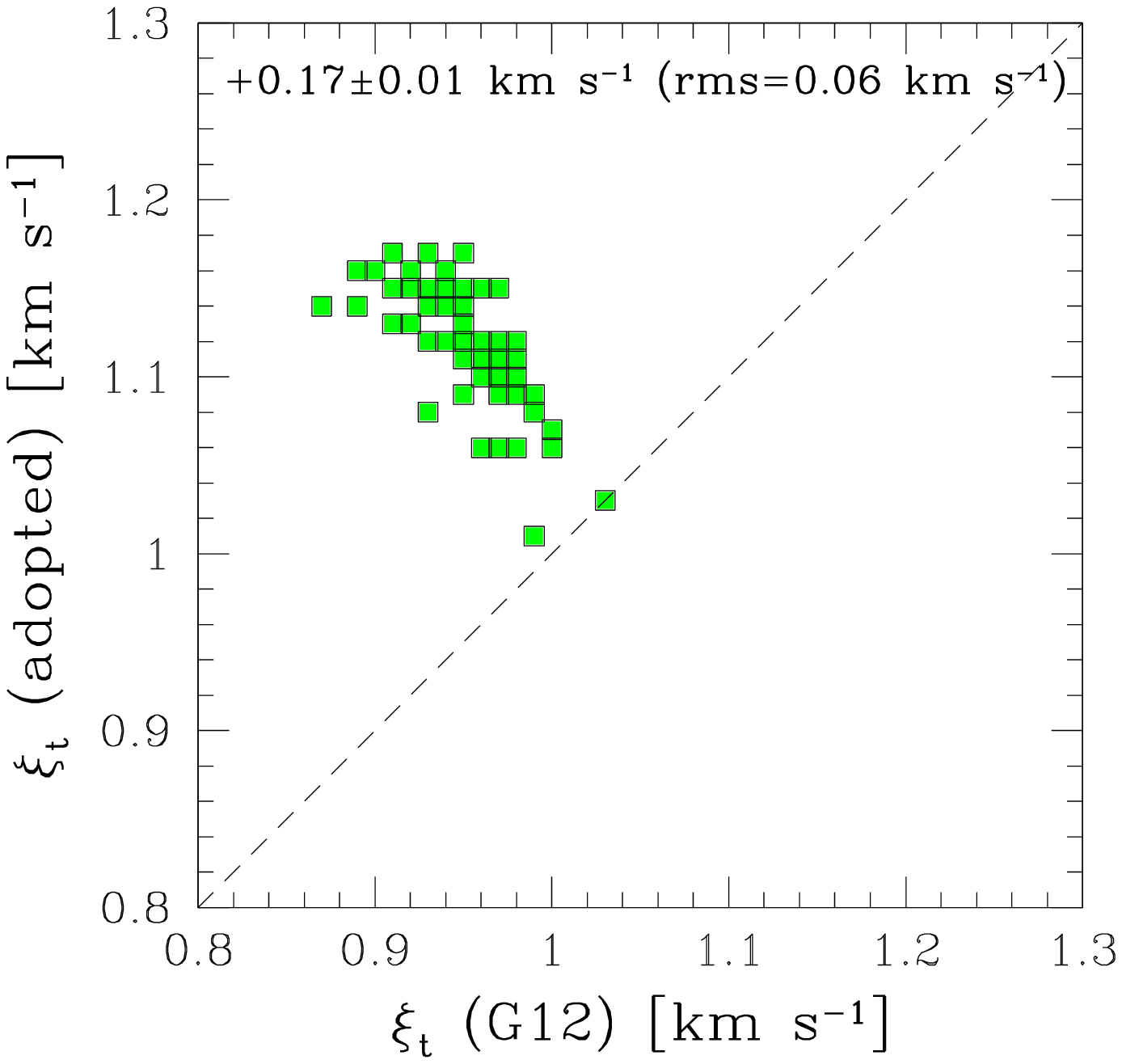}
    \caption{Adopted values for \teff, \logg, and \vmicro\ for the
      NGC\,1851 central field targets as a
      function of the values adopted in G12. The mean differences
      (adopted$-$G12) and associated rms are shown in each panel. }
    \label{fig:literatureATM}
   \end{figure*}

In Fig.~\ref{fig:literatureATM} we compare the adopted atmospheric
parameters with those of G12 for the central field of NGC\,1851.
Effective temperatures and surface gravities from G12 have been
determined by fitting two different isochrones for stars 
associated with the bright and faint SGB of NGC\,1851, using 
  different photometric catalogs than those used here (see G12 for details).
We note a systematic difference in \teff\ with the G12 values being
$\sim$200~K hotter, with a scatter of 59~K.
Only three stars of the internal field are in common with Lardo et
al. (2012), whose \teff\ have been determined from the Alonso
calibrations (Alonso et al. 1999). For two of these stars, our \teff\
are higher by $\sim$300~K, and one is slightly lower (by 40~K). 
These comparisons suggest that, although the \teff\ scale may be 
affected by systematics of a few hundreds of K, the internal error is
lower ($\sim$100~K).
The mean difference in \logg\ between the G12
and the adopted values is small, of 0.04~dex (rms=0.02~dex).
Microturbolent velocities in G12 are lower by 0.17~\kmsec\
(rms=0.06~\kmsec).
The different \vmicro\ values (as appear in
Fig.~\ref{fig:literatureATM}) are likely due to the different
relations used to estimate this parameter.
While our relation is a second order polynomial in
\teff\, \logg\ and metallicity, G12 used a linear relation in just
the surface gravity.

\begin{table*}
\caption{Adopted atmospheric parameters and chemical abundances for the NGC\,1851-halo and cluster (central field) stars. When available, we list the scatter of measurements derived from individual measurements. The full versions will be available online as supplementary material. \label{tab:abundances}}
\begin{tabular}{lcrccccrrrrrrrr}
\hline\hline
ID           &  \teff & err$_{\rm Teff}$ & \logg & \vmicro & [Fe/H]    & $\sigma$ & [C/Fe] & [Mg/Fe] & scatter & [Ca/Fe] & [Cr/Fe] &   [Sr/Fe] &  scatter &  [Ba/Fe] \\ \hline
  \multicolumn{15}{c}{HALO}  \\
      T198   &  5296 & 59      & 3.55   & 1.02      &  $-$1.36                 & 0.04 &  $-$0.27 &  0.53  &  0.3 &    0.43  &  $-$0.25 &$-$0.06  &   0.23   &       0.27 	\\
      T132   &  5966 & 107     & 4.29   & 1.11      &  $-$1.29                 & 0.09 &      --  &  --    &  --  &    0.37  &     --   &   0.33  &    --    &        --  	\\
      T150   &  5857 & 89      & 4.31   & 1.07      &  $-$1.34                 & 0.07 &      --  &  --    &  --  &    --    &     --   &    --   &    --    &       0.75 	\\
  \multicolumn{15}{c}{CLUSTER} \\
SGB-a.10     &  5923 &   1389  &  3.90  &   1.15    &  $-$1.05    & 0.06  &$-$0.47    &       0.31    &      0.11      &$-$0.02     &  $-$0.25  &$-$0.26   &    0.01 &   0.40        \\
SGB-a.12     &  5793 &     89  &  3.84  &   1.11    &  $-$1.25    & 0.11  &$-$0.32    &        --     &       --       &    --      &  $-$0.49  &$-$0.32   &     --  &   0.54	       \\
SGB-a.13     &  5716 &     81  &  3.78  &   1.10    &  $-$1.48    & 0.04  &$-$0.03    &       0.58    &       --       &   0.19     &     0.13  &   0.17   &    0.02 &   0.28	       \\
\hline
\end{tabular}
\end{table*}

\subsection{Abundance analysis}\label{sec:abb}

Chemical abundances were derived from a local thermodynamic
equilibrium (LTE)  analysis by using the latest version of the spectral 
analysis code MOOG, with no scattering included (Sneden 1973), by
using the $\alpha$-enhanced model
atmospheres of Castelli \& Kurucz (2004), whose parameters have been obtained 
as described in Sect.~\ref{sec:atm}.
All the target lines have been analysed by spectral
  synthesis, via an automatic $\chi^{2}$-minimisation between the
  synthetic and the observed spectra reported to same
  reference continuum.

We determined abundances for Fe, the neutron-capture ($n$-capture) elements
Sr and Ba, the light element C, the $\alpha$ elements Mg and Ca, and the iron-peak Cr. 
Iron abundances were derived from the Fe\,{\sc i} spectral features at 
4005.2, 4045.8, 4063.6, 4071.7, 4132.1, 4143.9, 4187.0, 4202.0, 4250.1, 
4260.5, 4271.2, 4383.6, 4404.8, 4415.1, 4476.0, 4482.2, 4528.6, 4556.1~\AA.
For the NGC\,1851-cluster stars, the S/N of the spectra 
was sufficient to allow us to synthesise all the Fe features.
In the case of the NGC\,1851-halo stars it was possible to infer abundances 
from all the lines only in the two RGBs. 
For the other stars we measured a subsample of the listed Fe features, 
typically a number of 7-10 spectral features.
For four stars (T128, T153, T134, T139) we could measure only 2-3 Fe features.
To ensure that the derived average metallicities were on the same scale, and avoid systematics depending on the analysed spectral feature for different stars, 
we corrected the abundances from each measured line for their systematic difference with the abundance obtained from the Fe\,{\sc i}~4404.5~\AA\ feature, that is measured in all the stars.

Limited by the relatively low
resolution and the small wavelength range of our spectra, we derived
Sr and Ba abundances only from the strong resonance transitions
Sr\,{\sc ii} 4077, 4215~\AA, and Ba\,{\sc ii} 4554~\AA. 
Both the Sr lines suffer from blends with other surrounding
transitions, 
mostly Fe features and other $n$-capture species (e.g. Dy and La) in the
case of the Sr\,{\sc ii} 4077~\AA. 
For all the contaminating elements we assumed a solar-scaled
  abundance. In the case of Sr, the blending with Dy and La transitions occurs
on the redward and blueward spectral region around the Sr\,{\sc ii}
4077~\AA\ line, respectively. The effect of these blending features has been
considered by performing spectral synthesis for a few representative
stars by varying Dy and La abundances relative to Fe.
In both cases, the effect on the [Sr/Fe] abundances is
quite small: [Sr/Fe] decreases by less than 0.05~dex by increasing 
the abundances for these two contaminating elements by $+$0.5~dex.
This variation should be regarded as a possible systematic error
affecting mostly the stars in our sample that have been enriched in
$n$-capture elements and have super-solar abundances relative to Fe for
these elements. We can argue that, in the case of NGC\,1851, the Dy
contamination may have a negligible effect on the star-to-star relative abundances.
This is because in the solar system Dy is expected to be mostly
sensitive to $r$-process nucleosynthesis (87.9\% from $r$-processes and
12.1\% from $s$-processes; see Table~10 in Simmerer et
al. 2004). If some stellar populations in NGC\,1851 have undergone
a substantial enrichment in $s$-processes, we may expect a lower
degree of chemical variations, if any,  in Dy.

Spectral synthesis in the analysis of our spectral lines (and particularly at
our moderate resolution) 
is needed to take these blends into account. Although the Ba\,{\sc ii}
4554~\AA\ is isolated from contaminating transitions, we also computed 
synthesis for the Ba spectral line, to take its isotopic splitting
into account. 

The linelists are based on the Kurucz line compendium, except the Ba 
transition for which we added hyperfine structure and isotopic data from 
Gallagher et al. (2010). For Sr our linelists neglect
hyperfine/isotopic 
splitting since the wavelength shifts are very small and Sr has one dominant isotope.

Barium has five major naturally occurring isotopes whose production
fractions in the rapid-process ($r$-process) and $s$-process are
significantly different (e.g., Kappeler et al. 1989). In particular,
abundances derived from the Ba\,{\sc ii} 4554~\AA\  transition are
very sensitive to the adopted $r/s$ ratio (e.g., Mashonkina \& Zhao
2006; Collet et al. 2009). 
In our previous analysis of M\,22 SGB stars we found that 
assuming a pure $r$-process isotopic ratio
(Arlandini et al. 1999), instead of a scaled solar-system Ba
abundance and isotopic fractions (Lodders 2003) has the effect of
decreasing the [Ba/Fe] abundances by $\approx$0.2~dex (Marino et al. 2012a).
Bearing in mind that this behaviour can be an issue in the 
analysis of NGC\,1851 that hosts stars with different contributions
from the $s$-process material, we decide to assume a scaled solar-system Ba
isotopic fractions for all the stars.  
However, we note that stars with lower Sr and Ba abundances may be
better reproduced by an $r$-process isotopic ratio and their Ba
abundances may be over-estimated in our analysis.

Carbon was measured from spectral synthesis of the 
CH ($A^2\Delta-X^2\Pi$) G-band heads near 4315~\AA\, assuming a solar
scaled oxygen abundance.
The molecular line data employed for CH were provided by B. Plez
(priv. comm.; some basic details of the linelist are given in Hill et
al. 2002). 
Given that we do not have information on the actual oxygen
  abundance of our stars, the assumption of solar scaled oxygen is
  reasonable as the [O/Fe] distribution in RGB stars in NGC\,1851
  spans a range that goes from [O/Fe]$\sim -$0.50~dex to [O/Fe]$\sim
  +$0.50~dex (Villanova et al.\,2010; Carretta et al.\,2010). The
  molecular equilibrium in stellar atmospheres generally affect the
  abundance of C, N, O. However, in our range of atmospheric
  parameters CO is not an abundant molecule, and the impact of [O/Fe] abundances varied by the entire
   range observed in giants (from $-$0.50 to $+$0.50~dex) 
  is negligible on the G-band.
Magnesium has been derived from the Mg\,{\sc i} lines
at $\sim$4057.5 and 4167.3~\AA, calcium from the Ca\,{\sc i} at
$\sim$4226.7~\AA, and chromium from the Cr\,{\sc i} line at 
$\sim$4254.3~\AA.

An internal error analysis was accomplished by varying the
temperature, gravity, metallicity, and microturbulence one by one, 
and re-determining the abundances for three NGC\,1851 halo stars and
three NGC\,1851 cluster stars spanning the observed range in
temperature. 
The parameters were varied by $\Delta$\teff=$\pm$100~K,
$\Delta$\logg=0.05~dex, $\Delta$[Fe/H]=$\pm$0.11~dex, 
and $\Delta$\vmicro=$\pm$0.2~\kmsec. 

The limited S/N of our spectra introduces significative internal
  uncertainties to our chemical abundances. To estimate these
  uncertainties we computed a set of 100 synthetic spectra for three
  inner field stars (SGB-a.9, SGB-a-St.24 and SGB-b-St.21) 
  and two halo stars (T186 and T198), whose atmospheric parameters
  are representative of the whole sample. 
  These set of synthetic spectra were calculated by using the
  best-fit inferred abundances, and were then degraded to the 
  S/N of the observed spectra. 
  We then analysed the chemical abundances of all these synthetic
  spectra at the same manner as the observed spectra. 
  The scatter that we obtain from the abundances from each spectral line for a
  set of synthetic spectra corresponding to a given star, represents a
  fair estimate of the uncertainty introduced by the fitting
  procedure, due to the S/N, the pixel size and the continuum
  estimate. 
  These uncertainties
  strongly depend on the S/N, and are higher for 
  halo stars with lower S/N. 
The mean errors in the chemical abundances from our fitting procedure
are then divided by the square root of the number of
available spectral lines to obtain an estimate of the uncertainty associated to each analysed element.
These errors are listed as $\sigma_{\rm fit}$ in Tab.~\ref{tab:err}. 
Since these
errors are random, the uncertainty is lower for those elements with a
large number of lines (e.g., Fe). For the other elements we have two
(or just one) lines, and this error contribution 
is higher. 
Of course, the larger uncertainties 
are found for the MS stars of the halo that have a lower S/N.

A list of the uncertainties in chemical abundances due to the various
considered sources is provided in Tab.~\ref{tab:err}, where double
entries in the errors for the halo stars mean that the estimated error
is different for RGB and MS stars.
The various errors were added in quadrature, resulting in typical
uncertainties of $\sim$0.10-0.20~dex\footnote{The fact that an error in the 
atmospheric parameters can affect in a different way a given element can 
generate spurious correlations between abundance ratios. As an example we verified through Montecarlo simulations 
that we expect a significant correlation between [Fe/H] and [Ca/Fe] due to the errors listed in Tab.~\ref{tab:err}.}, 
with Ba abundances having the
largest uncertainty of $\gtrsim$0.20~dex, that stems mostly from
uncertainties in \logg\ \vmicro\ and the limited S/N.
Iron abundance over hydrogen are mostly affected by uncertainties in
\teff, while the limited S/N translating into continuum errors, 
gives the major contribution to the other species.

\begin{figure*}
    \includegraphics[width=5.cm]{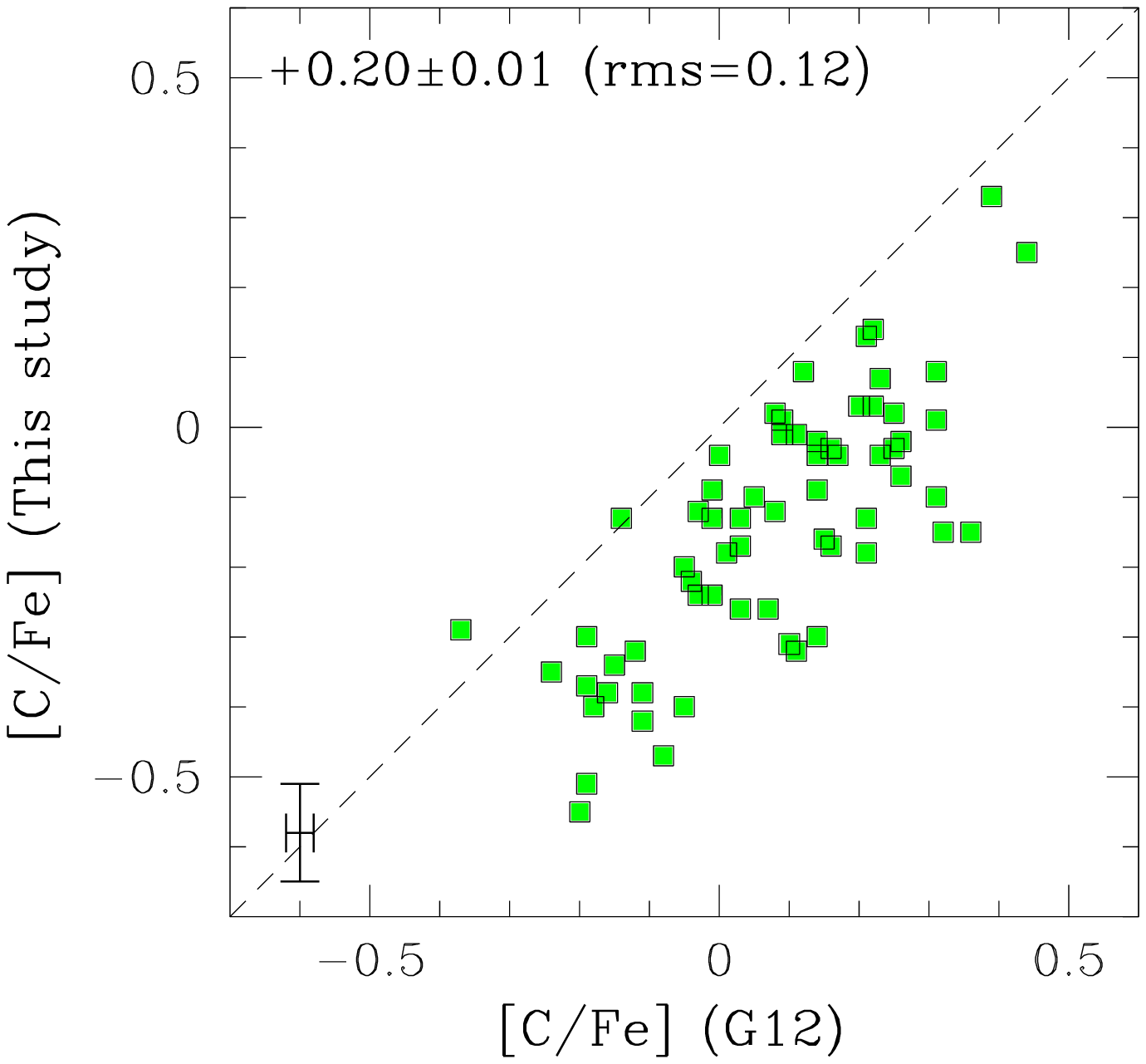}
    \includegraphics[width=5.cm]{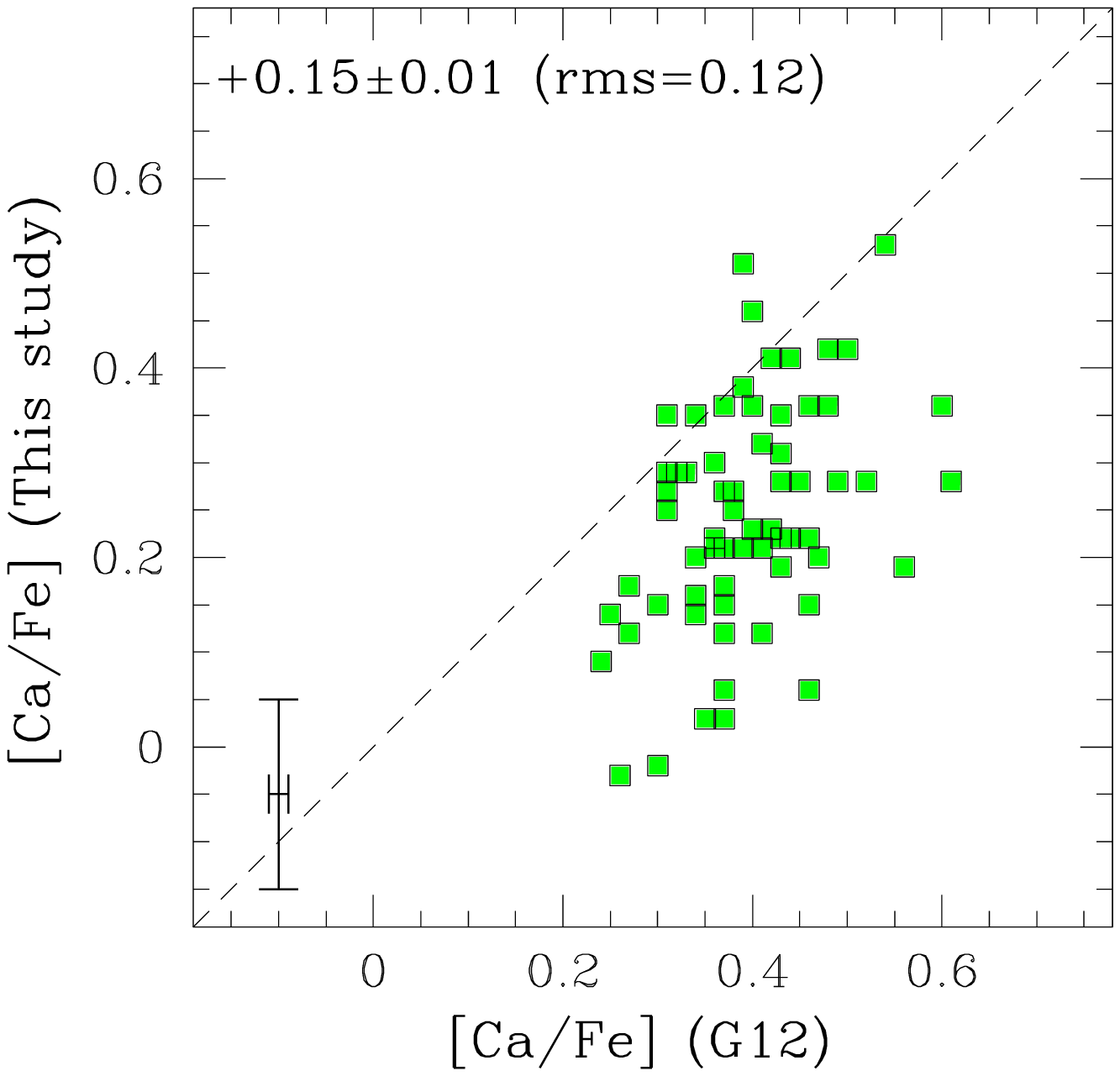}
    \includegraphics[width=5.cm]{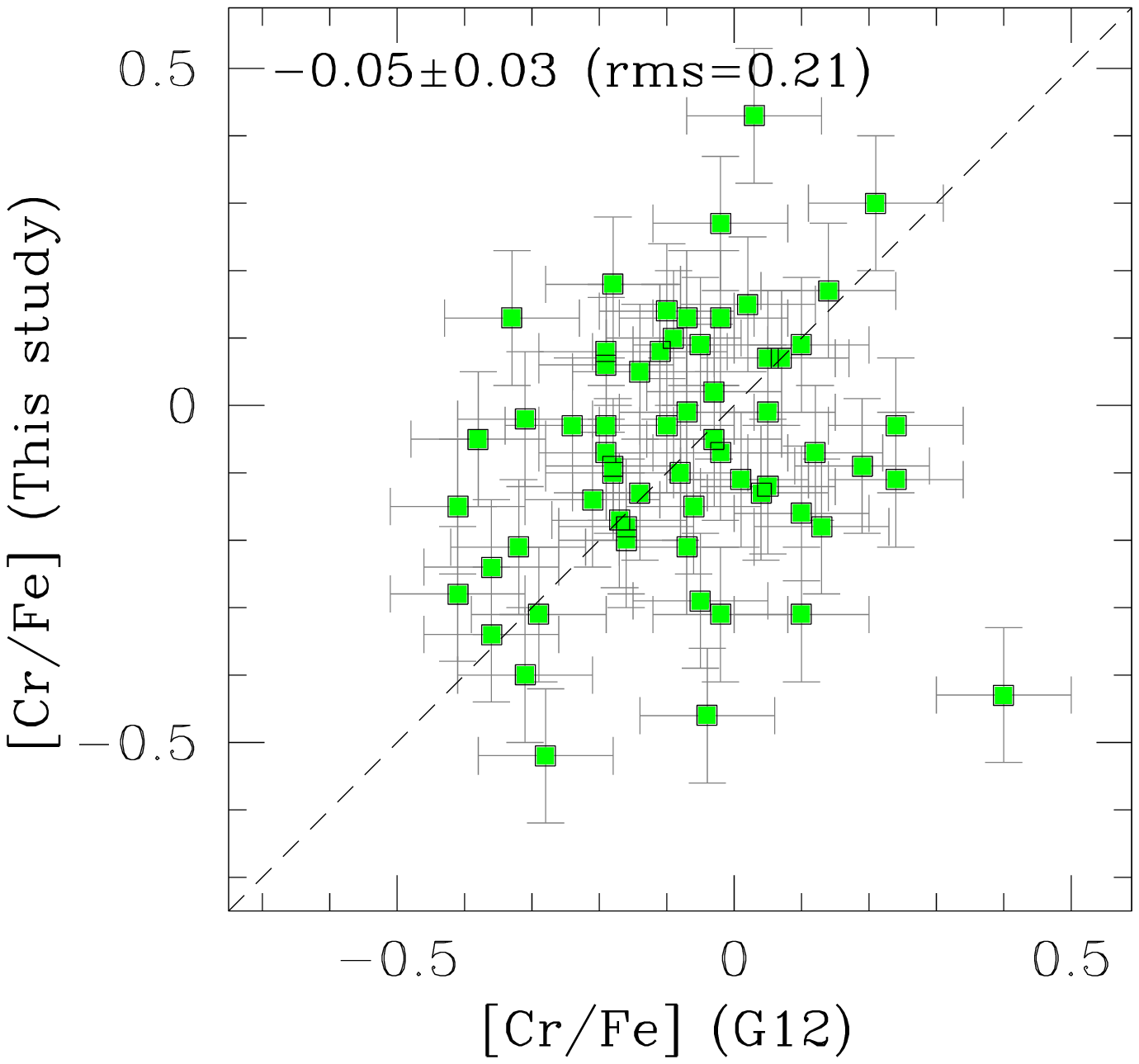}
    \includegraphics[width=5.cm]{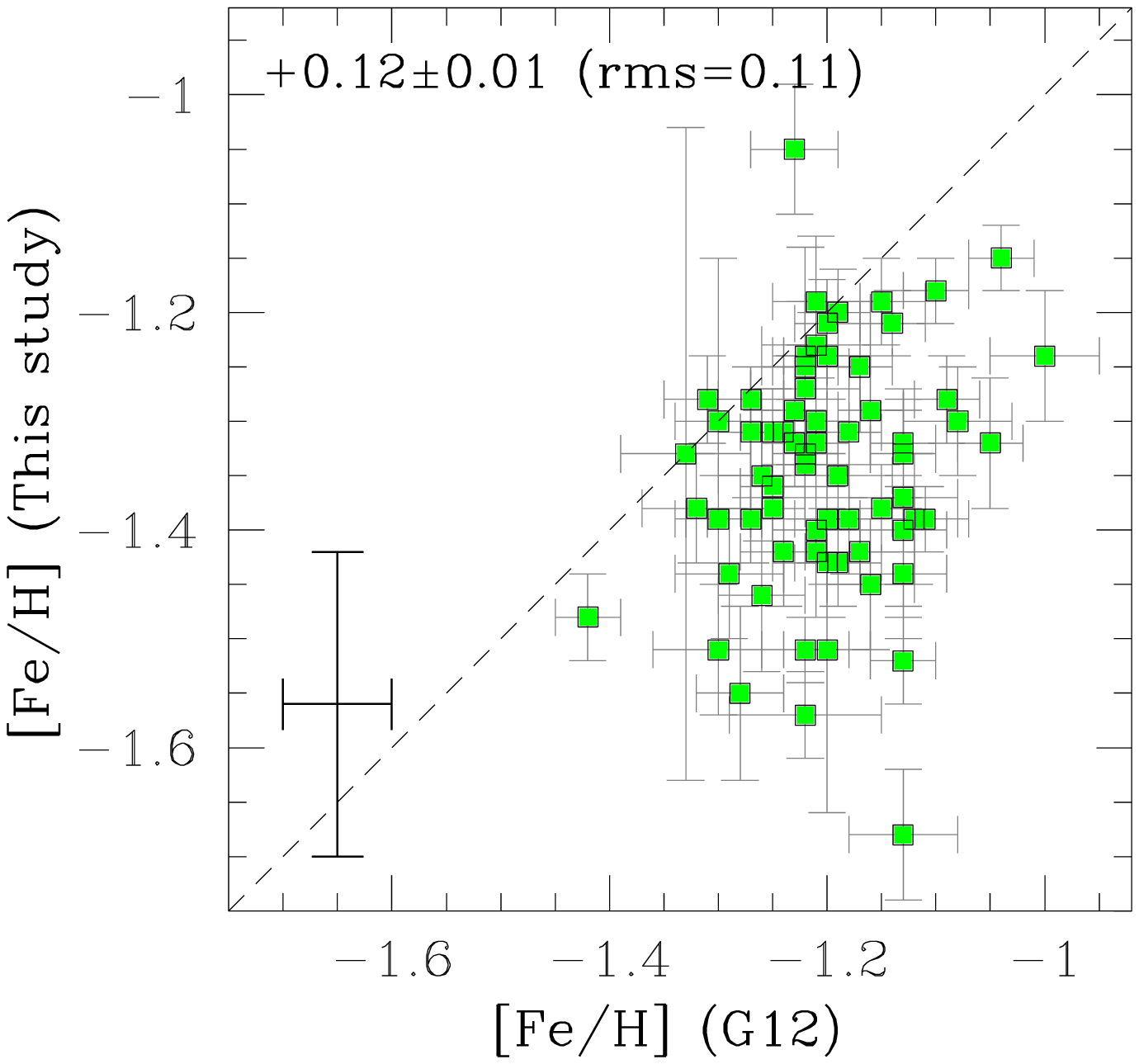}
    \includegraphics[width=5.cm]{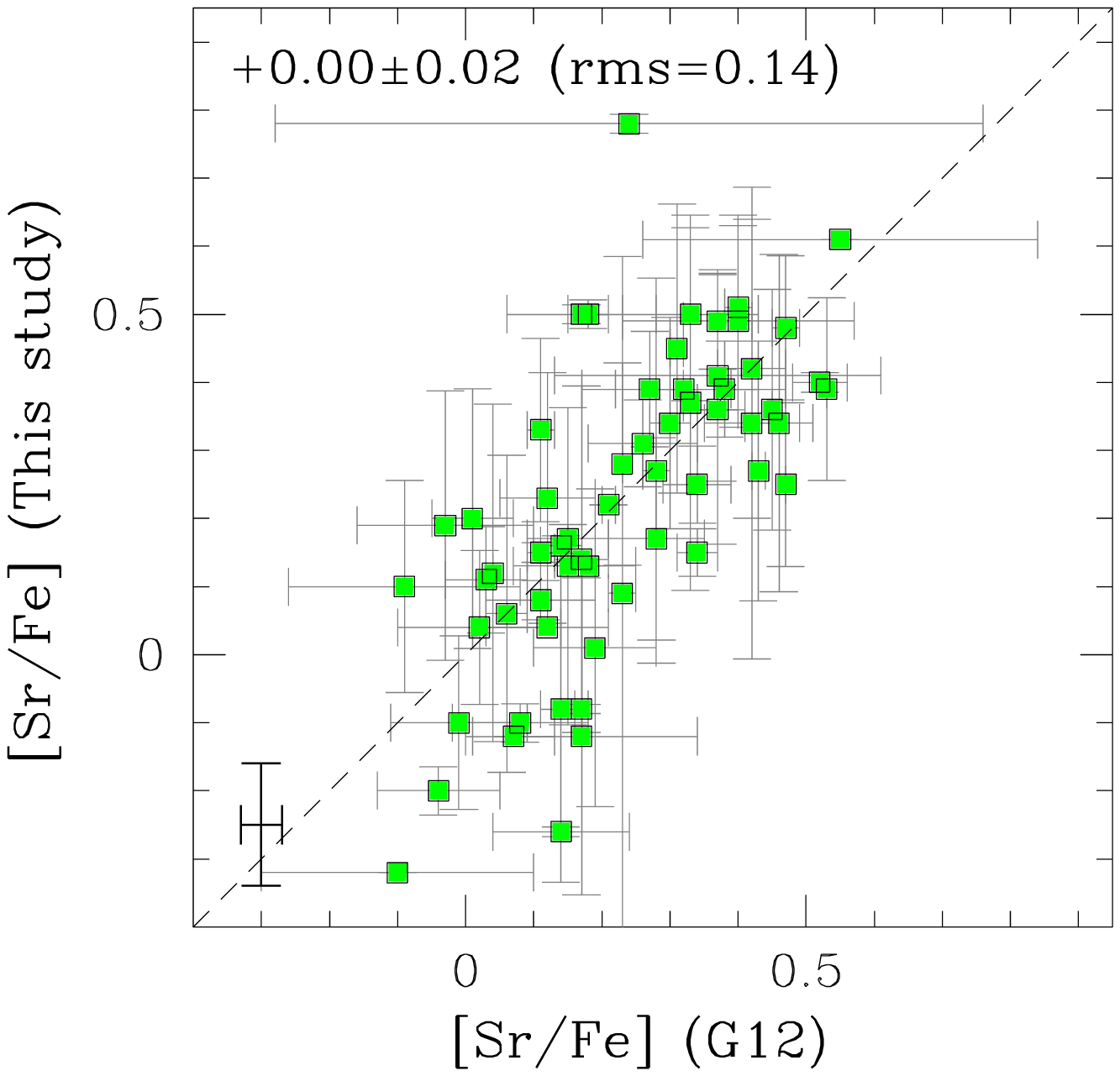}
    \includegraphics[width=5.cm]{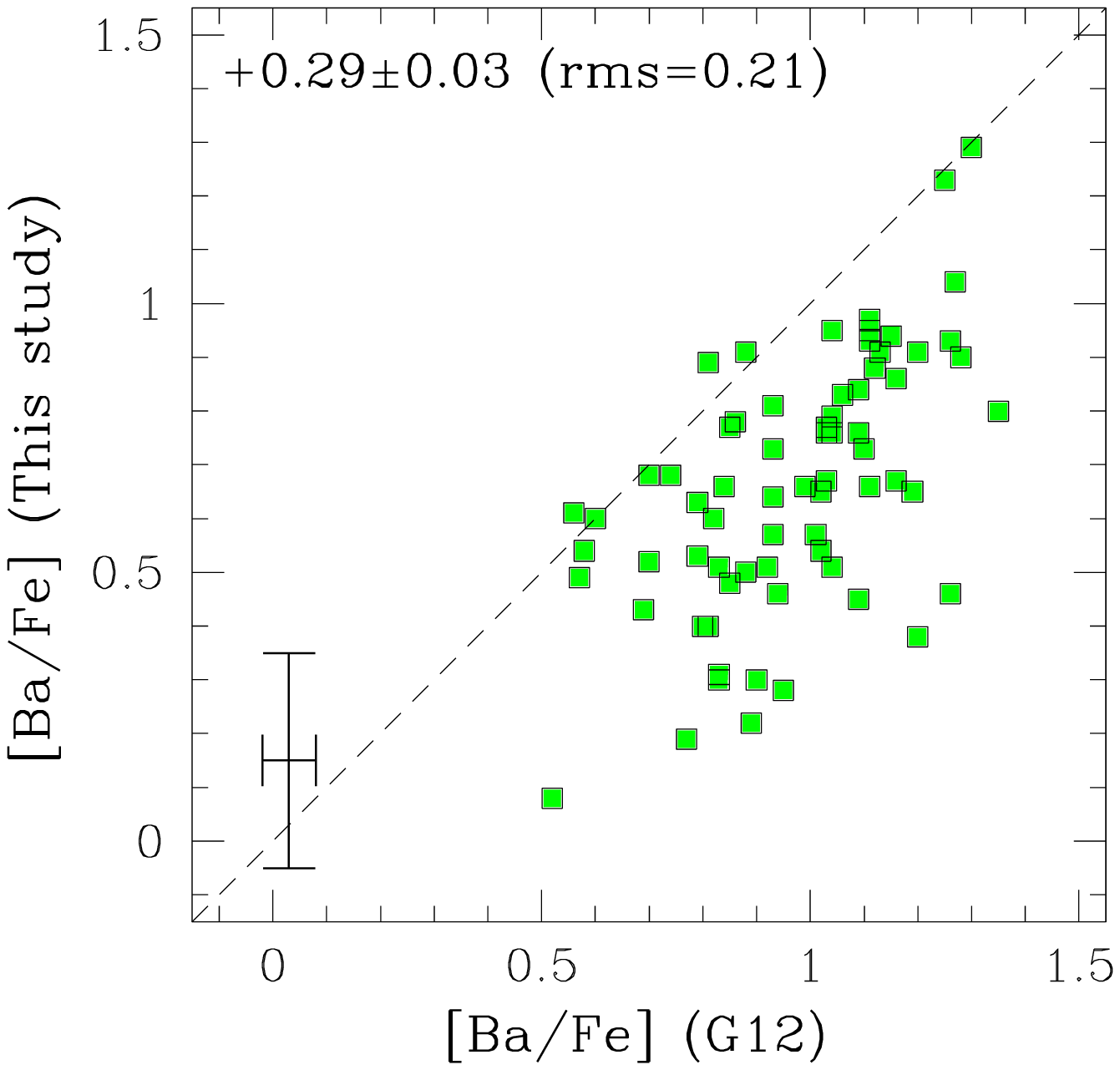}
    \caption{Chemical abundances inferred in this study for the
      NGC\,1851 central field targets as a
      function of those from G12. The mean differences
      (G12$-$adopted) and associated rms are shown in each panel. }
    \label{fig:literatureABB}
   \end{figure*}

A comparison of our chemical contents inferred for the central field
with those from G12 is shown in Fig.~\ref{fig:literatureABB}, for those species analysed
both in this study and in G12.
This comparison reveals some systematics between our abundances and theirs, the most of which can be explained by the systematic differences in the adopted atmospheric parameters discussed in Sect.~\ref{sec:atm}. 
As an example the difference in [Fe/H] of $\sim$0.20~dex can be entirely ascribed to the systematically higher 
\teff\ values in G12; while the large systematic in [Ba/Fe] of $\sim$0.30~dex are mostly due to our higher \vmicro\, and in minor part, to our slightly lower \logg.
We note that the systematic effects on stellar parameters seem to cancel each other in the cases of [Cr/Fe] and [Sr/Fe]; the systematics observed in [C/Fe] and [Ca/Fe] cannot be explained by differences in the atmospheric parameters, and may instead be due to different linelists and/or to possible systematics in the continuum placement.

The values of the observed rms for those elements with no significant internal variations can be used as a rough estimate of our internal errors, to be compared with the expected ones.
In Table~\ref{tab:medie} we list  the 68th percentile of the distributions of the inferred abundances, together with the median values for the NGC\,1851 halo stars and the NGC\,1851 cluster stars.
By comparing the 68th percentile values listed in the third column, with the expected total uncertainties (Table~\ref{tab:err}), we note that, 
in general the errors are in rough agreement. 
The element that clearly stands out 
 is strontium. We will discuss this point in the next section.

\begin{table*}
\caption{Median and 68th percentile ($\sigma_{\rm obs}$) of the chemical abundances for the NGC\,1851 halo and the cluster total, bSGB and fSGB samples. \label{tab:medie}}
\begin{tabular}{lcccccccccccc}
\hline\hline
              &  \multicolumn{2}{c}{HALO}     & &\multicolumn{2}{c}{CLUSTER}      &       &          \multicolumn{2}{c}{bSGB CLUSTER}  &&\multicolumn{2}{c}{fSGB CLUSTER} \\
              &   avg. & $\sigma_{\rm obs}$&    &   avg.                       & $\sigma_{\rm obs}$ &  &avg.&$\sigma_{\rm obs}$&&avg.&$\sigma_{\rm obs}$ \\\hline
{\rm [Fe/H]}  &   $-$1.35$\pm$0.02 & 0.09 &   &$-$1.33$\pm$0.01   &0.09   &   & $-$1.35$\pm$0.01 & 0.08 & & $-$1.30$\pm$0.02&0.11 \\
{\rm [C/Fe]}  &   $-$0.27$\pm$0.09 & 0.17 &   &$-$0.13$\pm$0.02   &0.18   &   & $-$0.09$\pm$0.02 & 0.15 & & $-$0.24$\pm$0.04&0.18 \\
{\rm [Mg/Fe]} &   $+$0.51$\pm$0.02 & 0.03 &   &$+$0.44$\pm$0.02   &0.16   &   & $+$0.41$\pm$0.02 & 0.15 & & $+$0.54$\pm$0.03&0.13 \\
{\rm [Ca/Fe]} &   $+$0.31$\pm$0.04 & 0.10 &   &$+$0.25$\pm$0.01   &0.11   &   & $+$0.25$\pm$0.02 & 0.10 & & $+$0.28$\pm$0.03&0.13 \\
{\rm [Cr/Fe]} &   $-$0.09$\pm$0.25 & 0.45 &   &$-$0.05$\pm$0.02   &0.18   &   & $-$0.05$\pm$0.03 & 0.18 & & $-$0.05$\pm$0.03&0.16 \\
{\rm [Sr/Fe]} &   $+$0.10$\pm$0.06 & 0.15 &   &$+$0.25$\pm$0.03   &0.23   &   & $+$0.13$\pm$0.03 & 0.21 & & $+$0.39$\pm$0.02&0.11 \\
{\rm [Ba/Fe]} &   $+$0.52$\pm$0.09 & 0.28 &   &$+$0.66$\pm$0.03   &0.24   &   & $+$0.57$\pm$0.03 & 0.17 & & $+$0.83$\pm$0.03&0.16 \\
\hline
\end{tabular}
\end{table*}

\section{The chemical content of the NGC\,1851 system}\label{sec:abb}

The mean chemical abundances for the NGC\,1851-RV like stars are
listed in Table~\ref{tab:medie}, along with those obtained for the central
field.
In the following, we first discuss results for the NGC\,1851 cluster
stars in the internal field, so that the chemical composition of the
NGC\,1851 halo stars can be compared with those of the central cluster. 

   \begin{figure}
    \includegraphics[width=8.4cm]{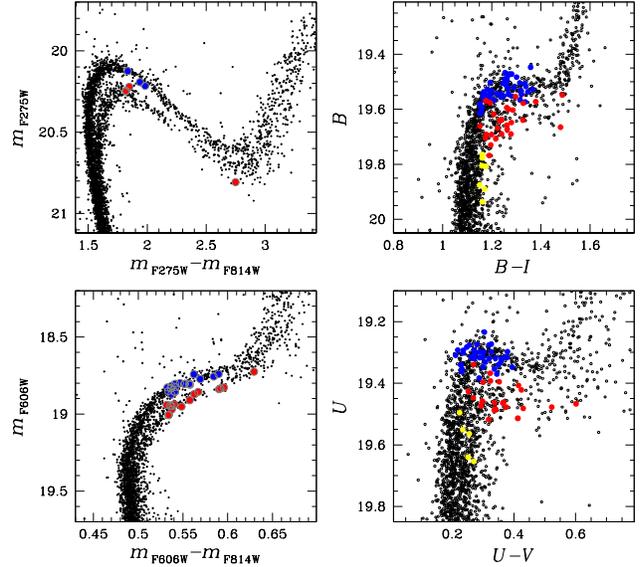}
    \caption{Distribution of the inner field spectroscopic targets 
      on the $m_{\rm F275W}$-$(m_{\rm F275W}-m_{\rm F814W})$ (from Piotto et al. 2012), 
      $m_{\rm F606W}$-$(m_{\rm F606W}-m_{\rm F814W})$ (from Milone et al. 2008),
      $B$-$(B-I)$, and $U$-$(U-V)$ CMDs (Milone et al. 2009).
      Stars assigned to the bSGB and fSGB have been represented 
      in blue and red, respectively, stars
      with more uncertain location respect to the double SGB have been coloured in yellow.}
    \label{fig:sgb_phot}
   \end{figure}

   \begin{figure}
   \includegraphics[width=8.5cm]{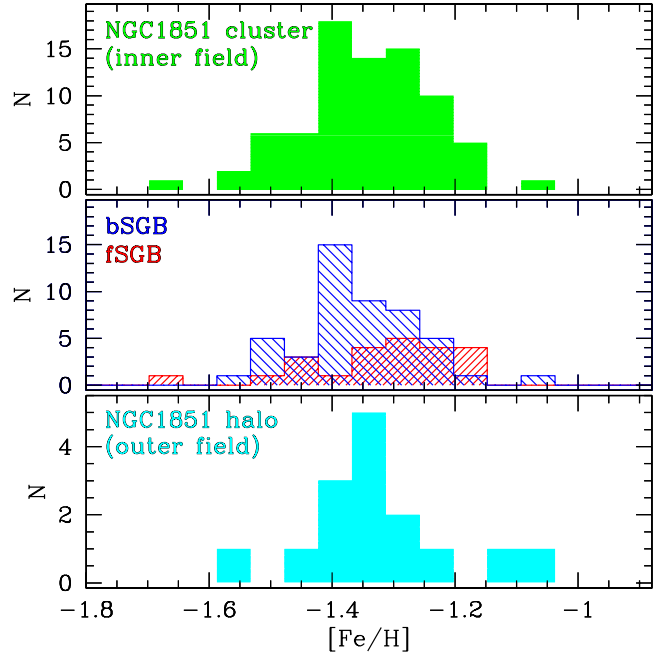}
    \caption{Distribution of the [Fe/H] abundances for the NGC\,1851
      stars in all the stars in the inner field (upper panel), the bSGB and fSGB (middle panel), and for the NGC\,1851-RV like halo stars (lower panel). 
          }
    \label{fig:histoCL}
   \end{figure}

   \begin{figure*}
    \includegraphics[width=5.6cm]{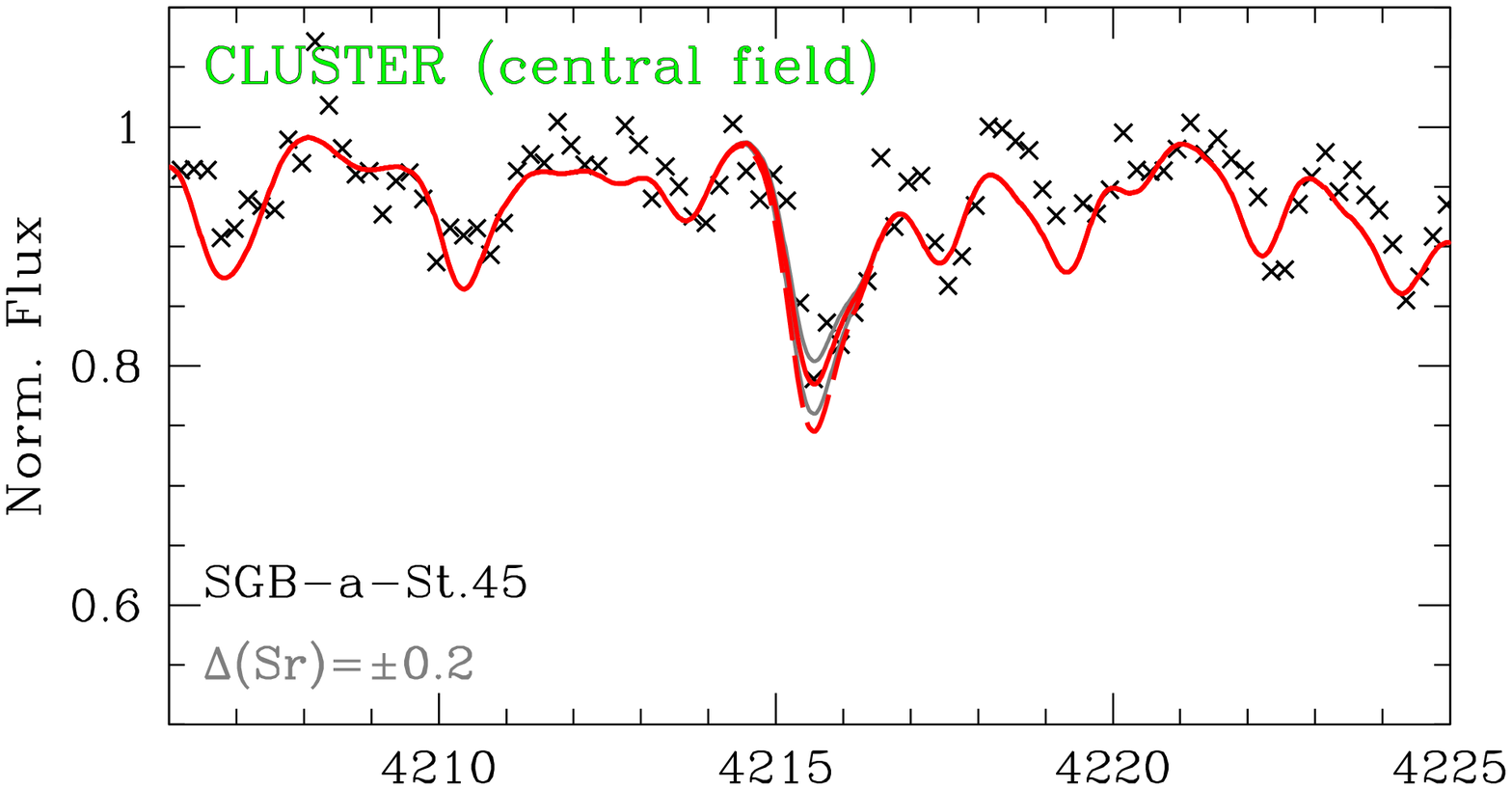}
    \includegraphics[width=5.6cm]{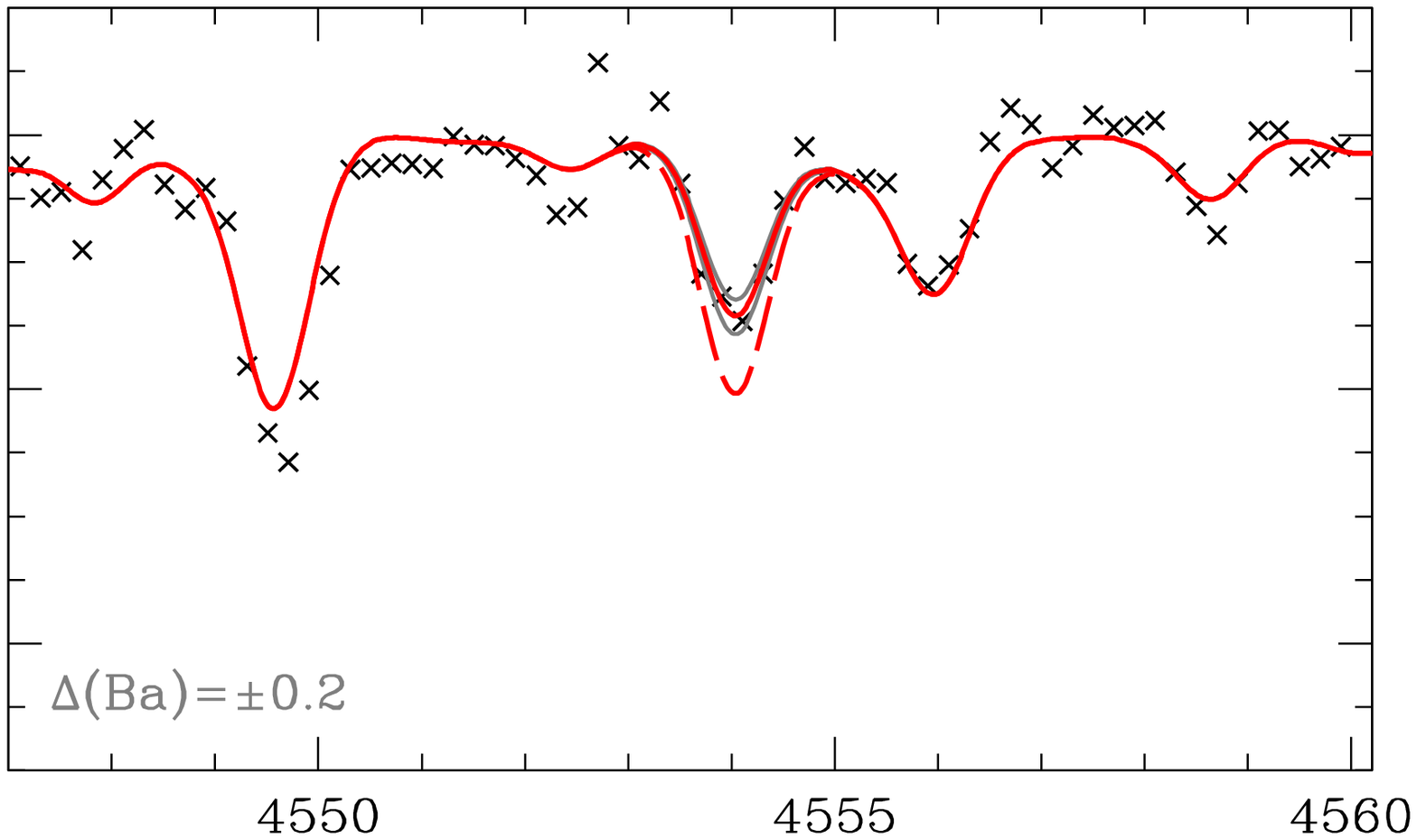}
    \includegraphics[width=5.6cm]{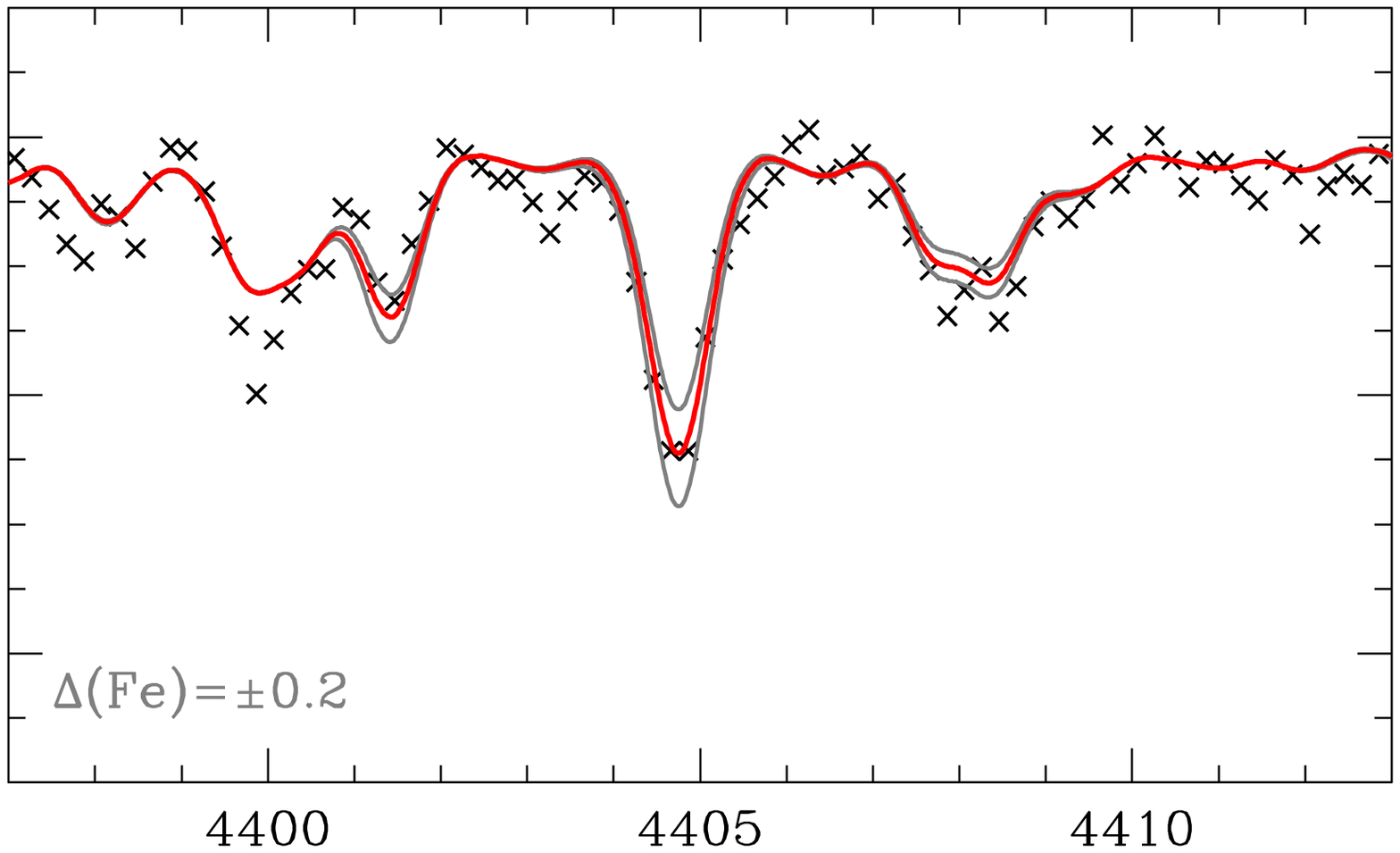}
    \includegraphics[width=5.6cm]{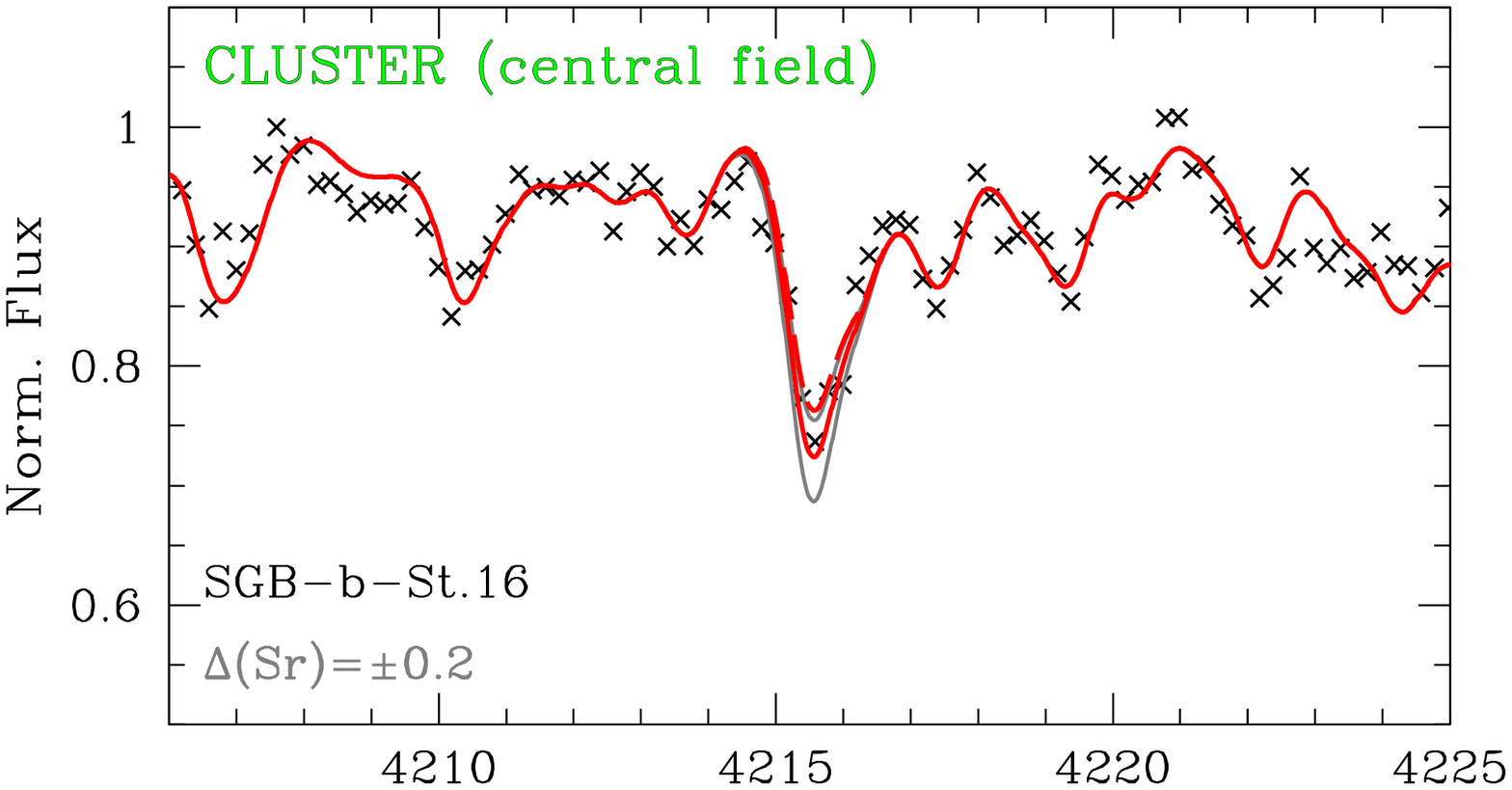}
    \includegraphics[width=5.6cm]{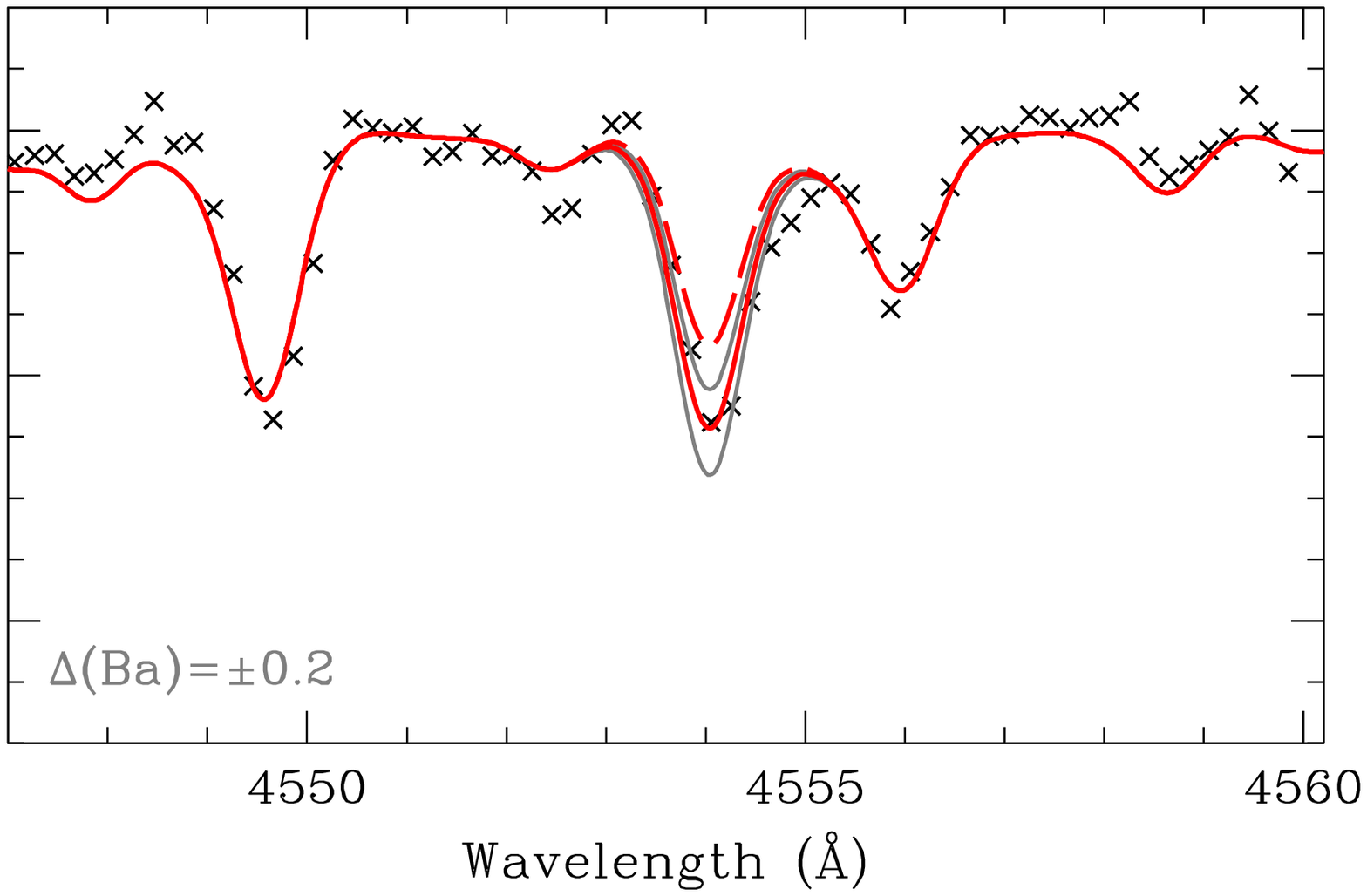}
    \includegraphics[width=5.6cm]{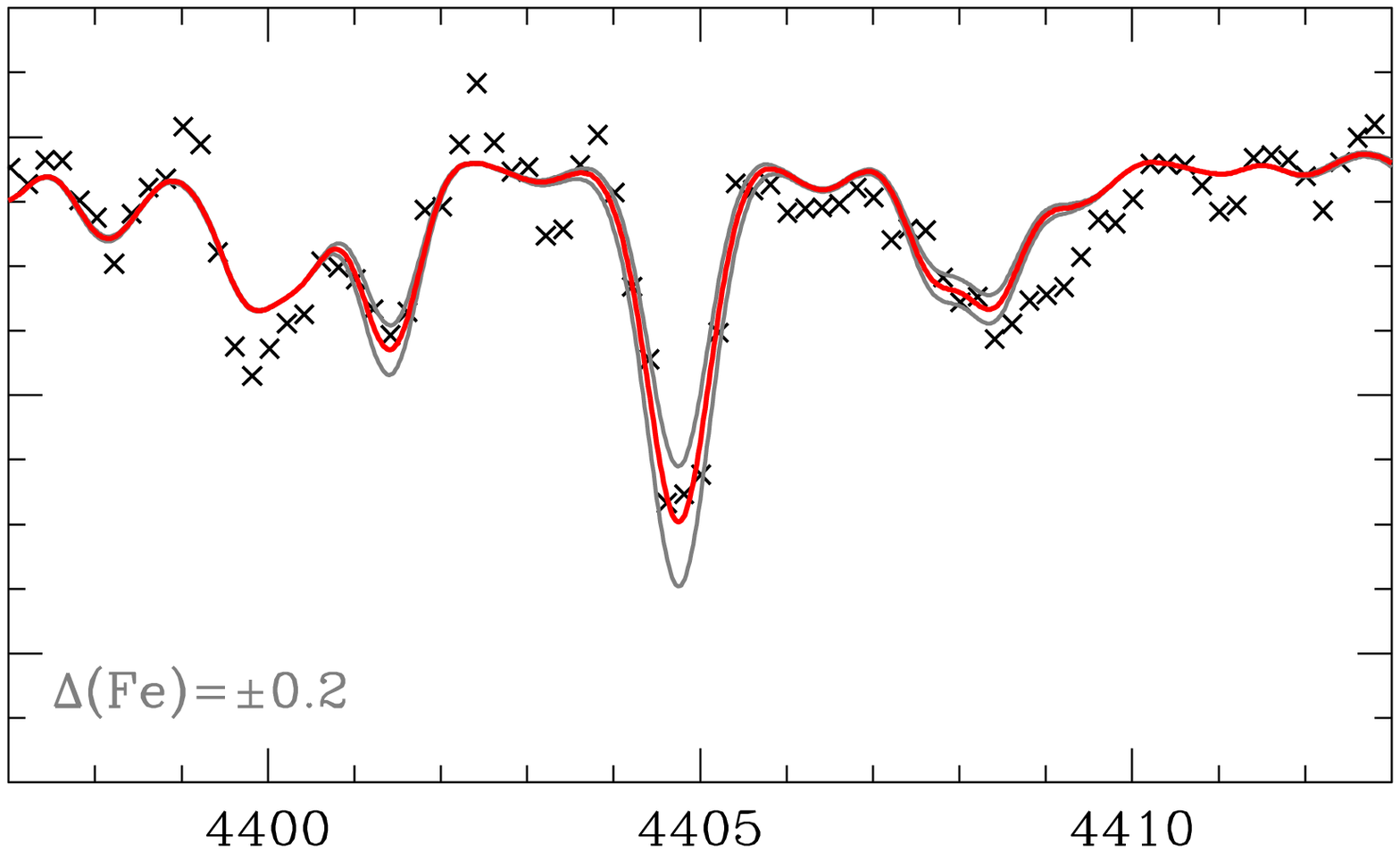}
    \caption{ Some examples of observed and synthetic spectra around some measured Fe, Sr, and Ba features for two SGB stars in the inner field of NGC\,1851. In the upper panel we plot the spectrum of one bSGB star (SGB-a-St.45), while in the lower panel the spectrum of one fSGB star (SGB-b-St.16). In each panel the points represent the observed spectrum. The red line is the best-fitting synthesis; to visualize the difference between the mean fSGB and bSGB abundances we represent with dashed red lines the synthesis corresponding to the average fSGB abundances (upper panel), and to the average bSGB ones (lower panel); the grey lines are the syntheses computed with abundances altered by $\pm$0.2~dex from the best value.}
    \label{fig:synthesis_example}
   \end{figure*}

\subsection{The inner field composition }\label{sec:abb_clu}

The chemical composition of NGC\,1851 is not {\it standard} for
a GC, as it shows internal variations in $s$-process elements, that is
a peculiarity of a few GCs such as those showing
split SGBs (e.g., M\,22 and $\omega$~Centauri).
A proper comparison between the external and internal field of
this GC requires accounting for these features.

Starting with the analysis of the cluster stars in the internal field, we have assigned
each target to the faint SGB (fSGB) or the bright SGB (bSGB), based on their position on the CMD. 
In Fig.~\ref{fig:sgb_phot} we show the position of these central field targets on
the CMD in various photometric bands. 
An inspection of this figure suggests that the separation 
between bSGB and fSGB is much more distinct by using $ACS$ and $WFC2$ 
images from $HST$ (left panels), than by using standard broad-band photometry 
from ground (right panels).
As the photometry from space is more precise,
we
use the $HST$ photometric information to assign each star to the bSGB or the fSGB, when available.
In Fig.~\ref{fig:sgb_phot} the bSGB and fSGB spectroscopic targets have 
been represented in blue and red, respectively.
These color codes will be used consistently in the following discussion.
In some cases, the association with one of the two SGB is not obvious 
for stars having only ground-based photometry, 
as some targets lie on the MS where the split is not
visible. 
So, we define 
a third group of stars whose position on the CMD is ambiguous 
(yellow dots in Fig.~\ref{fig:sgb_phot}), and we do not consider them in the comparison with the halo stars chemical composition.
In total we have identified 18 bSGB and 15 fSGB in the $HST$ field, and 32 bSGB and 14 fSGB in the ground-based photometry, with 7 uncertain stars.

To properly compare the abundances obtained in the halo with those of
the stellar groups observed in the cluster we derive the average
abundances of the bSGB and fSGB samples, as listed in Table~\ref{tab:medie}.
In the following discussion the comparison between the average abundances of the various analysed groups of stars (bSGB, fSGB and halo) has been performed by using as error the quadratic sum of the errors associated with the considered means.   

The [Fe/H] histogram of the two SGB groups is shown in the upper panel of Fig.~\ref{fig:histoCL}.
The total distribution of the entire sample, represented in green, is entirely within that expected from the error analysis of our relatively low resolution data.
By dividing the sample into bSGB and fSGB, our mean [Fe/H] values agree within observational errors, suggesting
that the two populations have the same average metallicity within $\sim$0.05~dex 
($\Delta_{\rm fSGB-bSGB}=+0.05\pm0.03$), 
confirming previous studies on the RGB based on higher quality data 
(Villanova et al. 2010; Carretta et al. 2011).
A smaller difference between the iron content of the two SGBs cannot be ruled out by our data.
We recall here that the trends in the [X/H] abundances
cannot be interpreted in support of any overall metallicity variation as they 
can result from our observational errors.

The Strontium-to-iron ratio, [Sr/Fe], exhibits a large star-to-star variation, exceeding the observational errors. 
By dividing stars in bSGB and fSGB the [Sr/Fe] mean abundance of the fSGB is larger than that inferred for the bSGB,
with a mean difference $\Delta$[Sr/Fe]$_{\rm bSGB-fSGB}$=$-$0.25$\pm$0.04  (a $\sim$6$\sigma$ significant difference). 
Despite the large errors on [Ba/Fe], the fSGB is also richer in barium 
by 0.26$\pm$0.04~dex (with a significance of more than 6~$\sigma$).  
In summary, our results on the internal field confirm previous findings 
on the presence of two groups of stars in NGC\,1851
with different abundances in those elements mostly 
produced in the $s$-processes, like Sr and Ba (e.g. Yong \& Grundahl 2008; Villanova et al.\,2010), and that the fSGB is 
populated by stars that have undergone some kind of enrichment from these processes (e.g. Gratton et al.\, 2012).
In this respect, NGC\,1851 is very similar to M\,22, which shows a bimodal distribution in $s$-elements (Marino et al.\,2009) and a fSGB populated by stars with higher $s$-element abundances (Marino et al.\,2012a). 

As a visual representation of the differences in Sr and Ba between the two SGBs,
in Fig.~\ref{fig:synthesis_example} we represent some portions of the analysed spectra around key 
spectral features for one bSGB star (SGB-a-St.45) and one fSGB star (SGB-b-St.16).
Superimposed on the observed spectra are the best-fit synthesis (solid red), two syntheses 
corresponding to Sr, Ba, and Fe abundances varied by $\pm$0.2~dex, and a synthesis computed
with the average abundance of the fSGB and the bSGB (dashed red) for SGB-a-St.45 and SGB-b-St.16, respectively.

The high dispersion in [C/Fe], significantly exceeding observational errors, suggests that the abundance in this element 
shows significant internal variations.
This is consistent with what was observed in other GCs, including the 
less complex ones (e.g., M~4, Marino et al. 2008), and mostly interpreted as 
due to some kind of intra-cluster pollution from material that has undergone H-burning at high temperature.
It is worth noting here that the high dispersion in [C/Fe] observed in the complete sample does not 
diminish by dividing stars in bSGB and fSGB, being only marginally lower for the bSGB (Tab.~\ref{tab:medie}).
The average [C/Fe] is higher in the bSGB, with a mean difference
  of $\Delta$([C/Fe])$_{\rm bSGB - fSGB}$=0.15$\pm$0.045, that is
  slightly higher than a 3~$\sigma$ level.
The chemical pattern of [C/Fe], its variation within different
  $s$-groups and the larger spread among the $s$-poor (bSGB) stars
  agrees with previous findings obtained first for M22 (Marino et
  al. 2011b; 2012a) and then for NGC\,1851 itself (Lardo et al.\ 2012;
  Gratton et al.\ 2012).

We note that [Mg/Fe] appears to be slightly higher in fSGB stars
  (at a level of $\sim$3.5~$\sigma$). Magnesium abundances for RGB
  stars in NGC\,1851 have been provided by Yong et al. (2008) from
  UVES spectra and Carretta et al. (2010) from GIRAFFE
  spectra. 
  The average [Mg/Fe] abundances for their sample of RGB stars is
  $+$0.38$\pm$0.03 ($\sigma$=0.07) and $+$0.38$\pm$0.01
  ($\sigma$=0.04) in Yong et al. and Carretta et al., respectively.
  Although a direct comparison with these two studies cannot be done,
  because they use different spectral Mg features and possible non-LTE
  corrections may apply differently. We note that 
  their values are consistent with the average abundance that we found for the
  total sample of SGB stars, [Mg/Fe]$_{\rm all SGB}$=$+$0.44$\pm$0.02
  ($\sigma$=0.16), and for the bSGB stars, [Mg/Fe]$_{\rm bSGB}$=$+$0.41$\pm$0.02
  ($\sigma$=0.15). On the other hand, the Mg abundance of the fSGB
  appears to be higher, but a similar effect has not been found in
  previous studies on the RGB. Given our large uncertainties
  associated with individual Mg abundance measurements, the presence
  of this difference should be viewed with caution and needs to be
  investigated further. 
None of the other species appears to show differences among the two
SGBs over a 3~$\sigma$ level.

\begin{table*}
\caption{Sensitivity of derived abundances to the atmospheric parameters and the fitting procedure. We reported the total formal error ($\sigma_{\rm total}$)  due to                                                  
  the atmospheric parameters plus errors in the fitting procedure ($\sigma_{\rm fit}$). \label{tab:err}}
\begin{tabular}{lcccccc}
\hline\hline
      &$\Delta$\teff &$\Delta$\logg&$\Delta$\vmicro&$\Delta$[A/H]& $\sigma_{\rm fit}$&$\sigma_{\rm total}$\\  
      & $\pm$100~K    & $\pm$0.05   & $\pm$0.20~\kmsec & $\pm$0.11~dex &     &                   \\
\hline
\\
\multicolumn{7}{c}{HALO}\\
\multicolumn{7}{c}{}\\
\\
$\rm {[C/Fe]}$  & $\pm$0.04           & $\pm$0.01 & $\pm$0.02 & $\mp$0.02           & $\pm$0.10/$\pm$0.16 & 0.11/0.17 \\
$\rm {[Mg/Fe]}$ & $\mp$0.08           & $\mp$0.01 & $\pm$0.01 & $\pm$0.01           & $\pm$0.13/$\pm$0.18 & 0.15/0.20 \\
$\rm {[Ca/Fe]}$ & $\pm$0.03           & $\pm$0.00 & $\pm$0.02 & $\mp$0.02           & $\pm$0.09/$\pm$0.13 & 0.10/0.14 \\
$\rm {[Cr/Fe]}$ & $\pm$0.17           & $\mp$0.01 & $\mp$0.02 & $\mp$0.02           & $\pm$0.18/$\pm$0.38 & 0.25/0.42 \\
$\rm {[Fe/H]}$  & $\pm$0.10           & $\mp$0.02 & $\mp$0.02 & $\mp$0.10/$\mp$0.07 & $\pm$0.02/$\pm$0.07 & 0.15/0.14 \\
$\rm {[Sr/Fe]}$ & $\mp$0.01           & $\pm$0.02 & $\mp$0.02 & $\pm$0.02           & $\pm$0.12/$\pm$0.20 & 0.13/0.20 \\
$\rm {[Ba/Fe]}$ & $\mp$0.07/$\mp$0.04 & $\mp$0.11 & $\mp$0.15 & $\pm$0.02           & $\pm$0.09/$\pm$0.21 & 0.22/0.29 \\
\\
\multicolumn{7}{c}{CLUSTER}\\
\multicolumn{7}{c}{}\\
$\rm {[C/Fe]}$  & $\pm$0.04   & $\pm$0.01  & $\pm$0.03  & $\mp$0.02 & $\pm$0.06 & 0.08 \\
$\rm {[Mg/Fe]}$ & $\mp$0.08   & $\mp$0.01  & $\pm$0.00  & $\mp$0.03 & $\pm$0.13 & 0.16 \\
$\rm {[Ca/Fe]}$ & $\pm$0.04   & $\mp$0.01  & $\pm$0.02  & $\mp$0.02 & $\pm$0.08 & 0.09 \\
$\rm {[Cr/Fe]}$ & $\pm$0.06   & $\pm$0.00  & $\mp$0.01  & $\mp$0.05 & $\pm$0.19 & 0.21 \\
$\rm {[Fe/H]}$  & $\pm$0.10   & $\mp$0.01  & $\mp$0.03  & $\mp$0.07 & $\pm$0.03 & 0.13 \\
$\rm {[Sr/Fe]}$ & $\mp$0.04   & $\pm$0.02  & $\pm$0.01  & $\pm$0.04 & $\pm$0.07 & 0.09 \\
$\rm {[Ba/Fe]}$ & $\mp$0.03   & $\mp$0.11  & $\mp$0.17  & $\pm$0.00 & $\pm$0.11 & 0.23 \\
\hline
\end{tabular}
\end{table*}

\subsection{The NGC\,1851-halo composition }\label{sec:abb_halo}

The chemical abundances and their averages
inferred for the halo stars are listed 
in Table~\ref{tab:abundances} and Table~\ref{tab:medie}, respectively.
Although we could infer metallicities for a relatively large number of halo stars (15), 
the abundances of the other elements were possible only in the stars with higher S/N spectra.

The [Fe/H] distribution of the NGC\,1851 halo stars is presented in the 
lower panel of Fig.~\ref{fig:histoCL}.
An inspection of this figure immediately suggests that 
the distribution for the halo stars spans a range similar to the one observed in the 
internal field. 
The average [Fe/H] for the halo stars is [Fe/H]=$-$1.35$\pm$0.02~dex (rms=0.09), and
its difference with the mean values obtained for the internal field are:
$\Delta$[Fe/H]$_{(\rm cluster-halo)}$=$+$0.02$\pm$0.03~dex; 
$\Delta$[Fe/H]$_{(\rm bSGB~cluster-halo)}$=$+$0.00$\pm$0.03~dex;
$\Delta$[Fe/H]$_{(\rm fSGB~cluster-halo)}$=$+$0.05$\pm$0.03~dex.
Indeed, we conclude that, within observational errors, the [Fe/H] in the stars analysed in the halo 
is consistent with
the mean abundance obtained for the NGC\,1851 cluster stars.

\begin{figure}
    \includegraphics[width=8.4cm]{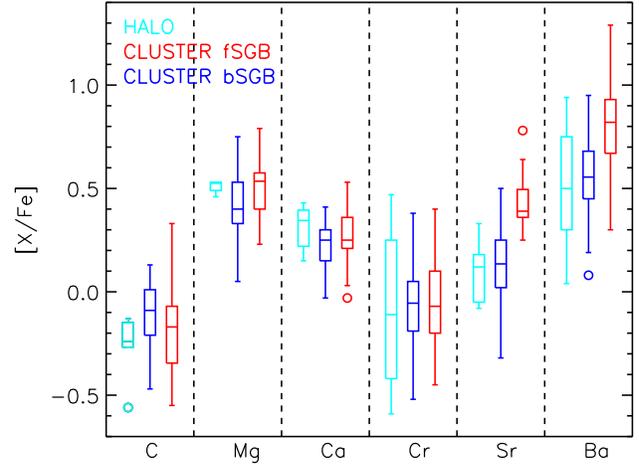}
    \caption{
Box and whisker plot of the bSGB (blue), fSGB (red) and halo (cyan) abundances. A boxed horizontal line indicates the interquartile range (the middle 50\% of the data) and median found for a particular element. The vertical tails extending from the boxes indicate the total range of abundances determined for each element, excluding outliers. Outliers (those 1.5 times the interquartile range) 
are denoted by open circles.
}
    \label{fig:all}
   \end{figure}

A comparison between all the abundances inferred for the halo with those obtained for the 
bSGB and fSGB in the inner field is shown in Fig.~\ref{fig:all}.
The first observation we make is that the [Sr/Fe] distribution for the halo is consistent with that shown by the  
bSGB, with just one star falling in the range spanned by the fSGB.
Although the internal error on [Ba/Fe] is much larger, the [Ba/Fe] range for the 
halo is also more similar to that spanned by bSGB stars than to that of the fSGB.
Note however that, in this case, some stars are also consistent with the fSGB range, but 
due to the large observational errors affecting barium abundances, 
we cannot draw strong conclusions on the distribution of 
this element alone.
To quantify the probabilities that the  abundance distributions in Sr and Ba obtained for the fSGB, bSGB in the inner field and in those for the outer field derive from the same parent distribution, we performed some KS tests. The KS probabilities that the bSGB and the fSGB abundances derive from the same distribution are 0.0000 both for Sr and Ba. Similarly low are the probabilities when we compare the fSGB and the halo, for which we obtain 0.0001 and 0.0047 for Sr and Ba, respectively. On the other hand, the probabilities that the bSGB and the halo Sr and Ba abundances derive from the same distribution are 0.8180 and 0.5622, respectively.

\begin{figure*}
    \includegraphics[width=5.4cm]{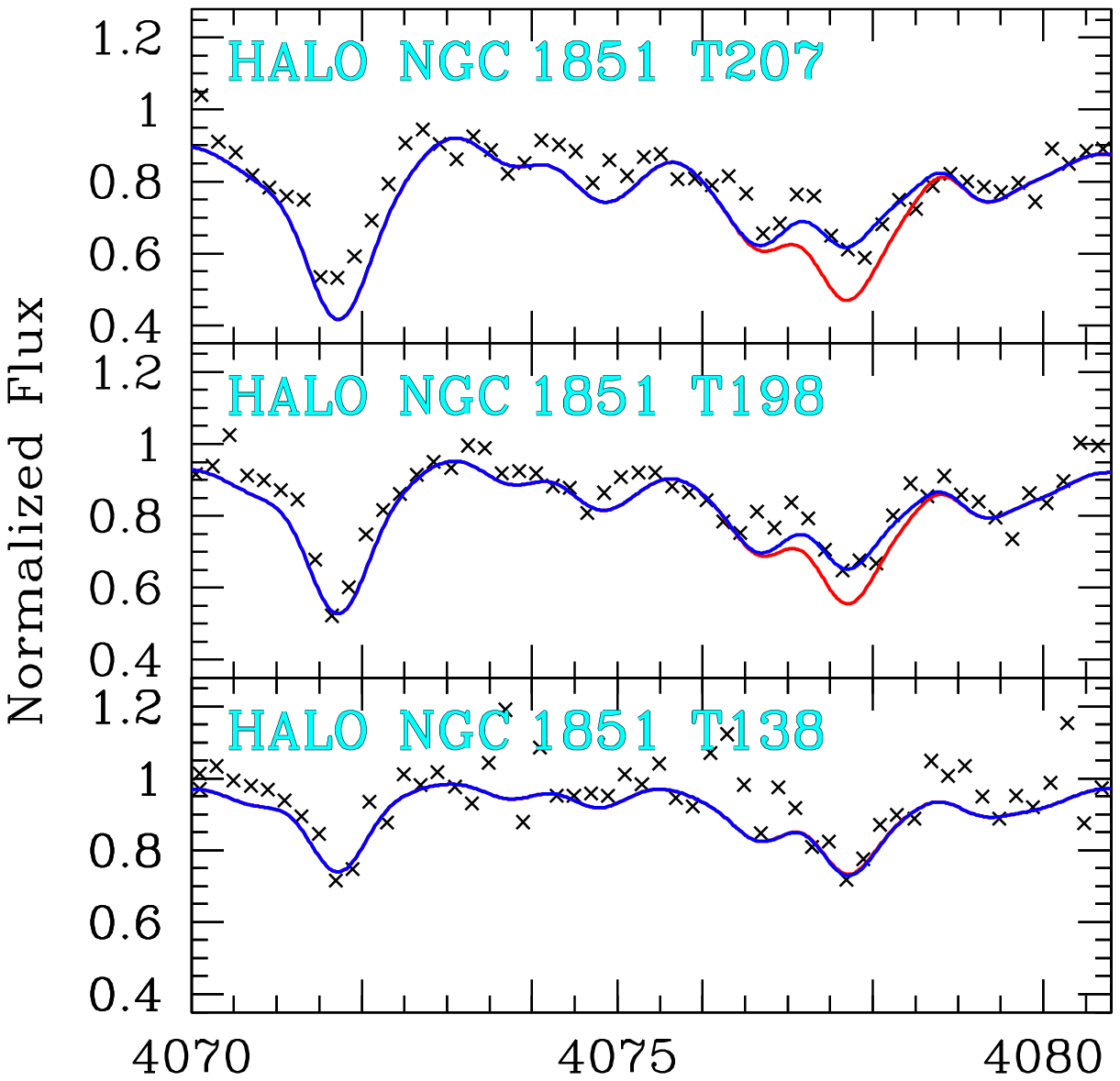}
    \includegraphics[width=5.4cm]{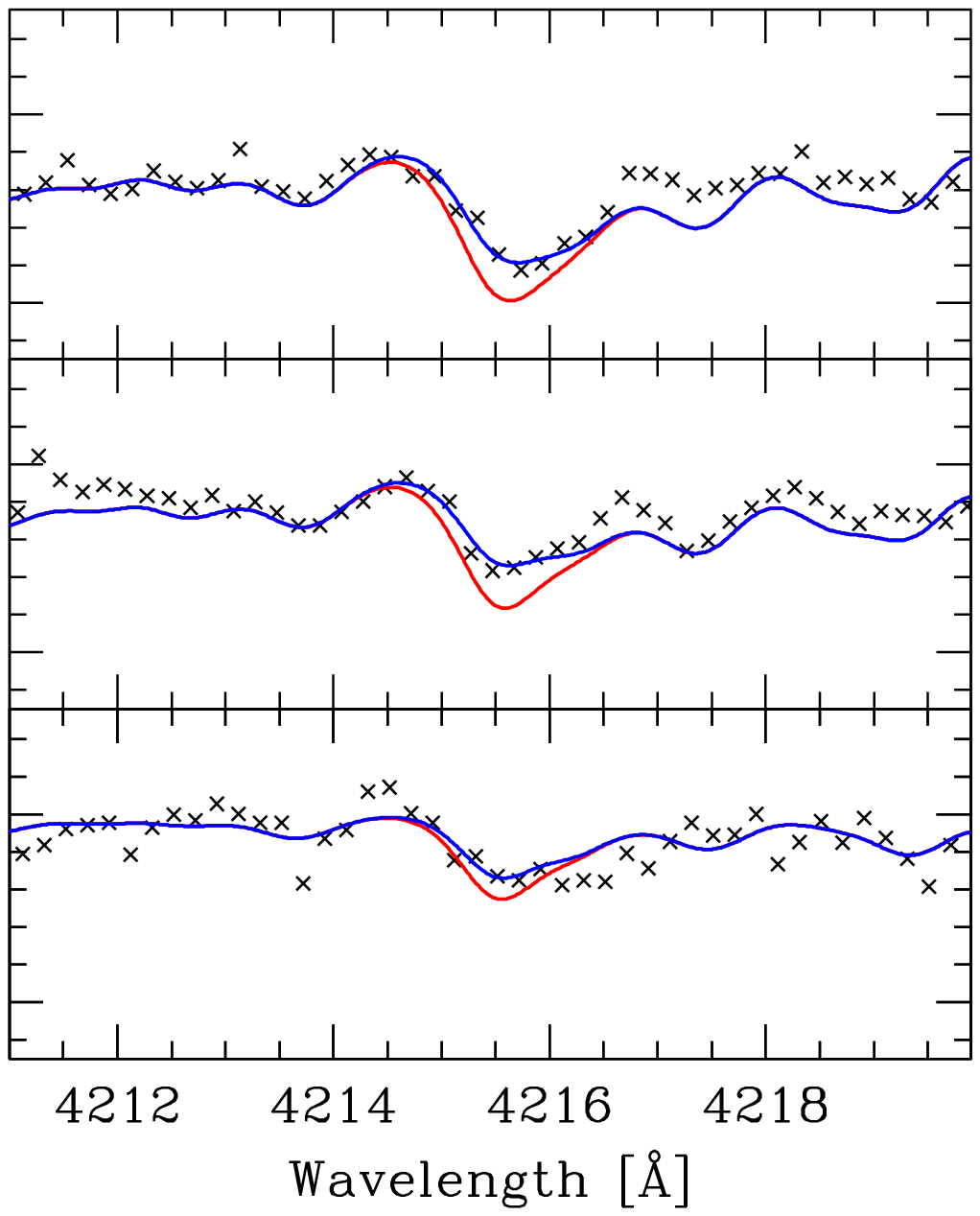}
    \includegraphics[width=5.4cm]{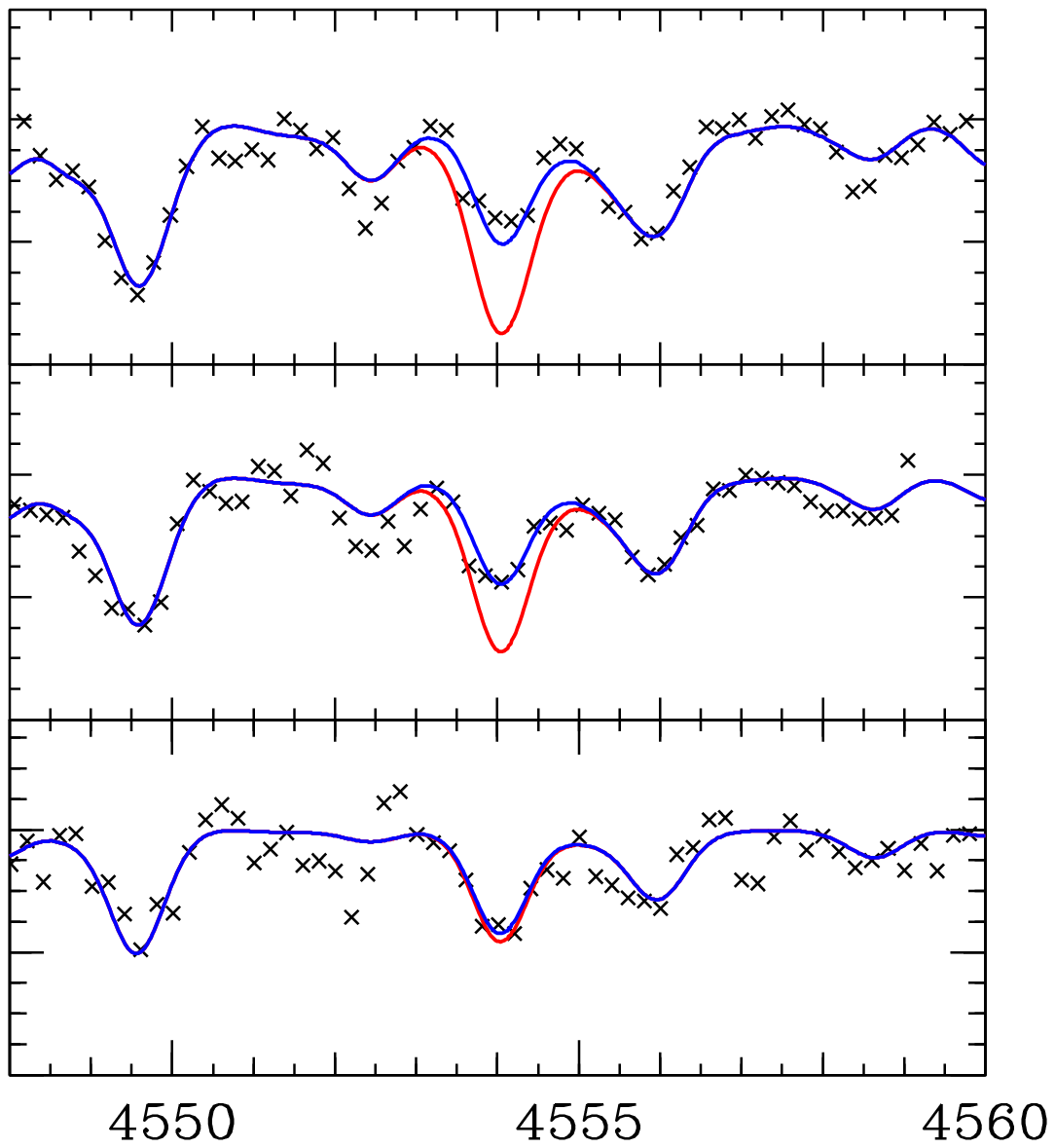}
    \caption{Some examples of observed and synthetic spectra around the measured
      Sr and Ba features for three NGC\,1851 halo stars (T207, T198 and T138). In each panel the
      points represent the observed spectrum, the blue 
      line is the best-fitting synthesis; and the red line is the syntheses computed with the mean abundances obtained for the fSGB in the NGC\,1851 internal field.}
    \label{fig:synthesis_examplehalo}
   \end{figure*}

\begin{figure}
    \includegraphics[width=8.cm]{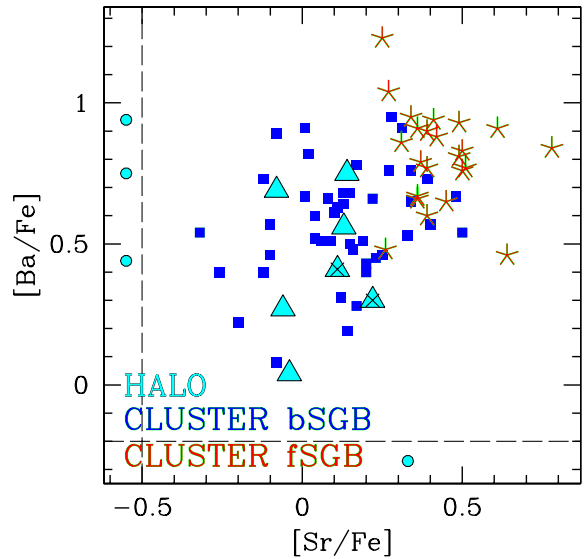}
    \caption{Barium as a function of strontium abundances relative to Fe for the NGC\,1851 system: halo stars are
      shown as cyan triangles, cluster bSGB and 
      fSGB have been represented in blue squares and red stars, respectively. The two halo stars with anomalous position on the CMD have been 
      indicated with black crosses; while stars with just Ba or Sr abundances have been represented with cyan dots.}
    \label{fig:sproc}
   \end{figure}

In Fig.~\ref{fig:synthesis_examplehalo} we show the spectral features for Sr and Ba in three halo stars, with the best-fit synthesis (blue) and the synthesis corresponding to the mean chemical abundances for the fSGB observed in the inner field (red).
This figure illustrates well the limit of our observations: 
e.g., while the chemical abundances of the two RGBs (T207 and T198)
is well distinct from the composition of the fSGB, 
the uncertainties associated with some MS stars (e.g., T138) are much larger.
In the particular case of T138, the star with more ambiguous position on the 
[Ba/Fe]-[Sr/Fe] plane (see Fig.~\ref{fig:sproc}), the abundances for barium 
and strontium are in fact only slightly lower than the fSGB average abundance.

Keeping in mind all the uncertainties we have, in particular for the MS stars, the most robust comparison between the halo and the cluster stars that we can make with our observations is by combining results for the two analysed $s$-process elements Sr and Ba.
In Fig.~\ref{fig:sproc} we show [Ba/Fe] as a function of [Sr/Fe] for the bSGB, the fSGB and the halo stars.
Stars with both Sr and Ba available clearly distribute in the same manner as the bSGB stars.
The two stars not lying on the NGC\,1851 sequence (see Sect.~\ref{subs:rvS10}) have been indicated with black crosses.

For completeness, in Fig.~\ref{fig:sproc} we plot stars with only Sr or Ba available as cyan dots separated by the horizontal and vertical dashed lines, respectively. 
We comment briefly these four stars.
As previously discussed, due to the large errors in [Ba/Fe] we cannot draw conclusions on the three stars with only Ba abundances available.
Indeed, the large uncertainties associated with the single Ba measurements for halo MS stars, that is about 0.29~dex as listed in Tab.~\ref{tab:err}, do not allow us to confidently assign the two out three stars with higher Ba to neither the $s$-rich or the $s$-poor group.
The location of the star with just Sr is inconclusive. 

We can conclude that the analysis of our sample of halo stars thus does not provide strong evidence for the presence of $s$-rich stars corresponding to the fSGB population, insofar as all the stars for which Ba and Sr abundances are available have values compatible with those of the bSGB.

On the probability of observing $s$-rich stars, considering that in the central field the fSGB contributes around 35\% of the stars (Milone et al. 2009), 
we would expect to observe 
1.75/5 (with Sr and Ba abundances available) at the high $s$-process composition of the fSGB.
We observe no such stars. As a further test, 
the binomial probability of observing none out of five stars in the $s$-poor group is 0.12, and none out of seven stars (if we include the two stars with anomalous position on the CMD) is 0.05.
These probablilties are small, however we remark here that they are not zero, and future analysis of larger sample of stars are necessary to increase the sample of halo stars with available chemistry.
As long as we refer to our analysed sample, the present results 
provide no strong evidence for the presence of $s$-rich stars, and indeed 
support a halo populated by just bSGB $s$-poor stars.
To make this statement more conclusive we need to improve the statistics in the future and eventually reduce our 0.12 probability to $\sim$0.00.
For these statistics we assume the bSGB and fSGB fractions based on $HST$ and ground-based photometry 
from Milone et al. (2009) that follow the SGB split from the cluster center out to 8\arcmin\ with no strong evidence 
for different radial distribution of the two branches. 
Beyond this distance out to the tidal radius, 
there are too few
SGB stars to identify the sequences and determine their contribution to the total cluster mass. 
On the other hand, Zoccali et al.\ (2009) found that the fSGB dramatically drops at a distance of $\sim$2.4\arcmin\ from the center.
We note that since we have detected $s$-rich/fSGB stars up to $\sim$5$\arcmin$ from the center, we confirm that there is no drop of fSGB stars at $\sim$2.4$\arcmin$.

\begin{figure*}
    \includegraphics[width=5.5cm]{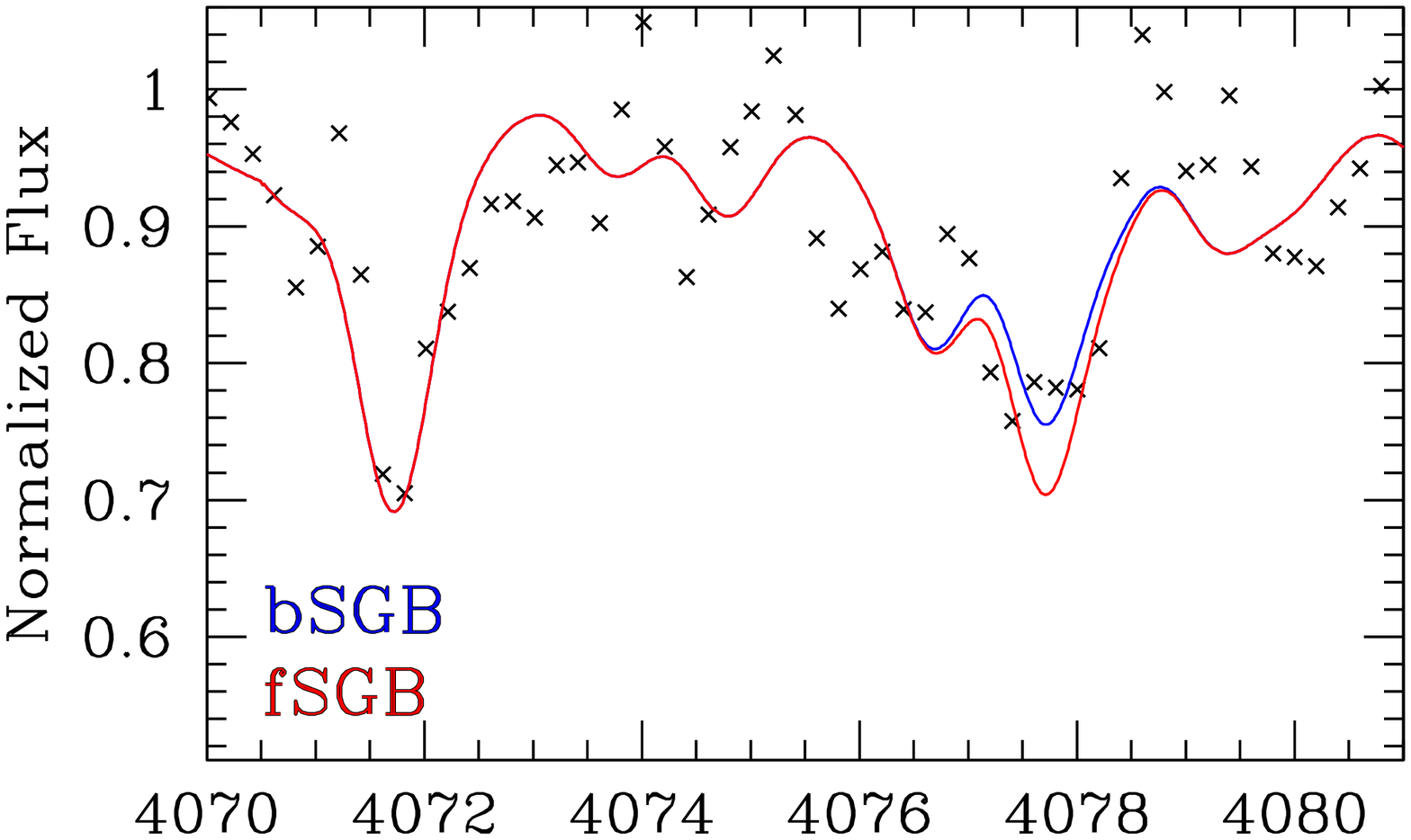}
    \includegraphics[width=5.5cm]{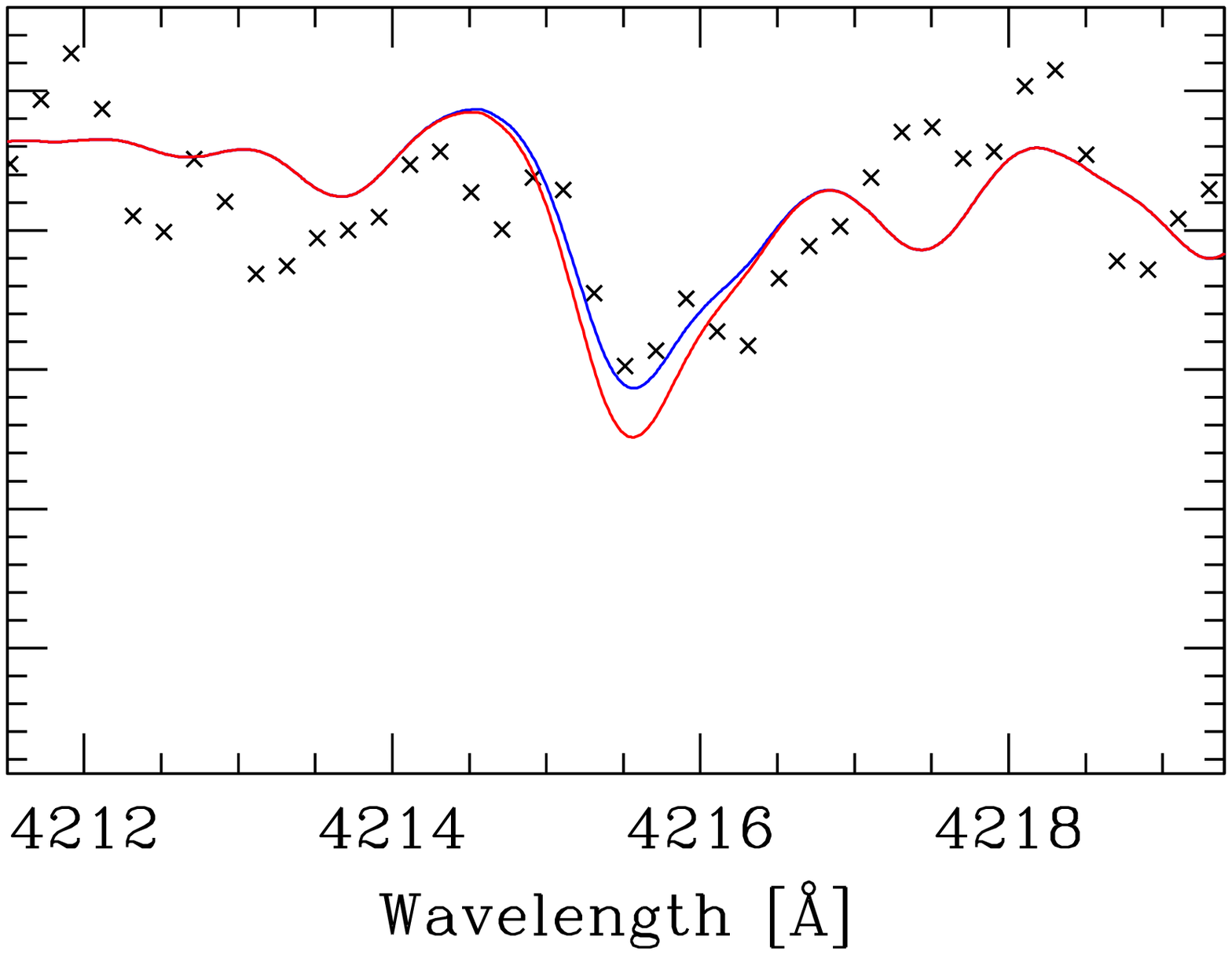}
    \includegraphics[width=5.5cm]{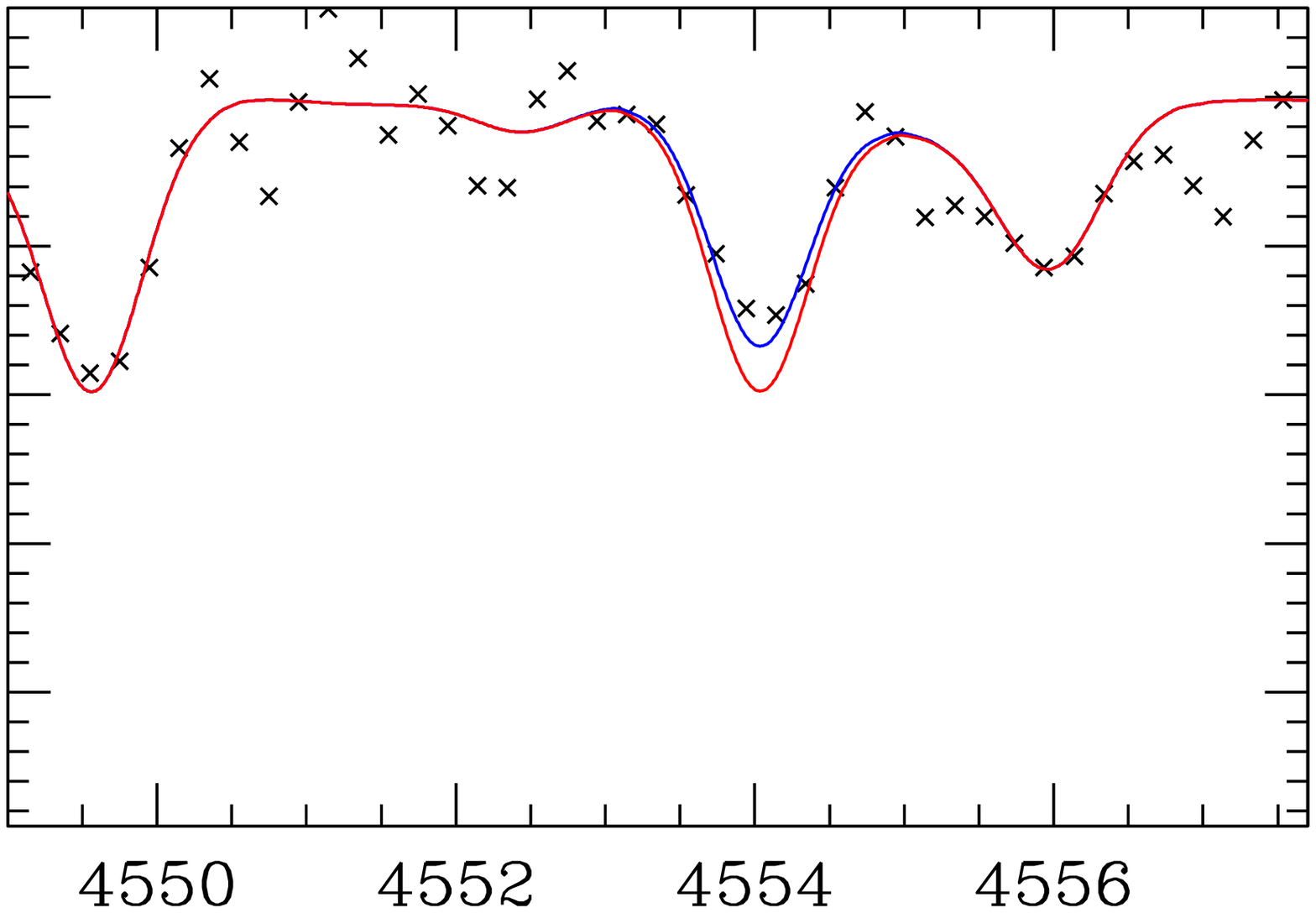}
    \caption{Observed spectrum obtained from the average of the MS halo stars for which the Sr and Ba individual abundances were not possible due to the low S/N. The observed spectrum, represented with black crosses, has been shown around the Sr (left and middle panels) and Ba (right panel) spectral features. The blue and red synthesis represent the average MOOG synthesis of the averaged stars for the mean Sr and Ba abundances inferred for the cluster bSGB and fSGB, respectively. }
    \label{fig:synthesis_avehalo}
   \end{figure*}

As abundances for Sr and Ba were possible for 7/23 stars, including two RGBs, we have 16 MS stars, including the four stars with only Sr or Ba abundance available, that do not have available abundances for both $n$-capture elements. To get chemical information for these 16 stars we combined their observed spectra by simply averaging them with the same weight, so that we obtained a spectrum with higher S/N. Then, we constructed two MOOG synthetic spectra by averaging those corresponding to the atmospheric parameters of each star: one with the mean Sr and Ba abundances obtained for the bSGB (mean bSGB synthetic spectrum), and the other with the mean Sr and Ba of the fSGB (mean fSGB synthetic spectrum). The comparison between the mean observed halo spectrum and the mean bSGB and fSGB synthetic spectra is shown in Fig.~\ref{fig:synthesis_avehalo}. The mean observed halo spectrum near the Sr and Ba spectral features is clearly best matched by the mean bSGB synthetic spectrum. The stars not analysed for individual abundances of Sr and Ba have predominantly the bSGB abundances.

Regarding the other elements, halo chemical composition agrees with that observed in the inner field of NGC\,1851.
As shown in Fig.~\ref{fig:all}, the distributions of all the elements in the halo are consistent with those observed in both the bSGB and fSGB, not being any significant difference between the abundances obtained for the two SGBs.
The halo stars are enhanced in $\alpha$ elements Ca and Mg, and roughly solar-scaled [Cr/Fe], although the 
rms in this latter species is high due to observational errors.
Carbon has been inferred for five stars, including T066 (not on the cluster sequence), the two red giants (T198 and T207), and two MS stars
(T186 and T207).
The comparison with the C abundances in the inner field can be done for just the two MS stars, as RGB stars have 
undergone the first dredge-up bringing carbon into the internal layers, resulting in lower surface abundances 
for this element.
The two MS stars T186 and T220 have [C/Fe]=$-$0.13 and $-$0.15, respectively, 
and the mean 
difference with the median values of the internal field are:
$\Delta$[C/Fe]$_{(\rm cluster-halo)}$=$+$0.01$\pm$0.03~dex; 
$\Delta$[C/Fe]$_{(\rm bSGB~cluster-halo)}$=$+$0.05$\pm$0.03~dex;
$\Delta$[C/Fe]$_{(\rm fSGB~cluster-halo)}$=$-$0.10$\pm$0.04~dex, that are consistent with both the bSGB and fSGB (see Tab.~\ref{tab:medie}). 

We conclude that, for all the analysed elements, the chemical composition of the halo is consistent with that observed in the cluster (as shown in Fig.~\ref{fig:all}). In particular the $s$-elements are consistent with the bSGB abundances. The similar abundance distributions for  all the elements with available measurements, is an additional signature for cluster membership, and more in general, for the existence of a halo surrounding NGC\,1851.

\section{The velocity dispersion radial profile}\label{subs:rvdisp}

Having established the existence of a halo of stars beyond the tidal radius of NGC\,1851, we can investigate the intrinsic RV dispersion ($\sigma_{\rm RV}$) of these outer stars in comparison with the inner field. 
To this aim we used the procedure described in Mackey et al.\,(2014; see their Sect.~4.2) that takes into account the contribution of observational errors to the RV dispersion. 
Briefly, we used a maximum likelihood technique, assuming that the measured RVs for our stars are normally distributed around the average value according to their measurement uncertainties and the intrinsic cluster velocity dispersion. We can obtain numerical estimates for the intrinsic cluster velocity dispersion by maximising the logarithm of the joint probability function for the observed RVs. The procedure has been done at three different bins of radial distances from the cluster center: namely in the central field of 3\arcmin$\times$3\arcmin\ covered by $HST$ photometry, in the inner field covered by ground-based photometry at a distance from the center between $\sim$4\arcmin\ and $\sim$9\arcmin, and for the stars in the outer field that covers the halo.

\begin{centering}
    \begin{figure}
     \includegraphics[width=8.2cm]{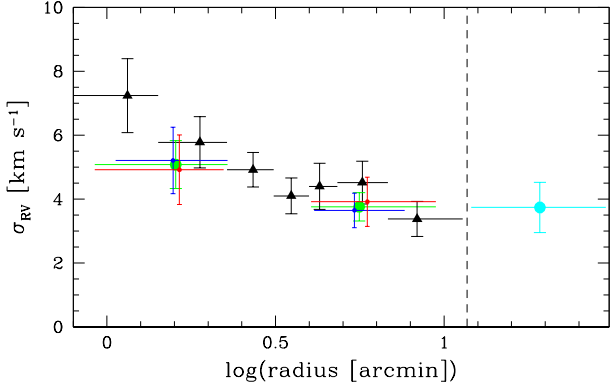}
     \caption{Intrinsic dispersions in RVs ($\sigma_{\rm RV}$) as a function of the 
       distance from the cluster center (in logarithmic units) for the stars in the 
       $HST$ central field (within 3\arcmin$\times$3\arcmin), the stars in the inner field 
       covered by ground-based photometry (between $\sim$4\arcmin\ and 9\arcmin\ from the cluster center), and 
       in the outer field (from the tidal radius up to $\sim$30\arcmin\ from the cluster center). 
       The color codes are as follows: green is for the inner field stars of NGC\,1851; 
       cyan for the halo of NGC\,1851; blue and red are the stars in the inner field divided in the bSGB and fSGB samples (see Sect.~\ref{sec:abb_clu}) respectively.
       The black triangles are values derived in Scarpa et al.\ (2011).
       The dashed line shows the location of the tidal radius.
     }
     \label{fig:RVdisp}
    \end{figure}
    \end{centering}

Figure~\ref{fig:RVdisp} shows the obtained intrinsic velocity dispersions in our three radial bins.
In the cluster field the velocity dispersion decreases with radius, from $\sim$5~\kmsec\ in $HST$ field, to $\sim$4~\kmsec\ between 4\arcmin\ and 9\arcmin. There is no significant difference between the bSGB and fSGB. 
Scarpa et al. (2011) determined the velocity dispersion radial profile from FLAMES spectra using high-resolution setups for 184 stars along the upper SGB and the lower RGB in the inner field of NGC\,1851. Their results are plotted, along with ours, in Fig.~\ref{fig:RVdisp}.
Our measurements agree with those reported by Scarpa and collaborators within 1~$\sigma$.
Out of the tidal radius we find a dispersion comparable with that observed in the region between 4\arcmin\ and 9\arcmin.

We note here that some physical and technical issues may affect our $\sigma_{RV}$ determination: 
\begin{itemize}

\item{Stellar masses: since most of the halo stars belong to the upper MS, while inner field stars are SGB, to properly compare the $\sigma_{RV}$ estimates at different radial distances we need to account for the different masses of the stars.
Lower mass stars should have a higher dispersion, typically, than higher mass stars. However, this effect is negligible in our sample, since the mass difference between upper MS and SGB is small. 
To quantify, under energy equipartition for two stars with masses $m_{1}$ and $m_{2}$ we have that $\sigma_{\rm RV_{2}}$=$\sigma_{\rm RV_{1}} \times \sqrt{m_{1}/m_{2}}$, that for a SGB with mass $m_1 \sim$0.8$M_{\odot}$ and a MS star with mass $m_2 \sim$0.7$M_{\odot}$ would be $\sigma_{\rm RV_2}$ larger by  $\sim$7\% than $\sigma_{RV_1}$; }
 
\item{ Field contamination: although we exclude that this is an issue
    in the inner field where we have much higher statistics, a low
    degree of field contamination ($\sim$2 stars out of 23) may affect
    our sample of NGC\,1851 RV-like stars in the halo, and may
      affect the value of $\sigma_{\rm RV}$. 
To estimate this effect we 
determined the radial velocity dispersion ($\sigma_{\rm RV}^{i,j}$) by excluding each pair of stars ({\it i,j}, with {\it i}=1,23 and {\it j}=1,23) from  
our NGC\,1851 halo sample. 
In the end, we obtained an average dispersion of 3.72~\kmsec\ (rms=0.22) with $\sigma_{\rm RV}$ ranging from 2.57 to 3.96~\kmsec;} 

\item { Binaries: binary contamination can inflate the dispersion.  As discussed in Sect.~\ref{sec:rv}, we do not expect a large fraction of binaries in NGC\,1851 (from Milone et al\,2012 the fraction of binaries in the central field outside the half-mass radius is $\sim$1.6\%). 
Of course, we cannot exclude that the halo stars may have, for some reason, 
a larger binary fraction.
For the NGC\,1851 halo sample we do not see any evidence for the presence of stars whose rms in RVs from different exposures significantly exceeds that introduced by the quality of the spectra (see Sect.~\ref{sec:rv} for more details);}

\item { Possible fiber-to-fiber systematic errors: 
while the plate-to-plate systematics are randomly distributed and are likely removed using our procedure, fiber-to-fiber effects may be there as the FLAMES fibers configuration that we used is the same for every exposure (that means each star is observed with the same fiber). Such effects are small, and never exceed $\sim$0.5~\kmsec\ (e.g., Sommariva et al.\, 2009).}

\end{itemize}

With due consideration of
 all the issues affecting the $\sigma_{\rm RV}$ estimates, we may speculate that 
the apparently continuous nature of the dispersion profile from the cluster to the halo provides 
yet more evidence that the halo is associated with the cluster.
We can also exclude a 
truncation in the observed $\sigma_{\rm RV}$ in the vicinity of the tidal radius, and that is contrary to the expectations from King-type theoretical models.
The bSGB and fSGB observed in the inner field are chemically different in the $s$-process elements, but they look kinematically similar.

\section{Discussion and concluding remarks}\label{sec:conclusions}

We have provided a spectroscopic analysis of the halo that surrounds NGC\,1851.
In particular we have studied the nature of this intriguing stellar structure
by analysing RVs and elemental abundances, providing
for the first time a chemical inventory in a halo surrounding a GC.

We have measured radial velocities and elemental abundances of
Fe, C, Mg, Ca, Cr, Sr and Ba in a sample of both halo and cluster stars, 
from which we draw the following conclusions: 

\begin{itemize}
\item{the RV distribution in the observed halo field shows a peak not expected from Galactic models at the characteristic RV of NGC\,1851, confirming the presence of a halo surrounding the cluster;}
\item{fifteen stars in the halo field exhibit radial velocities and metallicities consistent with the cluster field;}
\item{our observed RV distribution agrees (apart from the RV range of NGC\,1851) with that expected from Galactic models, suggesting that no other sub-structure, such as streams, is present in our field;}
\item{
the halo has the same metallicity distribution as NGC
1851. None of the NGC 1851 RV-like stars found outside the tidal
radius
shows a [Fe/H] content different from the range observed in the inner field;}
\item{the halo stars for which we could estimate abundances for Sr and Ba show abundances consistent with those observed on the
 bSGB, i.e., they have lower Sr and Ba compared with those of the fSGB;}
\item{our sample does not exhibit any strong evidence for the presence of stars with $s$-element abundances compatible with the fSGB.}
\end{itemize}

Within the multiple stellar population context, qualitatively, our observations are in agreement with a 
scenario in which: {\it (i)} the first stellar population of NGC\,1851 is made up of $s$-poor (bSGB) stars; 
{\it (ii)} the second generation is expected to form in the central region of the cluster possibly enriched in $s$-elements;
{\it (iii)} while the less radially concentrated first generation 
is lost early in the cluster evolution due to the expansion and stripping of the cluster outer layers resulting from early mass loss associated with first generation supernova ejecta (e.g., D'Ercole et al.\,2008).
If this scenario is correct, our observations constitute the
  first clear evidence that GCs lose primarily first-generation stars by
  evaporation mechanisms. The chemical composition of field stars in
  the halo, that is consistent with the first-generation stars in GCs,
  may mainly be due to the evaporation of first-generation GC stars
  into the field.
Although the central relaxation time of NGC\,1851 is small (of the order of $10^{7}$ yr), in the halo any dynamical time will be very long due to the low density.
So, while in the central part of the cluster the timescales are short enough that the population is mixed, 
in the external halo field, they are sufficiently long such that any initial gradient may have been preserved.

It is worth noticing that in the CMDs shown by
Milone et al. (2009) there is no obvious difference in the radial distributions of the bSGB and fSGB of NGC\,1851 up to $\sim$8$\arcmin$ from the center. However, their observations are limited to a small region well within the tidal radius of the cluster, and we cannot exclude different radial gradients at more external regions. 
If the fSGB is more centrally concentrated and the outer parts of the cluster are dominated by the bSGB, 
then since outer stars are more easily lost, that would be consistent with the bSGB dominance in the halo.  
We recall here that both the $s$-poor, and possibly also the $s$-rich groups in NGC\,1851 show 
internal variations in light elements (e.g., C, N, O, Na). 
The presence of sub-populations within each main $s$-group
makes the scenario more complex, challenging the identification of the sequence
of the various stellar bursts that gave origin to the 
multiple stellar populations in this GC. 
An exhaustive study of the radial distributions of the NGC\,1851 stellar groups should account also 
for the presence of these sub-populations.

The main question here is: why does NGC\,1851 
possess a stellar halo with these kinematic and chemical properties? 
In principle the presence of a halo may be simply consistent with a stellar system that 
is losing its external stars into the field. 
On the other hand, GCs are expected to follow a King profile, with the surface density drastically dropping at the tidal radius.
It has been well established that while King models succeed in describing the internal 
stellar surface-brightness profiles, they often fail in the outer regions of clusters, including the Milky Way GCs 
(e.g., McLaughlin \& van der Marel 2005; Carballo-Bello et al. 2012).
GCs surface-brightness profiles suggest that stellar clusters do not simply 
truncate on the King tidal radius, but have low density extended
halos.
All GCs should be stressed, to different
  extent, by interactions with the host galaxy. Hence, deviations from
  the King profile could not be surprising.
McLaughlin \& van der Marel (2005) have shown that the
 observed surface-brightness profiles of GCs is generally better reproduced by
a modified isothermal sphere based on an {\it ad hoc} stellar
distribution function developed for elliptical galaxies
by Wilson (1975).
The corresponding Wilson tidal radius of McLaughlin \& van der Marel (2005) is larger than that derived assuming a King profile, 
being $\sim$45\arcmin\ for NGC\,1851.
However, although we can formally fit the surface-brightness profiles of GCs in the regions beyond the
King radius adopting {\it ad hoc} analytical templates, 
in many cases it remains unclear why we observe such halo envelopes.
For NGC\,1851 the main question could hence not be why it possess a halo, but why it shows such a peculiarly uniform halo, without evident tidal features in the large photometric field analysed by Olszweski et al.\,(2009).

The NGC\,1851 halo may be a more interesting case, 
as it extends for 67\arcmin\ from the cluster center, that is much farther than any estimate for its tidal radius present in the literature.
Its very low density, as determined in Olswzeski et al., would make it easily perturbed by 
passages through the disk. 
According to Dinescu et al. (1999), NGC\,1851 is at a
  distance from the galactic center of 16.0~kpc, its orbit 
has apogalacticon and perigalacticon distances of $\sim$30~kpc and
$\sim$5~kpc, respectively, a high eccentricity of $\sim$0.7, and
passes through the disk of the Milky Way five times per Gyr.
Hence, it seems unlikely that an outer envelope of stars remains attached to the cluster for a Hubble time or, alternatively, that 
the cluster could shed sufficient stars to make such a huge envelope in half an orbit.  

Our spectroscopic sample of stars does not show any evidence of
  tidal streams. 
Although our observations allow us to draw this conclusion only for
a relatively small region of the halo, Olswzeski et al. did not find
any evidence for tidal streams in their large field of view.
This apparent lack of streams in the Olswzeski et al. analysis is puzzling. 
It is difficult to envisage a scenario that assumes that such an
extended low-density halo is bound to NGC\,1851, if the cluster has
been in orbit around the Milky Way for a significant portion of a Hubble time.  
However, even if it seems unlikely, we cannot 
exclude that, if the cluster has been in orbit around the Galaxy for a
long time, we are seeing a particular phase of the NGC\,1851 evolution:
e.g., a very extended envelope of stars that is in the process of
escaping from the cluster, but still bound to it. 
NGC\,1851 is currently at an intermediate distance from the
  galactic center between the apogalacton and perigalacticon, 
  and it is at a distance of 7.1~kpc from the Galactic plane (Djorgovski
1993), that is around the maximum distance from the plane allowed by
its orbit (Dinescu et al.\,1999). 
The halo may have formed after the last passage through the disk, and
will then be largely swept away next time the cluster crosses the disk and,
presumably, recreated as the cluster moves through the halo in the
half-orbit time before the next disk crossing. 
Future dynamical simulations may be enlightening in this regard,
  and should prove if such a scenario is plausible for the formation of the NGC\,1851 halo.

Alternatively, it is tempting to speculate that
we really are seeing a system that has been captured relatively recently 
and the envelope does represent the former dwarf galaxy. In this case, there just have not been enough orbits yet for the envelope to be completely stripped off by passages through the disk.  Hence it is possible that the halo population could look like the first generation in the cluster in the same way that the first generation in most globular clusters looks like the field halo population at the same [Fe/H]. 
If this hypothesis is correct, then we would expect that the halo population does not host stars enriched in Na because dwarf galaxies do not show Na-O anticorrelations. 

The first cluster proposed to be the nucleus of a dwarf tidally disrupted through Milky Way interactions was $\omega$~Centauri (Norris et al.\ 1996). 
Assuming a similar scenario, it would be less problematic to understand the large chemical variations, also in heavy elements, 
displayed by this cluster (e.g., Norris \& Da Costa 1995; Johnson \& Pilachowski 2010; Marino et al.\ 2011a).
Although less pronounced, NGC\,1851 shows chemical and photometric peculiarities similar to those of $\omega$~Centauri 
(e.g., Milone et al.\, 2008; Yong \& Grundahl 2008; Villanova et al.\ 2010; Carretta et al.\, 2010).
In contrast to NGC\,1851, for the case of $\omega$~Centauri
there is no evidence for stellar halo envelopes,
probably because its location relatively close the Galactic
  Centre and its tightly bound orbit.  These ensure that Galactic
  tidal forces at its location are relatively strong so that any
  loosely bound population is quickly removed preventing the survival
  of any such structure to the present-day (see Da Costa \& Coleman
  2008). If the accretion of $\omega$~Centauri has not occurred
  recently there has been time to remove the loosely bound
  outer envelope that could have originally surrounded the cluster.
On the other hand,
M\,54, that is the GC associated with the nucleus of the Sagittarius
dwarf galaxy, preserves an external envelope, that is the 
field population of the Sagittarius. 
Collectively, these facts may reinforce the conjecture that
{\it anomalous} GCs could represent the surviving nuclei of dwarf galaxies disrupted by tidal interactions with the Milky Way, 
and we can speculate that the halo surrounding NGC\,1851 is the remnant of
  the parent dwarf galaxy. 

To test the feasibility of this hypothesis, dynamical simulations taking into account the tidal interactions between the Milky Way and dwarfs are fundamental.
Recent simulations presented in Bekki \& Yong (2012) show that the halo in NGC\,1851 can be reproduced if this object formed in a dwarf galaxy environment, through the merger
of two clusters (corresponding to the $s$-rich and $s$-poor groups) that sink into the center of 
the host galaxy.
From their simulations, they expect to have three stellar populations in the halo:
the $s$-poor and the $s$-rich stars, plus a third population representing the 
field of the host dwarf.

Our results seem to not strongly favour
this idea. Assuming that each of the three populations should be 
equally represented in the halo, we have not found any clear evidence for the 
presence of either the $s$-rich stars or a third population differing from the metal distribution observed in NGC\,1851 to be
associated with a field population from the host galaxy.
If, following the Bekki scenario, we assume that the three populations are equally represented (e.g., fSGB=33\%; bSGB=33\%, and field=33\%), 
the probability to not observe any star with the Ba and Sr
  abundances of the fSGB population of NGC\,1851, even if not null, is
 low (see Sect.~5.2).
This suggests that,
even with our modest statistics, it is unlikely that there are three populations equally represented in the field, as in the Bekki \& Yong scenario. 

Furthermore, even if an internal
  metallicity  variation of the order of 0.10~dex in the halo (that we cannot see because of observational errors) cannot be ruled out, the metal distribution of the external field
 (15 stars) is very similar to that observed within the tidal
 radius. This favours the idea that the halo does not host stellar
 populations with metallicities too distinctive with respect to
 NGC\,1851, but we cannot completely exclude that the lack of such
 stars is instead introduced by photometric selection effects that can affect our sample.  
We note that Olswzeski et al.\ found evidence for the presence
of at least the two SGB populations (bSGB and fSGB) in the CMD of the NGC\,1851 halo.
However, it is possible that their fSGB may not correspond to the one observed in the cluster; given they 
found a significantly larger separation between the two SGBs than that
found out to 8\arcmin\ from the cluster center (Milone et al.\ 2009).

As previously discussed, the halo population we are actually observing could be either the $s$-poor population observed in the cluster or
the field population of the host galaxy, that may show 
the same metallicity and chemical properties of the cluster bSGB.
In the latter case, we don't have to see Bekki \& Yong's postulated 3rd population, as it may be the same as the cluster first generation and we may have to look at a lot of halo stars to see a different [Fe/H] distribution from that of the cluster first generation.
It is also worth noting that the dynamical simulations by Bekki \& Yong assume NGC\,1851 is 
the result of a merger between two clusters. 
It might be interesting to see if similar simulations assuming the NGC\,1851 various groups 
formed in a self-pollution scenario are able to better reproduce the observations.

The lack of a drop in the velocity dispersion profile is another piece of
  evidence that makes NGC\,1851 hardly compatible 
with the King model. A similar lack 
is intriguingly seen in M\,54, the
GC lying at the center of Sagittarius. 
This GC shows a velocity
dispersion profile that falls off with the distance from the cluster
center, and increases outside, due to either contamination from Sagittarius in the external part of the cluster or to tidal harassment from the host
  galaxy (Bellazzini et al.\,2008). 
In the hypothesis that the NGC\,1851 halo represents the parent galaxy
population, we would qualitatively expect a relatively high velocity dispersion for the halo.
On the other hand, 
if the halo is simply formed by stars evaporating from
NGC\,1851, it may be that the stars in the region outside the tidal
radius are undergoing some kind of heating, due to either tides and/or
unseen (dark) matter or modified Newtonian dynamics? We
  are not able to answer this question here.
We only note that, according to dynamical simulations
  presented in K{\"u}pper et al. (2010), the velocity dispersion of
  unbound escaping stars outside GC tidal radii should significantly increase and flattening
  due to tidal interactions with the
  Milky Way. The orbit of NGC\,1851 suggests that it is not at its
  perigalacticon and it is at its largest distance from the Galactic
  plane, 
  but according to K{\"u}pper et al.\, this effect,
  even if magnified at the perigalacticon, seems to be present at any location
  on the orbit. 
We note however that our sample of 23 stars outside the tidal radius may still be bound to the cluster.
Future investigations on the velocity dispersions in the halo of NGC\,1851 may give information on the nature of the NGC\,1851+halo system, assuming that measurements will be available at different radial locations in the halo and with better statistics. 

We conclude that the presence of such an extended  halo in NGC\,1851 
may suggest that the complex chemical enrichment in this GC 
has taken place in a dwarf whose nucleus is NGC\,1851.
Interestingly, the location of NGC\,1851 coincides with the disk of satellites, a relatively thin, highly inclined plane defined by the distribution of luminous Milky Way satellite galaxies (Kroupa et al.\,2005; Metz et al.\,2007; Pawlowski et al.\,2012; Pawlowski \& Kroupa\,2013).
It would be interesting in the future to see whether NGC\,1851 is co-orbiting with this structure or not.

It is conceivable that NGC\,1851's host galaxy was tidally disrupted by the collision and its stars were dispersed into the Milky Way halo in a similar way as what we observe with M54 and the Sagittarius dwarf today. The ultra low density halo of NGC\,1851 could therefore be the last trace of its ancient host.
Our spectroscopic analysis provides, for the first time, chemical abundances from low resolution data and
the identification of two low RGB halo stars lying on the fiducial of NGC\,1851, 
that may be useful to follow-up at higher resolution.
The present analysis is confined to a relatively small sample, that does not show strong evidence for  
$s$-enrichment. We stress however that in the future, we need to extend this analysis to larger fields to increase the statistics
and possibly confirm the lack/deficiency of fSGB stars in the halo that result from the present study.

\section*{acknowledgments}

\small
We warmly thank the referee Christian Moni Bidin for his 
helpful comments that improved the quality of the paper.
AFM, AD and MA have been supported by grants FL110100012 and DP120100991.
APM, HJ, GDC and JEN acknowledge the financial
support from the Australian Research Council through 
Discovery Project grant DP120100475.
\normalsize

\bibliographystyle{aa}

\begin{thebibliography}{}

\bibitem[Alonso et al.(1999)]{1999A&AS..140..261A} Alonso, A., Arribas, S., \& Mart{\'{\i}}nez-Roger, C.\ 1999, \aaps, 140, 261 

\bibitem[Alves-Brito et al.(2012)]{2012A&A...540A...3A} Alves-Brito, A., Yong, D., Mel{\'e}ndez, J., V{\'a}squez, S., \& Karakas, A.~I.\ 2012, \aap, 540, A3 

\bibitem[Anderson et al.(2008)]{2008AJ....135.2055A} Anderson, J., Sarajedini, A., Bedin, L.~R., et al.\ 2008, \aj, 135, 2055
\bibitem[Arlandini et al.(1999)]{1999ApJ...525..886A} Arlandini, C., K{\"a}ppeler, F., Wisshak, K., et al.\ 1999, \apj, 525, 886 
\bibitem[Bekki \& Freeman(2003)]{2003MNRAS.346L..11B} Bekki, K., \& Freeman, K.~C.\ 2003, \mnras, 346, L11 
\bibitem[Bekki \& Norris(2006)]{2006ApJ...637L.109B} Bekki, K., \& Norris, J.~E.\ 2006, \apjl, 637, L109 
\bibitem[Bekki \& Yong(2012)]{2012MNRAS.419.2063B} Bekki, K., \& Yong, D.\ 2012, \mnras, 419, 2063 
\bibitem[Bellazzini et al.(2008)]{2008AJ....136.1147B} Bellazzini, M., Ibata, R.~A., Chapman, S.~C., et al.\ 2008, \aj, 136, 1147 
\bibitem[Bellini et al.(2010)]{2010AJ....140..631B} Bellini, A., Bedin, L.~R., Piotto, G., et al.\ 2010, \aj, 140, 631 
\bibitem[Bessell \& Murphy(2012)]{2012PASP..124..140B} Bessell, M., \& Murphy, S.\ 2012, \pasp, 124, 140 
\bibitem[Carballo-Bello et al.(2012)]{2012MNRAS.419...14C} Carballo-Bello, J.~A., Gieles, M., Sollima, A., et al.\ 2012, \mnras, 419, 14 
\bibitem[Cardelli et al.(1989)]{1989ApJ...345..245C} Cardelli, J.~A., Clayton, G.~C., \& Mathis, J.~S.\ 1989, \apj, 345, 245 
\bibitem[Carretta et al.(2010)]{2010A&A...520A..95C} Carretta, E., Bragaglia, A., Gratton, R.~G., et al.\ 2010, \aap, 520, A95 
\bibitem[Cassisi et al.(2008)]{2008ApJ...672L.115C} Cassisi, S., Salaris, M., Pietrinferni, A., et al.\ 2008, \apjl, 672, L115 
\bibitem[Castelli \& Kurucz(2004)]{cas04} Castelli, F., \& Kurucz, R.~L.\ 2004, arXiv:astro-ph/0405087 

\bibitem[Collet et al.(2009)]{2009PASA...26..330C} Collet, R., Asplund, M., \& Nissen, P.~E.\ 2009, \pasa, 26, 330

\bibitem[Da Costa \& Coleman(2008)]{2008AJ....136..506D} Da Costa, G.~S., \& Coleman, M.~G.\ 2008, \aj, 136, 506 

\bibitem[Da Costa et al.(2009)]{2009ApJ...705.1481D} Da Costa, G.~S., Held, E.~V., Saviane, I., \& Gullieuszik, M.\ 2009, \apj, 705, 1481 

\bibitem[Da Costa et al.(2013)]{2013arXiv1312.5796D} Da Costa, G.~S., Held, E.~V., \& Saviane, I.\ 2013, arXiv:1312.5796 

\bibitem[D'Ercole et al.(2008)]{2008MNRAS.391..825D} D'Ercole, A., Vesperini, E., D'Antona, F., McMillan, S.~L.~W., \& Recchi, S.\ 2008, \mnras, 391, 825

\bibitem[Dinescu et al.(1999)]{1999AJ....117.1792D} Dinescu, D.~I., Girard, T.~M., \& van Altena, W.~F.\ 1999, \aj, 117, 1792 

\bibitem[Foreman-Mackey(2010)]{2010PhDT.......116F} Foreman-Mackey, D.~T.\  2010, Ph.D.~Thesis,  

\bibitem[Gallagher et al.(2010)]{2010A&A...523A..24G} Gallagher, A.~J., Ryan, S.~G., Garc{\'{\i}}a P{\'e}rez, A.~E., \& Aoki, W.\ 2010, \aap, 523, A24 

\bibitem[Gallart et al.(2003)]{2003AJ....125..742G} Gallart, C., Zoccali, M., Bertelli, G., et al.\ 2003, \aj, 125, 742 

\bibitem[Gilmore et al.(2012)]{2012Msngr.147...25G} Gilmore, G., Randich, S., Asplund, M., et al.\ 2012, The Messenger, 147, 25 

\bibitem[Goldsbury et al.(2011)]{2011AJ....142...66G} Goldsbury, R., Richer, H.~B., Anderson, J., et al.\ 2011, \aj, 142, 66 

\bibitem[Gratton et al.(2004)]{2004ARA&A..42..385G} Gratton, R., Sneden, C., \& Carretta, E.\ 2004, \araa, 42, 385 

\bibitem[Gratton et al.(2012)]{2012A&A...544A..12G} Gratton, R.~G., Villanova, S., Lucatello, S., et al.\ 2012, \aap, 544, A12 

\bibitem[Gray \& Corbally(1994)]{1994AJ....107..742G} Gray, R.~O., \& Corbally, C.~J.\ 1994, \aj, 107, 742 

\bibitem[Han et al.(2009)]{2009ApJ...707L.190H} Han, S.-I., Lee, Y.-W., Joo, S.-J., et al.\ 2009, \apjl, 707, L190 

\bibitem[Harris(1996)]{1996AJ....112.1487H} Harris, W.~E.\ 1996, \aj, 112, 1487 

\bibitem[Hauschildt et al.(1999a)]{1999ApJ...525..871H} Hauschildt, P.~H., Allard, F., Ferguson, J., Baron, E., \& Alexander, D.~R.\ 1999a, \apj, 525, 871 

\bibitem[Hauschildt et al.(1999b)]{1999ApJ...512..377H} Hauschildt, P.~H.,  Allard, F., \& Baron, E.\ 1999b, \apj, 512, 377 

\bibitem[Hill et al.(2002)]{2002A&A...387..560H} Hill, V., Plez, B., Cayrel, R., et al.\ 2002, \aap, 387, 560 

\bibitem[Johnson \& Pilachowski(2010)]{2010ApJ...722.1373J} Johnson, C.~I., \& Pilachowski, C.~A.\ 2010, \apj, 722, 1373 

\bibitem[Kappeler et al.(1989)]{1989RPPh...52..945K} Kappeler, F., Beer, H., \& Wisshak, K.\ 1989, Reports on Progress in Physics, 52, 945 

\bibitem[Kauffmann et al.(1993)]{1993MNRAS.264..201K} Kauffmann, G., White, S.~D.~M., \& Guiderdoni, B.\ 1993, \mnras, 264, 201 

\bibitem[King(1962)]{1962AJ.....67..471K} King, I.\ 1962, \aj, 67, 471 

\bibitem[Klypin et al.(1999)]{1999ApJ...522...82K} Klypin, A., Kravtsov, A.~V., Valenzuela, O., \& Prada, F.\ 1999, \apj, 522, 82 

\bibitem[Koch et al.(2004)]{2004AJ....128.2274K} Koch, A., Grebel, E.~K., Odenkirchen, M., Mart{\'{\i}}nez-Delgado, D., \& Caldwell, J.~A.~R.\ 2004, \aj, 128, 2274 

\bibitem[Kraft(1994)]{1994PASP..106..553K} Kraft, R.~P.\ 1994, \pasp, 106, 553 

\bibitem[Kroupa et al.(2005)]{2005A&A...431..517K} Kroupa, P., Theis, C., \& Boily, C.~M.\ 2005, \aap, 431, 517 

\bibitem[K{\"u}pper et al.(2010)]{2010MNRAS.407.2241K} K{\"u}pper, A.~H.~W., Kroupa, P., Baumgardt, H., \& Heggie, D.~C.\ 2010, \mnras, 407, 2241 

\bibitem[Ibata et al.(1994)]{1994Natur.370..194I} Ibata, R.~A., Gilmore, 
G., \& Irwin, M.~J.\ 1994, \nat, 370, 194 

\bibitem[Lardo et al.(2012)]{2012A&A...541A.141L} Lardo, C., Milone, A.~P., Marino, A.~F., et al.\ 2012, \aap, 541, A141 

\bibitem[Lardo et al.(2013)]{2013MNRAS.433.1941L} Lardo, C., Pancino, E., Mucciarelli, A., et al.\ 2013, \mnras, 433, 1941 

\bibitem[Lee et al.(2009)]{2009Natur.462..480L} Lee, J.-W., Kang, Y.-W., Lee, J., \& Lee, Y.-W.\ 2009, \nat, 462, 480 

\bibitem[Lodders(2003)]{2003ApJ...591.1220L} Lodders, K.\ 2003, \apj, 591, 1220 

\bibitem[McLaughlin \& van der Marel(2005)]{2005ApJS..161..304M} McLaughlin, D.~E., \& van der Marel, R.~P.\ 2005, \apjs, 161, 304 

\bibitem[Marino et al.(2008)]{2008A&A...490..625M} Marino, A.~F., Villanova, S., Piotto, G., et al.\ 2008, \aap, 490, 625 

\bibitem[Marino et al.(2009)]{2009A&A...505.1099M} Marino, A.~F., Milone, A.~P., Piotto, G., et al.\ 2009, \aap, 505, 1099 

\bibitem[Marino et al.(2011a)]{Na_OomegaCen} Marino, A.~F., Milone, A.~P., Piotto, G., et al.\ 2011a, \apj, 731, 64 

\bibitem[Marino et al.(2011b)]{CNO_M22} Marino, A.~F., Sneden, C., Kraft, R.~P., et al.\ 2011b, \aap, 532, A8 

\bibitem[Marino et al.(2012a)]{SGB_M22} Marino, A.~F., Milone, A.~P., Sneden, C., et al.\ 2012a, \aap, 541, A15 

\bibitem[Marino et al.(2012b)]{CNO_OmegaCen} Marino, A.~F., Milone, A.~P., Piotto, G., et al.\ 2012b, \apj, 746, 14 

\bibitem[Mashonkina \& Zhao(2006)]{2006A&A...456..313M} Mashonkina, L., \& Zhao, G.\ 2006, \aap, 456, 313 

\bibitem[Metz et al.(2007)]{2007MNRAS.374.1125M} Metz, M., Kroupa, P., \& Jerjen, H.\ 2007, \mnras, 374, 1125 

\bibitem[Milone et al.(2008)]{2008ApJ...673..241M} Milone, A.~P., Bedin, L.~R., Piotto, G., et al.\ 2008, \apj, 673, 241 

\bibitem[Milone et al.(2009)]{2009A&A...503..755M} Milone, A.~P., Stetson, P.~B., Piotto, G., et al.\ 2009, \aap, 503, 755 

\bibitem[Milone et al.(2012)]{2012A&A...540A..16M} Milone, A.~P., Piotto, G., Bedin, L.~R., et al.\ 2012, \aap, 540, A16 

\bibitem[Monelli et al.(2013)]{2013MNRAS.431.2126M} Monelli, M., Milone, A.~P., Stetson, P.~B., et al.\ 2013, \mnras, 431, 2126 

\bibitem[Moni Bidin et al.(2011)]{2011A&A...535A..33M} Moni Bidin, C., Mauro, F., Geisler, D., et al.\ 2011, \aap, 535, A33 

\bibitem[Moore et al.(1999)]{1999ApJ...524L..19M} Moore, B., Ghigna, S., Governato, F., et al.\ 1999, \apjl, 524, L19 

\bibitem[Norris \& Da Costa(1995)]{1995ApJ...447..680N} Norris, J.~E., \& Da Costa, G.~S.\ 1995, \apj, 447, 680 

\bibitem[Norris et al.(1996)]{1996ApJ...462..241N} Norris, J.~E., Freeman, K.~C., \& Mighell, K.~J.\ 1996, \apj, 462, 241 

\bibitem[Norris et al.(2013)]{2013ApJ...762...25N} Norris, J.~E., Bessell, M.~S., Yong, D., et al.\ 2013, \apj, 762, 25 

\bibitem[Odenkirchen et al.(2001)]{2001ApJ...548L.165O} Odenkirchen, M., Grebel, E.~K., Rockosi, C.~M., et al.\ 2001, \apjl, 548, L165 

\bibitem[Olszewski et al.(2009)]{2009AJ....138.1570O} Olszewski, E.~W., Saha, A., Knezek, P., et al.\ 2009, \aj, 138, 1570 

\bibitem[Pasquini et al.(2002)]{2002Msngr.110....1P} Pasquini, L., Avila, G., Blecha, A., et al.\ 2002, The Messenger, 110, 1 

\bibitem[Pawlowski \& Kroupa(2013)]{2013MNRAS.435.2116P} Pawlowski, M.~S., \& Kroupa, P.\ 2013, \mnras, 435, 2116 

\bibitem[Pawlowski et al.(2012)]{2012MNRAS.423.1109P} Pawlowski, M.~S., Pflamm-Altenburg, J., \& Kroupa, P.\ 2012, \mnras, 423, 1109 

\bibitem[Piotto et al.(2012)]{2012ApJ...760...39P} Piotto, G., Milone, A.~P., Anderson, J., et al.\ 2012, \apj, 760, 39 

\bibitem[Robin et al.(2003)]{2003A&A...409..523R} Robin, A.~C., Reyl{\'e}, C., Derri{\`e}re, S., \& Picaud, S.\ 2003, \aap, 409, 523 

\bibitem[Ryan et al.(1999)]{1999ApJ...523..654R} Ryan, S.~G., Norris, J.~E., \& Beers, T.~C.\ 1999, \apj, 523, 654 

\bibitem[Sarajedini et al.(2007)]{2007AJ....133.1658S} Sarajedini, A., Bedin, L.~R., Chaboyer, B., et al.\ 2007, \aj, 133, 1658 

\bibitem[Sbordone et al.(2011)]{2011A&A...534A...9S} Sbordone, L., Salaris, M., Weiss, A., \& Cassisi, S.\ 2011, \aap, 534, A9 

\bibitem[Scarpa et al.(2011)]{2011A&A...525A.148S} Scarpa, R., Marconi, G., Carraro, G., Falomo, R., \& Villanova, S.\ 2011, \aap, 525, A148 

\bibitem[Schlegel et al.(1998)]{1998ApJ...500..525S} Schlegel, D.~J., Finkbeiner, D.~P., \& Davis, M.\ 1998, \apj, 500, 525 

\bibitem[Simmerer et al.(2013)]{2013ApJ...764L...7S} Simmerer, J., Ivans, I.~I., Filler, D., et al.\ 2013, \apjl, 764, L7 

\bibitem[Smith et al.(2000)]{2000AJ....119.1239S} Smith, V.~V., Suntzeff, N.~B., Cunha, K., et al.\ 2000, \aj, 119, 1239 

\bibitem[Sneden(1973)]{1973ApJ...184..839S} Sneden, C.\ 1973, \apj, 184, 839

\bibitem[Sollima et al.(2012)]{2012MNRAS.426.1137S} Sollima, A., Gratton, R.~G., Carballo-Bello, J.~A., et al.\ 2012, \mnras, 426, 1137 

\bibitem[Sommariva et al.(2009)]{2009A&A...493..947S} Sommariva, V., Piotto, G., Rejkuba, M., et al.\ 2009, \aap, 493, 947 

\bibitem[Stetson(2000)]{2000PASP..112..925S} Stetson, P.~B.\ 2000, \pasp, 112, 925 

\bibitem[Stetson(2005)]{2005PASP..117..563S} Stetson, P.~B.\ 2005, \pasp, 117, 563 

\bibitem[Trager et al.(1993)]{1993ASPC...50..347T} Trager, S.~C.,  Djorgovski, S.,  \& King, I.~R.\ 1993, Structure and Dynamics of Globular Clusters, 50, 347 

\bibitem[Ventura et al.(2009)]{2009MNRAS.399..934V} Ventura, P., Caloi, V., D'Antona, F., et al.\ 2009, \mnras, 399, 934 

\bibitem[Ventura et al.(2014)]{2014MNRAS.437.3274V} Ventura, P., Criscienzo, M.~D., D'Antona, F., et al.\ 2014, \mnras, 437, 3274 

\bibitem[Villanova et al.(2010)]{2010ApJ...722L..18V} Villanova, S., Geisler, D., \& Piotto, G.\ 2010, \apjl, 722, L18 

\bibitem[Wilson(1975)]{1975AJ.....80..175W} Wilson, C.~P.\ 1975, \aj, 80, 175 

\bibitem[Yong \& Grundahl(2008)]{2008ApJ...672L..29Y} Yong, D., \& Grundahl, F.\ 2008, \apjl, 672, L29 

\bibitem[Yong et al.(2009)]{2009ApJ...695L..62Y} Yong, D., Grundahl, F., D'Antona, F., et al.\ 2009, \apjl, 695, L62 

\bibitem[Yong et al.(2014)]{2014arXiv1404.6873Y} Yong, D., Roederer, I.~U., 
Grundahl, F., et al.\ 2014, arXiv:1404.6873 

\bibitem[Zoccali et al.(2009)]{2009ApJ...697L..22Z} Zoccali, M., Pancino, E., Catelan, M., et al.\ 2009, \apjl, 697, L22 

\end{thebibliography}

\end{document}